\documentclass[a4paper,11pt]{article}
\usepackage{jheppub}
\usepackage[utf8]{inputenc}

\usepackage{mathtools, tensor}
\usepackage[dvipsnames]{xcolor}
\usepackage{enumitem}
\usepackage{microtype}
\usepackage{braket}
\usepackage{tipa}

\newcommand{\RGB}{R_{\text{GB}}}                
\newcommand{\cGB}{c_{\text{GB}}}                
\newcommand{\cRf}{c_{\text{R4}}}                
\newcommand{\cRt}{c_{\text{R3}}}               
\newcommand{\Mpl}{M_{\rm Pl}}                   
\newcommand{\lp}{\nabla^2_d}                    

\newcommand{\hh}{\text{\texththeng}}

\newcommand{\sd}{\text{d}}

\DeclarePairedDelimiter\abs{\lvert}{\rvert}     

\makeatletter
\DeclareFontFamily{OMX}{MnSymbolE}{}
\DeclareSymbolFont{MnLargeSymbols}{OMX}{MnSymbolE}{m}{n}
\SetSymbolFont{MnLargeSymbols}{bold}{OMX}{MnSymbolE}{b}{n}
\DeclareFontShape{OMX}{MnSymbolE}{m}{n}{
    <-6>  MnSymbolE5
   <6-7>  MnSymbolE6
   <7-8>  MnSymbolE7
   <8-9>  MnSymbolE8
   <9-10> MnSymbolE9
  <10-12> MnSymbolE10
  <12->   MnSymbolE12
}{}
\DeclareFontShape{OMX}{MnSymbolE}{b}{n}{
    <-6>  MnSymbolE-Bold5
   <6-7>  MnSymbolE-Bold6
   <7-8>  MnSymbolE-Bold7
   <8-9>  MnSymbolE-Bold8
   <9-10> MnSymbolE-Bold9
  <10-12> MnSymbolE-Bold10
  <12->   MnSymbolE-Bold12
}{}

\let\llangle\@undefined
\let\rrangle\@undefined
\DeclareMathDelimiter{\llangle}{\mathopen}%
                     {MnLargeSymbols}{'164}{MnLargeSymbols}{'164}
\DeclareMathDelimiter{\rrangle}{\mathclose}%
                     {MnLargeSymbols}{'171}{MnLargeSymbols}{'171}
\makeatother

\renewcommand{\[}{\left[} 
\renewcommand{\]}{\right]}
\renewcommand{\(}{\left(}
\renewcommand{\)}{\right)}

\def\be{\begin{equation}}
\def\ee{\end{equation}}
\def\ba{\begin{eqnarray}}
\def\ea{\end{eqnarray}}
\def\nn{\nonumber}

\def\p{\partial}
\def\d{\mathrm{d}}

\def\({\left(}
\def\){\right)}
\def\mpl{M_{\rm Pl}}

\title{Surfin' pp-waves with Good Vibrations: Causality in the presence of stacked shockwaves
}

\author[a]{Calvin Y.-R. Chen,}
\author[a,b]{Claudia de Rham,}
\author[a]{Aoibheann Margalit}
\author[a,b]{and Andrew J. Tolley}

\affiliation[a]{Theoretical Physics, Blackett Laboratory, Imperial College, London, SW7 2AZ, UK}
\affiliation[b]{Perimeter Institute for Theoretical Physics,
31 Caroline St N, Waterloo, Ontario, N2L 6B9, Canada}

\emailAdd{calvin.chen16@imperial.ac.uk}
\emailAdd{c.de-rham@imperial.ac.uk}
\emailAdd{a.margalit19@imperial.ac.uk}
\emailAdd{a.tolley@imperial.ac.uk}

\abstract{Relativistic causality constrains the $S$-matrix both through its analyticity, and by imposing lower bounds on the scattering time delay. These bounds are easiest to determine for spacetimes which admit either a timelike or null Killing vector. We revisit a class of pp-wave spacetimes and carefully determine the scattering time delay for arbitrary incoming states in the eikonal, semi-classical, and Born approximations. We apply this to the EFT of gravity in arbitrary dimensions. It is well-known that higher-dimension operators such as the Gauss-Bonnet term, when treated perturbatively at low energies, can appear to make both positive and negative contributions to the time delays of the background geometry. We show that even when multiple shockwaves are stacked, the corrections to the scattering time delay relative to the background are generically unresolvable within the regime of validity of the effective field theory so long as the Wilson coefficients are of order unity. This is in agreement with previously derived positivity/bootstrap bounds and the requirement that infrared causality be maintained in consistent low-energy effective theories, irrespective of the UV completion.}

\begin{document}

\maketitle

\setlength{\parskip}{1em}                   


\section{Introduction}

Causality is known to be a powerful constraint on relativistic quantum theories. For scattering in Minkowski spacetime, it is well-known to lead to powerful analyticity properties which have, in recent years, been put to significant use in deriving positivity/bootstrap bounds. 
Furthermore, relativistic causality also constrains correlation functions with similarly fruitful implications. 
For a recent review see \cite{deRham:2022hpx}.  
A closely related result, which has an equally long history, is that causality should impose lower bounds on the time delay incurred during a scattering process \cite{Eisenbud:1948paa,Wigner:1955zz}. 
This has been well-studied in non-relativistic theories \cite{DECARVALHO200283} and has, more recently, been applied particularly to pp-wave spacetimes \cite{Camanho:2014apa}. 
One of the advantages of computing time delays is that it does not require analyticity and is therefore potentially applicable to generic spacetimes and field theory backgrounds. 
In this regard, it has recently been shown that imposing a lower bound on scattering time delays \cite{CarrilloGonzalez:2022fwg,CarrilloGonzalez:2023cbf} can either be complementary to or reproduce many of the positivity bounds previously obtained by a subtle combination of analyticity and crossing symmetry arguments \cite{Tolley:2020gtv,Caron-Huot:2020cmc,
Sinha:2020win,Haldar:2021rri,Raman:2021pkf,Chowdhury:2021ynh}.

Causality in gravitational theories is well known to be more subtle, where already positivity/bootstrap bounds are known to be weakened \cite{Bjerrum-Bohr:2014lea,Alberte:2020jsk,Alberte:2020bdz,Alberte:2021dnj,Bern:2021ppb,Caron-Huot:2022ugt,Herrero-Valea:2022lfd,Noumi:2022zht,Henriksson:2022oeu,Chiang:2022jep,deRham:2022gfe,Caron-Huot:2022jli,Hamada:2023cyt,Aoki:2023khq}. 
This is in part because, in a gravitational effective field theory (EFT), the low-energy lightcones of different species are known to be affected by the background, including for gravitational waves themselves \cite{Drummond:1979pp,Hollowood:2007kt,Hollowood:2007ku,Hollowood:2008kq,Hollowood:2009qz,Hollowood:2010bd,Hollowood:2010xh,Hollowood:2011yh,Hollowood:2012as,Reall:2014pwa,Papallo:2015rna,Hollowood:2015elj,Goon:2016une,deRham:2020zyh,deRham:2020ejn,deRham:2019ctd,deRham:2021bll}.
Nevertheless, consistent low-energy EFTs which emerge from fundamentally Lorentz-invariant UV completions should still respect relativistic causality in some form --- even with gravity. 
A minimal requirement that seems to satisfy this is the asymptotic causality requirement of \cite{Camanho:2014apa,Hinterbichler:2017qyt,Bonifacio:2017nnt,Hinterbichler:2017qcl,AccettulliHuber:2020oou,Bellazzini:2021shn,Bittermann:2022hhy,Bellazzini:2022wzv}, which imposes positivity of the total time delay experienced by propagating states relative to the asymptotic Minkowski geometry --- more precisely any negativity should be within the confines of the uncertainty principle. 

\paragraph{Time Delay ---}On a spacetime with a timelike Killing vector $\partial_t$ and associated conserved energy $E$, such as in the stationary black hole background considered in \cite{Chen:2021bvg}, the Eisenbud-Wigner time delay \cite{Eisenbud:1948paa,Wigner:1955zz,Smith:1960zza,Martin:1976iw} can be identified with the $E$-derivative of the asymptotic scattering phase shift: $\Delta T = 2 \partial \delta/\partial E$.
In what follows, we shall instead be interested in pp-wave spacetimes, where $\partial_v$ is a Killing vector for the null time $v$ and the associated conserved quantity is momentum in the null $v$-direction $k_v$. 
It is then natural to define the analogue null time delay via $\Delta v = 2 \partial \delta/\partial k_v$, although we will give a more formal definition of both, applicable to arbitrary incoming states, in Section \ref{sec:ScatteringTimeDelay}. In both cases we shall formally define a time delay operator which is the quantum conjugate variable to the conserved energy operator and explain how this is not in contradiction with Pauli's theorem.

\paragraph{Asymptotic Causality ---} In a gravitational theory all species couple to gravity, so the generic form of the time delay has a contribution from the Shapiro delay due to the spacetime metric and an additional delay due to local non-gravitational interactions. When working at tree-level in a low-energy EFT, this may be split as
\be
\Delta T = \Delta T_{\rm Shapiro} + \Delta T_{\rm EFT} \, ,
\ee
with an analogous split for the null time delay $\Delta v $. 
The causal properties of an EFT operator can be inferred from its contribution to the time delay experienced by propagating metric perturbations. 
Asymptotic causality is the statement that $\Delta T$, the total time delay relative to a flat Minkowski geometry is positive, or more precisely is never resolvably negative
\be
\Delta T \gtrapprox - E^{-1} \, .
\ee
The allowed negativity is simply the reflection of the usual uncertainty principle and is explicitly built into the bound of Eisenbud and Wigner \cite{Eisenbud:1948paa,Wigner:1955zz}. 

\paragraph{Infrared Causality ---} A more precise proposal, dubbed infrared causality in \cite{Chen:2021bvg}, is that consistent EFTs should further require
\be
\Delta T_{\rm EFT} \gtrapprox - E^{-1} \, .
\ee
This was demonstrated for FLRW spacetimes in \cite{deRham:2019ctd,deRham:2020zyh} and spherically symmetric black hole spacetimes in \cite{Chen:2021bvg,deRham:2021bll}. 
In the latter case, it was shown in \cite{Chen:2021bvg} that infrared causality is the requirement that reproduces known positivity/bootstrap bounds, while asymptotic causality leads to weaker bounds. 
In the present work we will extend the discussion to pp-wave spacetimes which can be stacked to add multiple negative contributions to the time delay. 

\paragraph{Stacking Unresolvable Contributions ---} Many authors impose strict positivity $\Delta T>0$ on the grounds that, if a set-up were found which can give even a perturbatively small negative contribution to the time delay, it would be possible to stack these to make a negative and arbitrarily large and hence resolvable time delay. 
A common argument for this is that given in  \cite{goldberger1962concerning}: Consider a sequence of $N$ scattering events separated by times sufficiently large that we can safely assume the $n$'th scattering has taken place between $t_n$ and $t_{n+1}$. The total unitary evolution in interaction picture may be split into a time-ordered product 
\be
\hat U(t_{N},t_0) =\hat U(t_N,t_{N-1})  \hat U(t_{N-1},t_{N-2}) \dots =   {\cal T}\ \prod_{n=1}^N \hat U(t_n,t_{n-1}) \, .
\ee
Given the assumption that the times $t_{n}-t_{n-1}$ are sufficiently long relative to the scales of interactions, the states approach asymptotic states and we may regard $\hat U(t_n,t_{n-1}) $ as the $S$-matrix for the $n$'th scattering event: $\hat S_{n}= \hat U(t_n,t_{n-1})$.
Then, the total $S$-matrix is
\be
\hat S_{\rm total} =  {\cal T} \prod_{n=1}^N \hat S_n \, .
\ee
Assuming each scattering event is identical, $\hat S_n=\hat S_1$, then we have $\hat S_{\rm total} =(\hat S_1)^N$ so that the total time delay is simply
\be
\Delta T_{N} = N \Delta T_{1} \, .
\label{eq:NT1}
\ee
On first sight, this may seem to na\"ively suggest that one could make any small (unresolvable) time advance as large (and hence resolvable) as we like by simply making $N$ arbitrarily large. 
If correct, this argument would indicate that a resolvable (and hence physically observable) time advance can be produced by accumulating a large number of unresolvable (unphysical) effects.

As we shall see, there are multiple issues with the previous arguments. 
First of all, the previous result \eqref{eq:NT1} is only true over the time period for which we can neglect the process of quantum diffusion, \textit{i.e.} the dynamics of the free Hamiltonian $\hat H_0$. 
As we shall show explicitly in what follows, the key point is that \textit{e.g.} even when the interaction Hamiltonian $\hat H_1(t)=\hat H(t) -\hat H_0$ takes the form
\be
\hat H_{1}(t) = \sum_{n=1}^N \delta(t-a-t_{n-1}) \hat K \, ,
\ee
with $0<a<t_n-t_{n-1}$ for fixed $\hat K$, the $S$-matrix for each scattering (being defined in interacting picture) is
\be
\hat S_{n} = e^{i \hat H_0 (a+t_{n-1})} e^{-i \hat K}e^{-i \hat H_0 (a+t_{n-1})}  \, .
\ee
Given $\hat K$ and $\hat H_0$ necessarily do not commute, each scattering event cannot be regarded as identical and $\hat S_{\rm total} \neq (\hat S_1)^N$. 
The physical effect of the free evolution $e^{i \hat H_0 (a+t_{n-1})} $ is to diffuse the incoming state, and it is due to this diffusion that each scattering event is not identical to the previous one.
The quantum diffusion builds up, eventually putting a limit on the total time delay that can be accumulated\footnote{Mathematically we can accommodate this by considering the Hamiltonian $\hat H(t) = \hat H_0+\sum_{n=1}^N \delta(t-a-t_{n-1}) e^{-i \hat H_0 (a+t_{n-1})} \hat K e^{i \hat H_0 (a+t_{n-1})} $ so that the interacting picture interaction is $\hat H_1^{\rm int}(t) =\hat K$ and $\hat S_{\rm total}= e^{-iN\hat  K}$, but this cannot arise in a local theory and is tantamount to defining a theory by its $S$-matrix with no reference to locality.}. 
We shall see later precisely how to account for the effect of the quantum diffusion.

\paragraph{Null Time delay in Stacking Configurations \& Main Result ---} One of the central virtues of pp-wave geometries is that it is easy to superpose them, thereby allowing the construction of exact solutions which describe the previously discussed sequence of scattering events. 
In this paper we shall be primarily concerned with these stacked shockwaves or, more generally, smooth pp-wave geometries. 
In what follows we give a precise definition of the null time delay appropriate for arbitrary initial states and pp-wave spacetimes, which is more general than the approximate impact parameter expression whose applicability is restricted to eikonal scattering, and show how to compute this in the \textit{eikonal}, \textit{semi-classical}, and \textit{Born} approximations. 
We use this to investigate the background configurations for the gravitational EFTs considered in \cite{Camanho:2014apa} and carefully account for the regime of validity of the EFT. 
In physically sensible situations, we find that, in addition to asymptotic causality being satisfied, the generically stronger condition of infrared causality is also satisfied, provided we remain in the regime of validity of the EFT and assuming the Wilsonian coefficient of the Gauss-Bonnet (GB) term is bounded $|c_{\rm GB}|\lesssim \mathcal{O}(1)$\footnote{This complements the results found in  \cite{Caron-Huot:2022jli} using positivity bounds, where the coefficient of the GB-term was shown to be bounded by $|c_{\rm GB}|\lesssim \mathcal{O}(10)$ in five dimensions and $|c_{\rm GB}|\lesssim \mathcal{O}(5)$ in higher dimensions.}. 
This statement does not rely on a specific UV completion, say string theory  or a weakly coupled tree-level completion. 
It also holds despite the superficial device of being able to stack multiple shockwaves together to accumulate time advances. 
It is the combined effect of diffusion and scattering which is responsible.
We highlight in particular how, in the balancing shockwaves case (for which, classically, scattering is absent), the two effects combine in a non-trivial way to render any would-be time advance unresolvable.

\paragraph{Outline ---} The rest of this work is organised as follows: In Section~\ref{sec: pp-waves}, we review pp-wave spacetime solutions in the EFT of gravity in arbitrary dimensions. While pp-waves are exact vacuum solutions of any EFT of gravity independent of their precise shape, the same no longer holds once metric fluctuations (\textit{i.e.} gravitational waves) are considered. 
Ensuring that the EFT of gravity remains under control in the presence of small infinitesimal fluctuations leads to a precise regime of validity beyond which the EFT can no longer be trusted. 
With the EFT under control, we then introduce the standard Eisenbud-Wigner-Smith time delay in Section~\ref{sec:ScatteringTimeDelay} and its null analogue. 
We emphasize the role of scattering and quantum diffusion, and the relation with the standard eikonal approximation, which is typically considered in the literature. 
With this clarification in hand, we start by computing the time delay in the eikonal approximation in Section~\ref{subsec: pt source scattering}, establishing the scattering from a single source before considering a (possibly continuous) succession of point sources.
Then we show that, within the regime of validity of the EFT, the time advance remains unresolvable so long as the coefficient of the GB-term is at most of order unity.
In Section~\ref{subsec: balancing sources}, we turn our attention to the engineered situation where sources are carefully positioned so as to balance each other, leading to a local extremum in the potential between two sets of pp-waves. 
In this case, the instability of the potential leads to a maximum time advance which we prove is always either unresolvable or indistinguishable from the general relativity (GR) contribution within the regime of validity of the EFT. 
We then move beyond the eikonal approximation and consider scattering in the semi-classical approximation in Section~\ref{semiclassical1}, highlighting features that would be impossible to diagnose in the eikonal limit. 
For completeness, we also compute the time delay in perturbation theory in Section~\ref{sec: quantum calcs}. 
The diffusion and scattering are identified in this limit and we show how they compete with one another, leading to an unresolvable time advance in the regime of validity of the EFT. Section~\ref{sec:Conclusion} provides a summary of our main results.
Technical details on the perturbations are left to the Appendices~\ref{app: full field eqs}--\ref{app: expressions for a's}.

\paragraph{Conventions ---} We work in units where $\hbar=c=1$, and in mostly-plus signature $(-,+,\dots,+)$.
The pp-wave metric $g_{\alpha \beta}$ in so-called Brinkmann coordinates is given by
\begin{equation}\label{eq: pp wave metric}
    g_{\alpha \beta}\sd x^{\alpha} \sd x^{\beta} = 2\sd u \sd v + H(u,\mathbf{x}) \sd u^2 +\delta_{ij} \sd x^i \sd x^j,
\end{equation}
where $\mathbf{x}$ are coordinates on the $d = (D-2)$-dimensional Euclidean transverse subspace.
Tensors of the full $D$-dimensional manifold are indexed with letters from the Greek alphabet $\{\alpha,\beta,\dots\}$.
The Laplace-Beltrami operator is written as $\Box = g^{\alpha \beta}\nabla_{\alpha}\nabla_{\beta}$.
It will also be useful to define the Laplace-Beltrami operator as it would act only on a scalar, written as $\tilde{\Box} = g^{\alpha \beta}\partial_{\alpha}\partial_{\beta} = 2\partial_u\partial_v - H\partial_v^2 + \partial_i\partial_i$.
Tensors on the $d$-dimensional transverse Euclidean space are usually indexed with letters from the middle of the Roman alphabet $\{i,j,\dots\}$ when expressed in Cartesian coordinates.
No distinction is made between upper and lower indices on the transverse space in these coordinates.
The Laplacian on the transverse space is $\lp = \delta^{ij}\partial_i\partial_j \equiv \partial_i \partial_i$.

On occasion, we will discuss a particular pp-wave solution which is spherically symmetric in the transverse directions $H(u,\mathbf{x}) = H(u,r)$.
In that case, we will use spherical coordinates,
\begin{equation}\label{eq: flat metric in spherical coordinates}
    \delta_{ij}\sd x^i \sd x^j = \sd r^2 + r^2 \gamma_{ab} \sd \theta^a \sd \theta^b,
\end{equation}
with $r^2 = \mathbf{x}^2$ and $\gamma_{ab}$ the metric on the $(d-1)$-sphere.
Tensors on the $(d-1)$-sphere are indexed with letters from the start of the Roman alphabet $\{a,b,\dots\}$, with the covariant derivative represented by $\hat{D}_a$ and the Laplace-Beltrami operator by $\hat{\Delta}_{d-1} = \gamma^{ab}\hat{D}_a\hat{D}_b$.


\section{pp-waves in the EFT of gravity}\label{sec: pp-waves}
Every metric, with a choice of null geodesic, can be associated to a pp-wave metric via the Penrose limit process \cite{Penrose1976}.
This means that they can be used as analogues to study the physics of systems where the exact metric is unknown.
For example, a pp-wave metric \eqref{eq: pp wave metric} with appropriately localised singularities in $H(u,\mathbf{x})$ is analogous to a multi-black hole spacetime via an Aichelburg-Sexl boost \cite{Aichelburg1971}.
By this association, the (a)causality of an EFT operator on a pp-wave metric is actually reflective of its (a)causality in a much broader class of situations.

The pp-wave metric is an exact solution not only in GR but also to all orders in the EFT of gravity, since all higher curvature invariants vanish \cite{Horowitz:1989bv}. 
Remarkably this is true even for the Aichelburg-Sexl shockwaves despite their delta function singularity. 
Since all irrelevant operators vanish, there is na\"ively no constraint on the amplitude or shape of a pp-wave in order to be consistently treated within a given EFT.

However, this is no longer true as soon as the pp-wave metric is perturbed. 
A wave of arbitrarily small amplitude travelling in the opposite direction can potentially lead to an arbitrarily large curvature once it hits the wavefront of the shockwave, at which point it can no longer be described within the EFT.
This fact is enough to constrain not only the energy of gravitational perturbations, but also the energy of the background pp-wave metric on which they propagate, in a similar fashion to \cite{Chen:2021bvg}.
In this section, we will describe some properties of the background pp-wave metric before introducing perturbations and their dynamics. 
These can then be used to ascertain the constraints from requiring that a given EFT is under control, which in turn will impose restrictions on the form of the pp-wave metric. 
These constraints will play a crucial role in our later discussion of scattering time delays.

\subsection{Properties and examples of the pp-wave metrics}\label{subsec: properties of pp-waves}

\begin{figure}
    \centering
    \includegraphics[width=\linewidth]{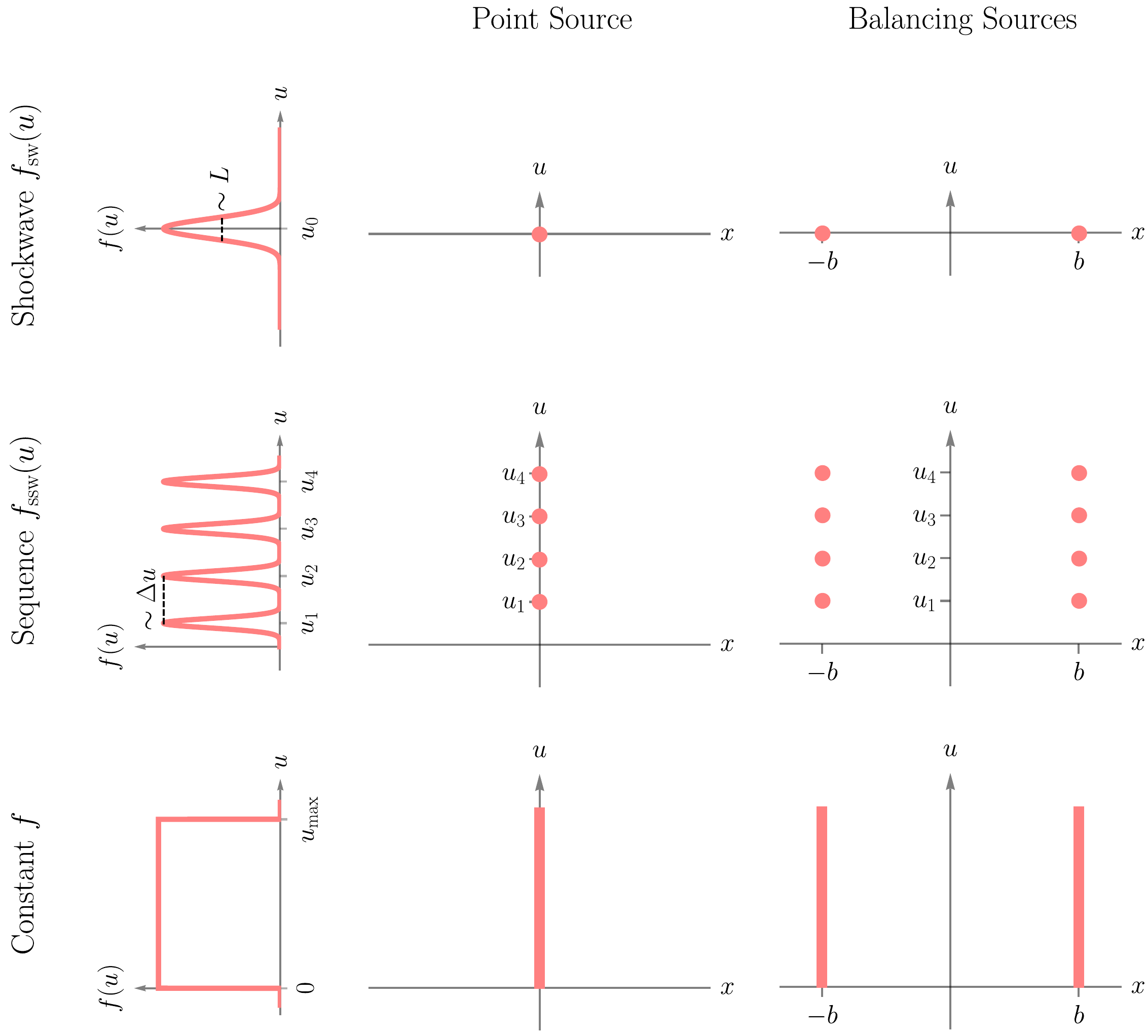}
    \caption{An illustration of the pp-wave sources referred to in the text. Each dot represents a Dirac delta function in the $u$-direction.}
    \label{fig: different sources}
\end{figure}

In what follows we shall exclusively be concerned with pp-waves of the form
\begin{equation}
    g_{\alpha \beta}\sd x^{\alpha} \sd x^{\beta} = 2\sd u \sd v + H(u,\mathbf{x}) \sd u^2 +\delta_{ij} \sd x^i \sd x^j \, .
\end{equation}
Up to symmetry, the only non-zero component of the Riemann tensor associated to the pp-wave metric are
\begin{equation}\label{eq: Riemann tensor of pp-wave metric}
    R_{uiuj}=-\frac{1}{2}\partial_{i}\partial_{j}H.
\end{equation}
It follows that the only non-zero component of the Ricci tensor is $R_{uu} = -\frac{1}{2}\lp H$ and the Ricci scalar automatically vanishes because the inverse metric component $g^{uu}$ is zero.
The metric \eqref{eq: pp wave metric} is thus a solution to the vacuum Einstein equations if $H$ is harmonic in transverse space $\lp H(u,\mathbf{x}) = 0$.
More generally, we will be interested in spacetimes sourced by point particles (analogous to black holes) for which the Laplacian of $H$ is non-zero only at a discrete set of points $\{\mathbf{b}_i\}$:
\begin{equation}\label{eq: Poisson equation for metric function}
    -\lp H(u,\mathbf{x}) = \frac{2 \pi^{\frac{d}{2}}}{\Gamma\left(\frac{d}{2}\right)}\sum_i f_i(u) \delta^{(d)}(\mathbf{x}-\mathbf{b}_i) \, ,
\end{equation}
the solution to which is
\begin{equation}\label{eq: generic vacuum pp-wave metric function}
    H(u,\mathbf{x}) = \sum_i\frac{f_i(u)}{\abs*{\mathbf{x}-\mathbf{b}_i}^{d-2}} \, .
\end{equation}
The functions of the $u$-coordinate $f_i(u)$ must be positive for such a source to satisfy the null energy condition, but are otherwise unspecified at this time. 
Example functions which will be useful in what follows are given in Fig.~\ref{fig: different sources}. 
In addition to the different configurations of shockwaves we will consider --- situations which are captured through different choices of functions $f(u)$ --- we also have the freedom to choose different sets of singular points  $\{\mathbf{b}_i\}$.
Throughout this paper, we will consider two simple cases:
\begin{enumerate}
    \item {\bf Point Sources:} A single or set of point sources located at the origin with spherical symmetry in the transverse directions,
    \begin{equation}\label{eq: point source metric function}
        H_{\text{pt}}(u,r) = \frac{f(u)}{r^{d-2}}\,,
    \end{equation}
    which will be discussed in detail in Section~\ref{subsec: pt source scattering}, and
    \item {\bf Balancing Case:}  Two equal (sets of) sources located at $\pm \mathbf{b}$, which we shall refer to as the ``balancing" case
    \begin{equation}\label{eq: balancing source metric function}
        H_{\text{bal}}(u,\mathbf{x}) = f(u)\left(\frac{1}{\abs*{\mathbf{x}-\mathbf{b}}^{d-2}} + \frac{1}{\abs*{\mathbf{x}+\mathbf{b}}^{d-2}}\right)\, 
    \end{equation}
    to be considered in Section \ref{subsec: balancing sources}.
\end{enumerate}
For the most part, the arguments in this paper apply for arbitrary functions $f(u)$. 
There are however some specific examples we will refer to for illustrative purposes:
\begin{enumerate}
    \item[a)] The single ``{\bf Shockwave}",
    \begin{equation}\label{eq: delta-shockwave solution}
        f_{\text{sw}}(u) = \frac{4 \Gamma\left(\frac{d-2}{2}\right)}{\pi^{\frac{d-2}{2}}} G P_u \delta(u),
    \end{equation}
    corresponds to a point particle moving ultra-relativistically in the $v$-direction with momentum $P_u$, and features prominently in the discussion on causality in \cite{Camanho:2014apa}.
    \item[b)] The ``{\bf Sequence of shockwaves}",
    \begin{equation}\label{eq: sequence of shockwaves solution}
        f_{\text{ssw}}(u) = \sum_{n=0}^{N-1} f_{\text{sw}}(u-u_n),
    \end{equation}
    corresponds to series of shockwaves which occur one after another in the $u$-direction. 
    This is the situation one would typically refer to as ``stacking" or surfing on a multitude of shockwaves, carefully placed one after the other in the ``hope" of accumulating time delay/advance. 
    \item[c)] A {\bf constant} $f(u) = f$, which can be seen as the limit of $f_{\text{ssw}}$ as the number of shocks grows very large $N\rightarrow \infty$ while the $u$-distance between them goes to zero $u_n - u_{n+1} \rightarrow 0$. 
    In terms of optimising the amount of time delay/advance, this situation is typically more optimal than the previous one, but can easily be considered within the same formalism. 
    \end{enumerate}

\subsection{EFT of gravity}
Schematically, the $D$-dimensional low-energy EFT of gravity is of the form
\begin{equation}\label{eq: generic EFT action}
    S_{\text{EFT}} = \int \text{d}^D x \sqrt{-g} \, \Mpl^{D-2}\left(\frac{1}{2} R + \Lambda^2\sum_{m\ge 0,\, n\ge 2} c_{mn} \left(\frac{\nabla}{\Lambda}\right)^m \left(\frac{\text{Riemann}}{\Lambda^2}\right)^n\right),
\end{equation}
where $\Lambda \lesssim \Mpl$ is the cut-off energy and the Wilson coefficients $c_{mn}$ are dimensionless.
Terms of the form shown in the sum may arise either from tree-level effects, for instance of higher-spin ($s\geq 2$) states with mass/energy scale $\sim \Lambda$, or from loop corrections involving any spin with mass/energy scale $\ll \Lambda$.
The nuanced difference between these two possible origins is discussed in \cite{deRham:2019ctd}, but is unimportant from a low-energy perspective aiming to remain agnostic to the precise details of the UV completion.

In $D\geq 5$, the leading-order EFT in vacuum is
\begin{equation}\label{eq: EGB action}
    S_{\text{EFT}} = \int \text{d}^{D}x \sqrt{-g} \, \Mpl^{D-2}\left(\frac{1}{2} R + \frac{\cGB}{\Lambda^2} \RGB^{2} + \dots\right),
\end{equation}
where $R_{\text{GB}}^2 = R_{ \alpha \beta \sigma \rho}^2 - 4 R_{\alpha \beta}^2 + R^2$ is the GB-term. 
In specific UV completions $\cGB$ may vanish, but $\cGB$ is generically non-zero and the entire discussion of this paper assumes $\cGB \neq 0$. 
If it does vanish, then the cubic curvature terms will be leading corrections whenever supersymmetry is broken. These are considered in \cite{deRham:2020ejn,AccettulliHuber:2020oou,deRham:2021bll}. 

The vacuum Einstein-Gauss-Bonnet (EGB) equations are
\begin{equation}\label{eq: EGB background eqs}
    \mathcal{E}_{\alpha \beta} \coloneqq G_{\alpha \beta} + \frac{2 \cGB}{\Lambda^2} B_{\alpha \beta} = 0,
\end{equation}
where
\begin{equation}
    B_{\alpha \beta} =  4 R_{\sigma \alpha \beta \rho} R^{\sigma \rho} + 2 R\indices{_{\alpha}^{\sigma \rho \kappa}}R_{\beta \sigma \rho \kappa}- 4 R_{\alpha \sigma}R\indices{_{\beta}^{\sigma}}+ 2 R R_{\alpha \beta} -\frac{1}{2}  \RGB^2\, g_{\alpha \beta}.
\end{equation}
For the pp-wave metric \eqref{eq: pp wave metric} with Riemann tensor \eqref{eq: Riemann tensor of pp-wave metric}, $B_{\alpha \beta} = 0$.
The solutions discussed in the previous section \eqref{eq: generic vacuum pp-wave metric function} -- \eqref{eq: balancing source metric function} are thus also solutions to EGB-theory, as expected.

\subsection{Metric perturbations and their dynamics}\label{subsec: metric perturbations}
We will now introduce metric perturbations $h_{\alpha \beta}$ on top of background pp-waves.
It is their dynamics which will determine the causal properties of the GB-operator in Sections \ref{subsec: pt source scattering} -- \ref{sec: quantum calcs}.
Perturbations spoil the exactness of the pp-wave metric to any order in the EFT.
We will soon see how this gives rise to constraints on both the energy of the perturbations and the profile of the background function $H(u,\mathbf{x})$ (captured by the function $f(u)$).
This will play an important role for causality.

The field equations for the metric perturbations $h_{\alpha \beta}$ in this EFT are given by
\begin{equation}
    \delta \mathcal{E}_{\alpha \beta} = 0.
\end{equation}
The full set of equations is given in appendix~\ref{app: full field eqs}.
In lightcone gauge,
\begin{equation}
    h_{v\alpha} = 0 \quad \forall \, \alpha,
\end{equation}
the $\delta \mathcal{E}_{v \alpha}$ equations reduce to two constraint equations
\begin{align}
        &h_{ii} = 0,\label{eq: constraint 1: traceless}\\
        &\partial_v h_{iu} + \partial_j h_{ij} = 0.\label{eq: constraint 2: hui}
\end{align}
Taking the trace of the $\delta \mathcal{E}_{ij}$ equation and applying the above two constraints produces a third constraint,
\begin{equation}\label{eq: constraint 3: huu}
    \partial_v h_{uu} + \partial_i h_{iu} = -4\frac{d-2}{d}\frac{\cGB}{\Lambda^2}\partial_i\partial_j H \partial_v h_{ij},
\end{equation}
so that the $h_{uu}$ and $h_{ui}$ components are all specified in terms of $h_{ij}$ as long as $\partial_v h_{\alpha \beta} \neq 0$.
This leaves only the $D(D-3)/2$ components of the traceless $h_{ij}$ as the dynamical degrees of freedom.
Their equation of motion is given by
\begin{align}
\begin{split}\label{eq: eom for hij}
    &\tilde{\Box}h_{ij} - 8 \frac{\cGB}{\Lambda^2}\partial_v^2 X_{ij} = 0,\\
    &X_{ij} = \frac{1}{2}\left(h_{ik}\partial_j\partial_k H + h_{jk}\partial_i\partial_k H\right) - \frac{1}{d}\delta_{ij} h_{kl}\partial_k\partial_l H \, .
\end{split}
\end{align}
Unlike the background solution, the equation for the perturbations is modified by the GB-operator in the action.
In fact, it will generically receive corrections from any effective gravitational operator.
Demanding that these corrections are under control will define the regime of validity of the EFT in the next section.

The decoupled ``master variables" are linear combinations of the $h_{ij}$ depending on the form of $H(u,\mathbf{x})$.
In the case of the point source \eqref{eq: point source metric function}, spherical symmetry means the equations \eqref{eq: eom for hij} are all immediately decoupled when expressed in spherical coordinates.
The single caveat is that not all of the diagonal components of $h_{ij}$ are independent because of the tracelessness condition \eqref{eq: constraint 1: traceless}.
Choosing $h_{DD}$ as the dependent component, the master variables are
\begin{equation}\label{eq: master variables for point source}
    \Phi \in \{h_{rr},h_{ra},h_{ab} \text{ for } a\neq b, h_{aa} - h_{DD} \text{ (no sum)}\},
\end{equation}
where $\{a,b,\dots\}$ label the angular directions.
Their master equations are
\begin{equation}\label{eq: master equation for point source}
    \tilde{\Box}\Phi + A \frac{\cGB}{\Lambda^2}\frac{\partial_r H}{r}\partial_v^2 \Phi = 0,
\end{equation}
where $A$ is an integer depending on the master variable under consideration
\begin{equation*}
    A \in \{8(d-2),4(d-2),-8,-8\} \,.
\end{equation*}
In the balancing source case \eqref{eq: balancing source metric function}, it is less straightforward to identify the master variables.
Some details are given in appendix~\ref{app: master variables in balancing background} but are not crucial for what follows.
Suffice to say, the master equations will take a similar form to \eqref{eq: master equation for point source}.

\subsection{The EFT regime of validity}\label{subsec: EFT regime of validity}
The GB-term in the EGB effective theory \eqref{eq: EGB action} under consideration is just the lowest-order term in an infinite series of effective operators \eqref{eq: generic EFT action}.
Generically, any number of these terms could be present when UV degrees of freedom are integrated out to obtain a low-energy action.
The full EFT action is an expansion in powers of the spacetime curvature (``Riemann") and its derivatives, compared to the EFT cut-off energy $\Lambda$.
The expansion is only under control in regions where that ratio is small \cite{deRham:2020zyh,Chen:2021bvg}.
In other words, the EFT is only valid in the regime where
\begin{equation}\label{eq: schematic regime of validity}
    \left(\frac{\nabla}{\Lambda}\right)^{m} \left(\frac{\text{Riemann}}{\Lambda^2}\right)^n \ll 1.
\end{equation}
What is meant by this schematic expression on the left-hand side is all possible scalar contractions built out of $m$ covariant derivatives and $n$ powers of the Riemann tensor.
The background pp-wave metric seemed to evade validity constraints by exactly solving the EFT to any order, so that the left-hand side of \eqref{eq: schematic regime of validity} simply vanishes.
However, in the perturbed spacetime, the ``Riemann" tensor in \eqref{eq: schematic regime of validity} refers to the sum of the background Riemann tensor and its perturbation.
The perturbed Riemann tensor has a much richer structure than the background Riemann tensor, with none of its non-trivial components vanishing.
Hence there is a non-trivial EFT regime of validity for the pp-wave spacetime as soon as metric perturbations are allowed to propagate.

The perturbations to the Riemann tensor are given by second derivatives of the metric perturbations $h_{\alpha \beta}$.
Considering perturbations of momentum $k^{\alpha}$, we may represent the contribution of the derivatives acting on the perturbations by powers of $k^{\alpha}$.
To establish the regime of validity, it is then sufficient to consider all possible scalar invariants of the form
\begin{equation}\label{eq: schematic regime of validity with perturbations}
    \left(\frac{\nabla}{\Lambda}\right)^{m} \left(\frac{\bar{R}}{\Lambda^2}\right)^n\left(\frac{k}{\Lambda}\right)^p \ll 1 \, ,
\end{equation}
where $\bar{R}$ represents the background Riemann tensor \eqref{eq: Riemann tensor of pp-wave metric}.
There are a number of contractions which would render the left-hand side of \eqref{eq: schematic regime of validity with perturbations} trivially zero, or lead to a much weaker constraint.
These are discussed in appendix~\ref{app: determining the EFT regime of validity}.
The upshot is that the strongest EFT validity bounds come from contractions of the form
\begin{subequations}\label{eqs: non-trivial EFT bounds together}
    \begin{align}
        &(k\cdot \nabla)^m(k\cdot \bar{R} \cdot k)^{2n} \ll \Lambda^{2m + 8n},\label{eq: non-trivial EFT bound 1}\\
        &(\nabla \cdot \nabla)^m(k\cdot \bar{R} \cdot k)^{2n} \ll \Lambda^{2m + 8n}.\label{eq: non-trivial EFT bound 2}
    \end{align}
\end{subequations}
A non-trivial example of \eqref{eq: non-trivial EFT bound 1} with $m=n=1$, written in terms of components, is
\begin{equation*}
    k_{\alpha}\nabla^{\alpha}(k^{\beta} R_{\beta \gamma \delta \sigma}k^{\delta}k_{\kappa}R^{\kappa \gamma \rho \sigma}k_{\rho}) \ll \Lambda^{10} \, .
\end{equation*}
Similarly, $k_{\alpha}$ may be replaced with $\nabla_{\alpha}$ to obtain an example of \eqref{eq: non-trivial EFT bound 2}.
Taking the $n\rightarrow \infty$ limit of \eqref{eqs: non-trivial EFT bounds together} gives us our first EFT regime of validity bound:
\begin{equation}
    (k_v \partial_i \partial_j H k_v)^2 \ll \Lambda^8 \, , \quad \quad \text{or}
   \quad \quad \partial_i \partial_j H \ll \frac{\Lambda^4}{k_v^2} \, .
\end{equation}
In particular, for the spherically symmetric point source, it can be written as
\begin{equation}
    \partial_r^2 H \sim \frac{\partial_r H}{r} \ll \frac{\Lambda^4}{k_v^2}.
\end{equation}
This represents a bound on the strength of the source at any given distance from its singularity in transverse space.
Taking the $m\rightarrow\infty$ limit of \eqref{eq: non-trivial EFT bound 1} gives
\begin{equation}
    k^{\alpha}\nabla_{\alpha} \ll \Lambda^2,
\end{equation}
where $\nabla$ is understood to be acting on a scalar.
In particular, for non-zero momentum in the $v$-direction $k_v$, we obtain a bound on the $u$-derivative of $H$:
\begin{equation}
    \frac{\partial_u H}{H} \ll \frac{\Lambda^2}{k_v}.
\end{equation}
Lastly, taking the $m\rightarrow\infty$ limit of \eqref{eq: non-trivial EFT bound 2} gives
\begin{equation}
    \tilde{\Box} \ll \Lambda^2,
\end{equation}
where $\tilde{\Box}$ is understood to be acting on a scalar.
Isolating the transverse Laplacian results in a distance-resolution scale (say, for the point source),
\begin{equation}
    r \gg \Lambda^{-1}.
\end{equation}

Below we gather the three EFT regime of validity bounds.
On the left, they are written in their generic form as they would apply without specifying a background pp-wave metric.
In the middle, they are specialised to the point-source metric \eqref{eq: point source metric function}.
On the right-hand side, they are further specified for a particular impact parameter $\mathbf{x} = \mathbf{b} = b \mathbf{\hat{b}}$.
\begin{subequations}\label{eqs: collated EFT RoV bounds}
\begin{flalign}
\qquad & \text{General} & & \text{Point Source(s)} & & \mathbf{x} = \mathbf{b} = b \mathbf{\hat{b}} \notag \\
\hline \notag \\[-15pt]
    \qquad &\Box \ll \Lambda^2  & &r \gg \Lambda^{-1} & &b \gg \Lambda^{-1} \qquad \label{eq: EFT RoV distance bound}\\
    \qquad &\partial_i \partial_j H \ll \frac{\Lambda^4}{k_v^2} & &\frac{\partial_r H}{r} \ll \frac{\Lambda^4}{k_v^2} & &\frac{f(u)}{b^{d}} \ll \frac{\Lambda^4}{k_v^2} \qquad \label{eq: EFT RoV strength bound}\\
    \qquad &k^{\alpha}\nabla_{\alpha} \ll \Lambda^2 & &\frac{\partial_u H}{H} \ll \frac{\Lambda^2}{k_v} & &\frac{f'(u)}{f(u)} \ll \frac{\Lambda^2}{k_v} \label{eq: EFT RoV u-derivative bound}\qquad
\end{flalign}
\end{subequations}
The top line \eqref{eq: EFT RoV distance bound} is relatively intuitive and represents the lower limit on the distance scales we may probe before exiting the regime of the validity of the EFT.
As mentioned previously, the middle line \eqref{eq: EFT RoV strength bound} represents a bound on the strength of the pp-wave source at a given distance from the source.
Here the bound on $f(u)$ should be understood not in a point-wise sense but rather in an averaged sense, 
\begin{eqnarray}
\int_{u-\lambda_u}^{u+\lambda_u}  |f(\tilde u)|^p \d \tilde u \ll \left(\frac{\Lambda^{4}b^{d}}{k_v^{2}}\right)^{p}\lambda_u\,,
\end{eqnarray}
for any positive power $p$ and where $\lambda_u=P_u^{-1}$ is the wavelength integrated over in the $u$-direction. 
The last line \eqref{eq: EFT RoV u-derivative bound} bounds how quickly $f(u)$ may vary in $u$-time and should also be understood in its averaged sense,
\begin{eqnarray}
\int_{u-\lambda_u}^{u+\lambda_u}  |f'(\tilde u)|^p \d \tilde u \ll \frac{\Lambda^{2p}}{k_v^{p}} \int_{u-\lambda_u}^{u+\lambda_u}  |f(\tilde u)|^p \d \tilde u \,. \label{eq:averagefprime}
\end{eqnarray}
The latter two bounds depend not only on the EFT cut-off $\Lambda$, but also on the energy $k_v$ of the gravitational wave probing the spacetime.
This underlines the role that metric perturbations play in establishing an EFT regime of validity on a pp-wave spacetime. 

Notably, shockwaves (or sequences therefore) lead to divergences in the profile $f(u)$ and its derivative, and are hence object that cannot be probed within the regime of validity of the EFT without smoothing out as we discuss in Section~\ref{regulatedshockwave}.

\subsubsection{Regime of validity by explicit example}\label{subsec: RoV by example}
The above arguments to establish a regime of validity rely only on generic assumptions about the EFT series expansion, but were largely schematic.
In this section, we will consider an explicit example of a truncated EFT expansion.
Demanding that the higher-dimension operators in the expansion are subdominant to the GB-operator will concretely reproduce the pp-wave regime of validity given in \eqref{eqs: collated EFT RoV bounds}.

Consider the following effective action which contains \eqref{eq: EGB action} as its leading-order part supplemented by generic dimension-6 and -8 curvature operators:
\begin{eqnarray}\label{eq: EFT action with higher-dim ops}
    S_{\text{EFT}} = \int \text{d}^{D}x \sqrt{-g} \, \Mpl^{D-2}\Bigg(\frac{1}{2} R &+& \frac{\cGB}{\Lambda^2} \RGB^{2}
    + \frac{\cRt}{\Lambda^{4}} R_{\alpha \beta \gamma \delta} R^{\gamma \delta \sigma \rho}R\indices{_{\sigma \rho}^{\alpha \beta}}\\
    &+& \frac{\cRf}{\Lambda^{6}} R_{\alpha \beta \gamma \delta} R^{\gamma \delta \sigma \rho}R_{\sigma \rho \kappa \lambda}R^{\kappa \lambda \alpha \beta} + \dots \Bigg) \, .\notag
\end{eqnarray}
The full field equations for the background and perturbations are provided in appendix~\ref{app: higher-dim EFT}.
The pp-wave metric \eqref{eq: pp wave metric} of course remains an exact (background) solution to \eqref{eq: EFT action with higher-dim ops}.
Choosing lightcone gauge as before, the traceless components of the metric perturbations in the transverse directions $h_{ij}$ remain the dynamical degrees of freedom.
With the addition of the new operators, it is less straightforward to decouple the equations of motion and identify the master variables.
In the case of the spherically symmetric point source, one can perform a scalar-vector-tensor (SVT) decomposition on the transverse space with the result that the tensor modes $\Phi_T$ are immediately decoupled.
For our purposes, it is enough just to consider their master equation up to corrections of $\mathcal{O}(\Lambda^{-8})$:
\begin{align}
\begin{split}
        0 = \,\, &\tilde{\Box} \Phi_{T} - \frac{\kappa_T^2+2}{r^2}\Phi_T - 8\frac{\cGB}{\Lambda^{2}} \frac{\partial_{r}H}{r} \partial_{v}^{2} \Phi_{T} + 24 \frac{\cRt}{\Lambda^{4}}\partial_v^2 \left[d\frac{\partial_r H}{r^2}\left(\partial_r \Phi_T + \frac{\Phi_T}{r}\right)\right.\\
        &\left.-\frac{\partial_{u}\partial_{r}H}{r} \partial_{v} \Phi_{T}\right] + 16\frac{\cRf-12\cGB\cRt}{\Lambda^{6}} \left(\frac{\partial_{r}H}{r}\right)^{2} \partial_{v}^{4} \Phi_{T} \, ,
\end{split}
\end{align}
where $\kappa_T^2$ is the eigenvalue of the tensor spherical harmonic on the $(d-1)$-sphere.
This equation can be expressed in $v$-momentum space by replacing $\partial_v \rightarrow i k_v$.
For the GB-operator to truly be the leading-order term in the EFT expansion, the contributions of the higher-order operators to this equation of motion ($\propto \cRt, \cRf$), specifically to the potential, should be subdominant to the contribution from the GB-operator itself ($\propto \cGB$).
Assuming that the Wilson coefficients $\cRt$ and $\cRf$ are  $\mathcal{O}(\cGB)$, then control of the EFT amounts to:
\begin{alignat}{3}
    & 24 d \frac{\cRt}{\Lambda^{4}}\frac{\partial_r H}{r^3}\partial_v^2\Phi_T \ll 8\frac{\cGB}{\Lambda^{2}} \frac{\partial_{r}H}{r} \partial_{v}^{2} \Phi_{T} \quad && \implies \quad && r \ll \Lambda^{-1} \, , \\
    & 16 \frac{\cRf}{\Lambda^6} \left(\frac{\partial_{r}H}{r}\right)^{2} \partial_{v}^{4} \Phi_{T} \ll 8\frac{\cGB}{\Lambda^{2}} \frac{\partial_{r}H}{r} \partial_{v}^{2} \Phi_{T} \quad && \implies \quad && \frac{\partial_{r}H}{r} \ll \frac{\Lambda^4}{k_v^2} \, , \\
    & 24 \frac{\cRt}{\Lambda^{4}} \frac{\partial_{u}\partial_{r}H}{r} \partial_{v}^3 \Phi_{T}
    \ll 8\frac{\cGB}{\Lambda^{2}} \frac{\partial_{r}H}{r} \partial_{v}^{2} \Phi_{T} \quad && \implies \quad && \frac{\partial_{u}H}{H} \ll \frac{\Lambda^2}{k_v} \, ,
\end{alignat}
which is exactly the regime of validity of the EFT \eqref{eqs: collated EFT RoV bounds} obtained in the previous section.
Note, it was not strictly necessary to include the dimension-8 operator $\sim \text{Riemann}^4$ since the subleading correction from the dimension-6 operator produced the same term.
Its inclusion illustrates that there is nothing particular about our choice of higher-dimension operators in reproducing the regime of validity.
There are many possible higher-dimension operators whose presence in the EFT of gravity would constrain the parameters of the spacetime in the same way.

\subsubsection{Regulating the shockwave in the EFT of gravity} \label{regulatedshockwave}
The main result from the previous two sections is that any calculation performed within the effective theory of gravity should only be trusted within the regime in which the set of bounds \eqref{eqs: collated EFT RoV bounds} are satisfied.
The pp-wave solutions provided in Section \ref{subsec: properties of pp-waves} need to be revisited in light of this.
In particular, the singular shockwave solution $f_{\text{sw}}(u)$ \eqref{eq: delta-shockwave solution} is clearly against the spirit of the bound \eqref{eq: EFT RoV u-derivative bound}, or rather its average \eqref{eq:averagefprime}, on $f'(u)/f(u)$.
To bring it within the remit of the EFT, the shockwave may be regulated by expressing it as a Gaussian,
\begin{equation}
    f_{\text{sw}}^{\text{reg}}(u) =  \frac{\alpha}{\sqrt{2\pi L^2}}\exp\left[-\frac{u^2}{2L^2}\right] \, ,
\end{equation}
where $\alpha = 4\Gamma((d-2)/2)/\pi^{(d-2)/2}\, G P_u$.
Interpreting \eqref{eq: EFT RoV u-derivative bound} in an averaged sense, as indicated in \eqref{eq:averagefprime}, we then infer a bound on the width of the pp-wave that can be considered
\begin{equation}
    L \gg \frac{k_v}{\Lambda^2} \, .
\end{equation}
This means that the more peaked the source (smaller $L$), the ``slower" the GW needs to be (smaller $k_v$) in order to probe the spacetime within the regime of the validity of the EFT. 
To be clear, this does not mean that $k_v/\Lambda^2$ is the width of the wave in the UV completion. 
We should interpret this in the following sense: When we construct a low-energy EFT in the Euclidean we naturally coarse-grain over modes with momenta larger than $\sqrt{k^2} \sim \Lambda$, so that no correlation function can resolve distances smaller than $\Lambda^{-1}$. 
In the Wilsonian picture of renormalisation, this scale is our choosing and need not correspond to the actual scale of new physics. 
The arguments of the previous section similarly imply that we cannot resolve distances in $u$ smaller than $k_v/\Lambda^2$. 
It could well be that in the UV completion, the shockwave is actually more localised/peaked than this (\textit{i.e.} has a width shorter than $k_v/\Lambda^2$). However, it will always be perceived to a low-energy observer as having a width of at least $k_v/\Lambda^2$. 

The sequence of shockwaves may be regulated in an analogous way as $f_{\text{ssw}}^{\text{reg}} = \sum_i f_{\text{sw}}^{\text{reg}}(u-u_i)$.
Since each regulated shock now has a finite spread $\sim L$ and height $\sim \alpha L^{-1}$, they become indistinguishable as the distance between them shrinks, and approach a constant source $f(u) = f \sim \alpha L^{-1}$.
Conversely, if they are to remain distinguishable, there must be a minimum $u$-separation between them $\Delta u \sim u_{i+1}-u_i \gg L$.

\section{Scattering time delay}\label{sec:ScatteringTimeDelay}
In this section, we review the formal definition of scattering time delays and derive expressions for it in various representations that will be useful for future calculations. 
In particular we give a generic semi-classical expression to complement the more familiar eikonal expression which is frequently used in the analysis of pp-wave spacetimes. 

\subsection{Eisenbud-Wigner-Smith time delay}

In the elastic scattering region, the eigenvalues of the single-particle $S$-matrix $\hat{S}$ define the asymptotic phase shifts $e^{2i\delta}$. 
For example, in the case of spherically symmetric non-relativistic scattering, the $S$-matrix diagonalises into partial waves and each partial wave phase shift $\delta_{\ell}(E)$, which is a function of the incoming energy $E$, determines an eigenvalue of the $S$-matrix. 
Eisenbud and Wigner \cite{Eisenbud:1948paa,Wigner:1955zz} proposed that for non-relativistic scattering, associated with each partial wave, we can define the Eisenbud-Wigner time delay
\be
\Delta T_{\ell} = \frac{2 \p \delta_{\ell}}{\p E} \, ,
\ee
by differentiating the partial wave phase shifts with respect to the energy.
Qualitatively, this describes the delay of a scattering wave packet which is peaked around an energy $E$. 
In situations with less symmetry, it is in general difficult to construct the eigenvalues and eigenstates, so it is preferable to have a more general definition of what we mean by the time delay. 
This is provided by the Hermitian time delay operator, known as the Wigner-Smith operator \cite{Smith:1960zza,Martin:1976iw}, which is defined in terms of the $S$-matrix as
\begin{equation}\label{eq: WS time delay op}
    \widehat{\Delta T} = -\frac{i}{2} \hat{S}^{\dagger}\frac{\p \hat{S}}{\p E}+\frac{i}{2} \frac{\p \hat{S}^{\dagger}}{\p E} \hat{S} .
\end{equation}
Connection can be made with the Eisenbud-Wigner definition of the time delay by considering the elastic region, for which $\hat S^{\dagger} \hat S=1$, and considering the expectation value of $\widehat{\Delta T}$ on an eigenstate of $\hat{S}$. 
Denoting these by $\ket{\delta}$, then differentiating $\hat S | \delta \rangle=e^{2i \delta} | \delta \rangle$ we have
\be
\frac{\partial \hat S}{\partial E} | \delta \rangle +(\hat S-e^{2i \delta}) \frac{\partial  | \delta \rangle }{\partial E} = e^{-2i\delta} \frac{\partial (e^{2i\delta})}{\partial E}  | \delta \rangle  \, ,
\ee
from which we infer
\begin{equation}
 \bra{\delta}   \widehat{\Delta T}\ket{\delta} =2\frac{\partial\delta}{\partial E} \, ,
\end{equation}
which is exactly the Eisenbud-Wigner time delay.

The operator \eqref{eq: WS time delay op} can be formally extended to the full Fock space/QFT Hilbert space as follows: Since the full $S$-matrix $\hat{\mathbb{S}}$ commutes with the asymptotic Hamiltonian $\hat H_0$ for time-translation invariant systems, we may write it in terms of its spectral decomposition 
\be
\hat{\mathbb{S}} = \int \d E \sum_{I,J} \, |E,I \rangle {\mathbb{W}}_{IJ}(E)\langle E,J | \, ,
\ee
where $|E,I \rangle$ denote a complete set of multi-particle energy eigenstates of $\hat H_0$
\be
\int \d E \sum_I | E,I\rangle \langle I , E | = \hat 1 \, , \quad \langle E, I | E', J \rangle = \delta(E-E') \delta_{IJ} \, .
\ee
The time delay operator defined on the full Fock space is then
\ba \label{timedelaydefinition}
  \widehat{\Delta {\mathbb{T}}}   &=&-\frac{i}{2} \int \d E \sum_I\, \hat{\mathbb{S}}^{\dagger} \,  |E,I \rangle \frac{\partial}{\partial E} \( \langle I , E |  \hat{\mathbb{S}}\) +\frac{i}{2} \int \d E \sum_I\, \frac{\partial}{\partial E}  \( \hat{\mathbb{S}}^{\dagger} |E,I \rangle  \)  \langle I , E |  \hat{\mathbb{S}} \nn  \\
  &=& -i \int \d E \, \sum_I \frac{\partial}{\partial \epsilon} \left[ \hat{\mathbb{S}}^{\dagger}| E-\frac{\epsilon}{2}, I \rangle \langle E+\frac{\epsilon}{2}, I| \hat{\mathbb{S}} \right] \bigg|_{\epsilon=0}\, .
\ea
Given that $ \hat H_0$ commutes with $ \hat{\mathbb{S}}$ and  $ \hat H_0| E\pm \epsilon/2, I \rangle= (E \pm \epsilon/2)| E\pm \epsilon/2, I \rangle$ then formally we have
\be
[\hat H_0 , \widehat{\Delta {\mathbb{T}}} ] = i \int \d E \,   \frac{\partial}{\partial \epsilon} \left[ \epsilon \,  \hat{\mathbb{S}}^{\dagger}| E-\frac{\epsilon}{2}, I \rangle \langle E+\frac{\epsilon}{2}, I| \hat{\mathbb{S}} \right] \bigg|_{\epsilon=0}=i\, ,
\ee
which shows that the time delay is the operator appropriately conjugate to $\hat H_0$. Note that this expression only holds at a formal level since, according to Pauli's theorem \cite{Pauli}, there can be no self-adjoint operator conjugate to a Hamiltonian with an energy spectrum bounded below. 
In short, if there were then it would be possible to construct a unitary operator $e^{- i \Delta_E  \widehat{\Delta T}}$ which could translate energy eigenstates to arbitrarily negative energies $e^{- i \Delta_E  \widehat{\Delta T}}|E \rangle = |E-\Delta_E \rangle$.

However, it has long been known that is is possible to construct non self-adjoint operators which have the above commutation relation, see for example \cite{Lippmann:1966zz,razavy1967quantum}. 
These operators are mildly singular (usually requiring an inverse) and are Hermitian, but not self-adjoint due to the need to exclude singular points.\footnote{For a review of these and related issues see \cite{Aharonov:1961mka,branson1964time,razavy1969quantum,olkhovsky1974time,narnhofer1980another,Bolle:1974ni,Dodonov:2015wha,galapon1999consistency}. A more formal way to deal with this is to introduce positive-operator-valued observables \cite{giannitrapani1997positive,busch1994time,PhysRevA.66.044101,busch2002time}.} 
Nevertheless, the allowed domain contains all the scattering states of interest and so we can nevertheless compute expectation values of these operators. 
Concretely this can be achieved as follows: Let $\ket{{\Delta}_I(E)}$ denote a Fock space eigenstate of $\hat {\mathbb{S}}$ which is also an eigenstate of $\hat H_0$ with eigenvalue $E$ normalized such that $\langle  \Delta_I(E) | \Delta_J(E') \rangle=\delta_{IJ} \delta(E-E')$, and 
\be
\hat {\mathbb{S}}\ket{{\Delta}_I(E)}=e^{2i {\Delta}_I(E)}\ket{{\Delta}_I(E)} \, .
\ee
The full $S$-matrix may be written in spectral form as
\be \label{spectralS}
\hat {\mathbb{S}} = \int_0^{\infty} \d E \sum_I \ket{{\Delta}_I(E)} e^{2i {\Delta}_I(E)}  \bra{{\Delta}_I(E)} \, .
\ee
A generic state is similarly written as
\be
|\psi \rangle =  \int_0^{\infty} \d E  \sum_I  \,  \psi_I(E) \ket{{\Delta}_I(E)} \, ,
\ee
with $\psi_I(E) = \langle  \Delta_I(E) | \psi \rangle$.
The matrix elements of the time delay \eqref{timedelaydefinition} are explicitly
\be
\langle \psi |     \widehat{\Delta \mathbb{T}} | \phi \rangle =  \int_0^{\infty} \d E  \sum_I \left[-\frac{i}{2} \( \psi_I^*(E) \frac{\partial \phi_I(E)}{\partial E} - \frac{\partial \psi_I^*(E)}{\partial E}\phi_I(E) \right)+ 2 \frac{\d \Delta_I}{\d E} \psi_I^* \phi_I   \right] \, ,
\ee
which is manifestly Hermitian. Now since
\be
\hat H_0 | \psi \rangle =  \int_0^{\infty} \d E   \sum_I \,  E \psi_I(E) \ket{{\Delta}_I(E)} \, ,
\ee
which amounts to $\psi_I(E) \rightarrow E \psi_I(E)$, it is easy to see that
\ba
\langle \psi |    [\hat H_0,\widehat{\Delta \mathbb{T}}] | \phi \rangle &=&
-\frac{i}{2} \int_0^{\infty} \d E  \sum_I \Bigg[ \big(E \psi_I^*(E)\big) \frac{\partial \phi_I(E)}{\partial E} - \frac{\partial \big(E \psi_I^*(E)\big)}{\partial E}\phi_I(E) \nn \\
&&\phantom{\int_0^{\infty} \d E  \sum_I \Bigg[}
-  \psi_I^*(E) \frac{\partial \big(E \phi_I(E)\big)}{\partial E} + \frac{\partial \psi_I^*(E)}{\partial E} \big(E \phi_I(E)\big) \Bigg]  \nn \\
&=&i\int_0^{\infty} \d E  \sum_I  \psi_I^*(E) \phi_I(E) =i \langle \psi |  \phi \rangle \, ,
\ea
which confirms we have the correct definition of the time delay.

For the squared operator, we na\"ively have
\ba
&& \langle \psi |     \widehat{\Delta \mathbb{T}} {\,}^2 | \phi \rangle = \int_0^{\infty} \d E \sum_I \langle \psi |     \widehat{\Delta \mathbb{T}} | \Delta_I(E) \rangle \langle \Delta_I(E) |  \widehat{\Delta \mathbb{T}}| \phi \rangle \\
&=&  \int_0^{\infty} \d E \sum_I \left[ \(2 \frac{\d \Delta_I}{\d E} \)^2\psi_I^* \phi_I  -2 i  \frac{\d \Delta_I}{\d E} \(\psi_I^* \frac{\partial \phi_I}{\partial E}- \frac{\partial \psi_I^*}{\partial E}\phi_I  \) \right]  \nn \\
&+& \int_0^{\infty} \d E \sum_I \(  \frac{\partial \psi_I^*}{\partial E} + \frac{1}{2} \psi_I^*(0) \delta(E)  \)\( \frac{\partial \phi_I}{\partial E} + \frac{1}{2} \phi_I(0) \delta(E) \) \, .
\ea
The singular delta functions reflect the problem of defining integrals on the half line. 
However these expressions are well-defined for all states for which $ \phi_I(0)$ and $ \psi_I(0)$ vanish. 
Since states of zero energy are not of concern in scattering questions, we may define for all intents and purposes,
\be
\langle \psi |     \widehat{\Delta \mathbb{T}}{\,}^2 | \phi \rangle = \int_0^{\infty} \d E \sum_I \left[ \(2 \frac{\d \Delta_I}{\d E} \)^2\psi_I^* \phi_I  -2 i  \frac{\d \Delta_I}{\d E} \(\psi_I^* \frac{\partial \phi_I}{\partial E}- \frac{\partial \psi_I^*}{\partial E}\phi_I  \) +  \frac{\partial \psi_I^*}{\partial E}  \frac{\partial \phi_I}{\partial E} \right] \, ,\nn
\ee
which is again manifestly Hermitian. With this definition, we note that
\be
\langle \psi |    [\hat H_0, \widehat{\Delta \mathbb{T}}{\,}^2] | \phi \rangle = 2 i \langle \psi |    \widehat{\Delta \mathbb{T}} | \phi \rangle \, ,
\ee
which confirms the tenor of the commutation relations. 
We can repeat this procedure at higher orders, finding (with sufficient constraints on the behaviour of the wavefunctions at $E=0$) well-defined expressions for the expectation values. 
However, due to the increasing constraints, it is clear that there is no formal unitary operator $e^{ i \Delta_E \widehat{\Delta T}}$ that can be defined on the full Hilbert space, and this is how we evade the implications of the Pauli theorem \cite{giannitrapani1997positive,busch1994time,PhysRevA.66.044101,busch2002time}.

In the literature, the time delay is often associated only with the phase factor that occurs in the impact parameter representation of scattering amplitudes $\Delta T \sim 2\frac{\partial \delta(b,E)}{\partial E}$ or equivalently $\Delta v \sim 2\frac{\partial \delta(b,k_v)}{\partial k_v}$. 
We stress that this is only an approximation since the $S$-matrix does not diagonalise in impact parameter basis.
Further, this definition fails to connect with the well-understood situation for small multipoles, which was the original interest of Eisenbud and Wigner \cite{Eisenbud:1948paa,Wigner:1955zz}. 
By contrast, the Wigner-Smith operator is meaningful and matches precisely with the Eisenbud-Wigner definition for partial waves and the impact parameter definition when the latter is a good approximation. 

\subsection{Null time delay}

Given the symmetries of pp-wave spacetimes, it is natural to generalise the usual concept of a time delay to that of a null time delay conjugate to the conserved null momentum $\hat P_v$. 
We aim to define an operator $\widehat{\Delta {\mathbb{V}}}$ which acts on the asymptotic states and is conjugate to the asymptotic null momentum $\hat P_v^0$.

In Heisenberg picture, the quantised free-field metric fluctuations satisfy the master equation \eqref{eq: master equation for point source}. 
Given the null Killing vector it is natural to split the Hermitian fields into creation and annihilation operators
\be\label{sol1}
\hat \Phi(u,v,{\bf x})  = \int_0^{\infty} \frac{\d k_v}{2 \pi} \sum_I \left[ \hat a_{k_v,I} e^{i k_v v} \psi_I(u,{\bf x})+  \hat a_{k_v,I}^{\dagger} e^{-i k_v v} \psi_I^*(u,{\bf x})\right] \, ,
\ee
with $k_v>0$ by definition.
The complex single-particle wavefunctions $ \psi_I(u,{\bf x})$ satisfy the master equation \eqref{eq: master equation for point source} obtained from performing a Fourier transform in the $v$-coordinate $\partial_v \rightarrow i k_v$. 
This takes the form of a non-relativistic Schr\"odinger equation
\begin{equation}\label{eq: master eq as Schr eq}
    i \frac{\partial}{\partial u}  \psi(u,\mathbf{x})  = -\frac{1}{2 k_v} \lp \, \psi(u,\mathbf{x})  + V(u,\mathbf{x}) \psi(u,\mathbf{x}) \, ,
\end{equation}
where the $u$-coordinate plays the role of time, $\lp = \partial_i \partial^i$ is the Laplacian on the $d$-dimensional transverse Euclidean space, and the potential term arises from the curvature of the spacetime.
In this language, we can treat the metric perturbations as a particle of mass $k_v$ being scattered off a potential $V$ sourced by the pp-wave metric. 
For instance, assuming the potential vanishes asymptotically, we can label the incoming states by the incident transverse momenta ${\bf k}$, and so the incident field may be written as
\be 
\lim_{u \rightarrow -\infty }\hat \Phi(u,v,{\bf x})  = \int_0^{\infty} \frac{\d k_v}{2 \pi }  \frac{1}{\sqrt{2 k_v}} \int \frac{\d^d k}{(2 \pi)^d} \left[ \hat a^{\rm in}_{k_v,{\bf k}} e^{i k_v v} e^{i {\bf k} \cdot {\bf x}-i \frac{{\bf k}^2 }{2 k_v} u }+  {\hat a}^{\rm in}_{k_v,{\bf k}}{}^{\dagger}e^{-i k_v v} e^{-i {\bf k} \cdot {\bf x}+i \frac{{\bf k}^2 }{2 k_v} u } \right] \, ,
\label{indefinition}
\ee
with a similar expression for the outgoing field
\be 
\lim_{u \rightarrow \infty }\hat \Phi(u,v,{\bf x})  = \int_0^{\infty} \frac{\d k_v}{2 \pi }   \frac{1}{\sqrt{2 k_v}}  \int \frac{\d^d k}{(2 \pi)^d} \left[ \hat a^{\rm out}_{k_v,{\bf k}} e^{i k_v v} e^{i {\bf k} \cdot {\bf x}-i \frac{{\bf k}^2 }{2 k_v} u } +  {\hat a}^{\rm out}_{k_v,{\bf k}}{}^{\dagger} e^{-i k_v v} e^{-i {\bf k} \cdot {\bf x}+i \frac{{\bf k}^2 }{2 k_v} u } \right] \, .
\label{outdefinition}
\ee
The relation between the in- and out-fields is given by the full Fock space $S$-matrix $\hat {\mathbb{S}} $
\be 
\hat a_{\rm out} = \hat {\mathbb{S}}{}^{\dagger} \hat a_{\rm in} \hat {\mathbb{S}} \, ,
\label{Sinout}
\ee
and the normalisation is 
\be
[\hat a^{\rm in}_{k_v,{\bf k}} ,{\hat a}^{\rm in}_{k'_v,{\bf k}'}{}^{\dagger}]= (2\pi)^d \delta^{(d)}({\bf k-k'}) (2 \pi) \delta(k_v-k_v') \, .
\ee
In writing \eqref{Sinout} we understand that \eqref{indefinition} and \eqref{outdefinition} apply also to the fully interacting Heisenberg fields in the LSZ sense.
One of the central virtues of the pp-wave spacetimes is that, despite being ``time-dependent", the equations for fluctuations are first-order in (null) time.
This means we do not need to worry about the pair-creation characteristic of time-dependent spacetimes\footnote{There is no Bogoliubov transformation between in- and out- creation/annihilation operators since the in- and out- vacuum states are identical $|\rm in \rangle = | \rm out \rangle = |\rm vac \rangle$,} and since $k_v$ is conserved, the $S$-matrix diagonalises on states of definite $k_v$, so the vacuum satisfies $\hat {\mathbb{S}}|\rm vac \rangle =|\rm vac \rangle $. 
In the elastic limit, in which we neglect particle creation, the $S$-matrix projected onto single-particle states is then simply the naive $S$-matrix $\hat{S}$ inferred from the non-relativistic Schr\"odinger equation,
\be
\langle {\rm vac}|  \hat a_{k_v,{\mathbf{k}}'}^{\rm out}   {\hat a}^{\rm in}_{k_v,{\bf k}}{}^{\dagger} |{ \rm vac} \rangle = \langle {\rm vac} | \hat a_{k_v,{\bf k}'}^{\rm in} \hat{\mathbb{S}} {\hat a}^{\rm in}_{k_v,{\bf k}}{}^{\dagger} |{ \rm vac} \rangle  =   \langle {\bf k}'| \hat S | {\bf k} \rangle (2 \pi) \delta(k_v'-k_v)\,,
\ee
with normalisation $  \langle {\bf k}'| {\bf k} \rangle =(2\pi)^d \delta^{(d)}({\bf k-k'}) $. 
In other words, in the absence of particle creation, we may view the entire dynamics from the perspective of the single-particle Hilbert space. 
More generally, single-particle unitarity holds in the sense
\be
\hat {\mathbb{S}}^{\dagger} \hat {\mathbb{S}} = \hat 1 \Rightarrow \hat S^{\dagger} \hat S \le \hat 1 \, .
\ee

To define the null time delay we repeat the derivation of the Wigner-Smith time delay almost verbatim. 
Formally, given the full $S$-matrix
\be
\hat{\mathbb{S}} = \int_0^{\infty}  \d k_v \sum_{I,J} \, |k_v,I \rangle {\mathbb{W}}_{IJ}(k_v)\langle k_v,J | \, ,
\ee
with $|k_v,I \rangle$ a complete set of $\hat P^0_v$ eigenstates with eigenvalue $k_v$, we define the null time delay operator as
\be \label{Deltavdef}
  \widehat{\Delta {\mathbb{V}}}  = -i \int_0^{\infty}  \d k_v \, \sum_I \frac{\partial}{\partial \epsilon} \left[ \hat{\mathbb{S}}^{\dagger}| k_v-\frac{\epsilon}{2}, I \rangle \langle k_v+\frac{\epsilon}{2}, I| \hat{\mathbb{S}} \right] \bigg|_{\epsilon=0}\, .
\ee
This operator is automatically conjugate to the asymptotic null momenta
\be
\[  \hat P_v^0 , \widehat{\Delta {\mathbb{V}}} \] = i \, .
\ee
Projected onto the single-particle Hilbert space this is then
\begin{equation}\label{eq: WS time delay op 2}
    \widehat{\Delta v} = -\frac{i}{2} \hat{S}^{\dagger}\frac{\p \hat{S}}{\p k_v}+\frac{i}{2} \frac{\p \hat{S}^{\dagger}}{\p k_v} \hat{S} \, ,
\end{equation}
for which
\begin{equation}
 \bra{\delta}   \widehat{\Delta v}\ket{\delta} =2\frac{\p\delta}{\p k_v} \, .
\end{equation}
Like the $S$-matrix, the time delay operator should be understood as acting on the asymptotic states. 
Generic asymptotic states are not eigenstates of the $S$-matrix, but the expectation value of the Wigner-Smith operator can always be used as a measure of the time delay.

\subsection{Delay of wave packets}

To understand why the Eisenbud-Wigner-Smith operator \eqref{timedelaydefinition} and its null extension \eqref{Deltavdef} physically corresponds to a time delay, consider a generic incoming one-particle state, which can be written as a superposition of eigenstates of the single-particle $S$-matrix with a given $k_v$\footnote{We shall make this argument for the null time delay, the extension to the usual case being trivial.}
\be \label{wave packet}
|g, {\rm in} \rangle =\int_0^{\infty} \frac{\d k_v}{2\pi} \sum_I g_I(k_v){\hat a}^{\rm in}_{k_v,I}{}^{\dagger} | {\rm vac} \rangle =\int_0^{\infty} \frac{\d k_v}{2\pi} \sum_I g_I(k_v)|\delta_I(k_v) \rangle \, .
\ee
Here, $ {\hat a}^{\rm in}_{k_v,I}{}^{\dagger} $ is the creation operator for a single-particle $S$-matrix eigenstate $|\delta_I(k_v) \rangle$ and the wave packet is encoded in the profile function $g_I(k_v)$. 
Consider the same state with an additional null time delay $\Delta v$, which can be inferred from translating \eqref{sol1} by $v \rightarrow v-\Delta v$ \cite{Bellazzini:2022wzv}
\be
| g,{\rm out} \rangle =e^{i \hat P_v \Delta v} | g,{\rm in} \rangle=\int_0^{\infty}\frac{\d k_v}{2\pi}  \sum_I  g_I(k_v) e^{i k_v \Delta v}{\hat a}^{\rm in}_{k_v,I}{}^{\dagger} | \rm vac \rangle \, .
\ee
Given
\be
 \langle {\rm vac} | \hat a_{k_v',I'}^{\rm in} \hat{\mathbb{S}} {\hat a}^{\rm in}_{k_v,I}{}^{\dagger} |{ \rm vac} \rangle  = 2 \pi \delta_{II'} e^{2 i \delta_I(k_v)} \delta(k_v'-k_v) \, ,
\ee
then
\be
\langle g, {\rm out} |  \hat{\mathbb{S}} |g, {\rm in} \rangle = \int_0^{\infty}\frac{\d k_v}{2\pi} \sum_I |g_I(k_v)|^2 e^{2i \delta_I(k_v)-i k_v \Delta v } \, .
\ee
If we assume the incoming state is peaked at some value $\bar k_v$ and that further $g_I(k_v)$ is dominated by a single $S$-matrix eigenstate $I$, then the amplitude will be maximised when the phase is stationary --- this is precisely when
\be
\Delta v = \frac{2 \partial \delta_I(\bar k_v)}{\partial \bar k_v} \, ,
\ee
and we recover the pp-wave generalisation of the Eisenbud-Wigner time delay. 
In this sense, the outgoing state $ \hat{\mathbb{S}} | g, {\rm in} \rangle $ is well-approximated by an asymptotic state close to indistinguishable from the incoming state were it not for the presence of the time delay. 
The expectation value of the Fock space Wigner-Smith operator is
\be
\langle g,{\rm in} |    \widehat{\Delta {\mathbb{V}}} | g,{\rm in} \rangle = \int_0^{\infty}\frac{\d k_v}{2\pi}  \sum_I  |g_I(k_v)|^2  \frac{2 \partial \delta_I(k_v)}{\partial k_v} \, ,
\ee
which is the appropriately weighted average of the Eisenbud-Wigner time delay for each $S$-matrix $\hat P_v$ eigenstate.

More precisely, given a distribution which, near its peak, is well-approximated as a Gaussian of width $\Delta k_v  \ll \bar k_v$ for a single $S$-matrix eigenstate
\be
g_I(k_v) \approx \frac{1}{\pi^{1/4}\Delta k^{1/2}} e^{-\frac{1}{2 \Delta k_v^2}(k_v-\bar k_v)^2} \, ,
\ee
expanding the phase around $\bar k_v$, the magnitude of the amplitude may be approximated as
\be
|\langle g, {\rm out} |  \hat{\mathbb{S}} |g, {\rm in} \rangle | \sim \frac{1}{\(1+[\delta_I''(\bar k_v) \Delta k_v^2]^2\)^{1/4}} e^{-\frac{1}{4} \frac{\Delta k_v^2 (2 \delta_I'(\bar k_v)-\Delta v)^2}{\(1+[\delta_I''(\bar k_v) \Delta k_v^2]^2\)}} \sim e^{-\frac{1}{4} \Delta k_v^2 (2 \delta_I'(\bar k_v)-\Delta v)^2} \, ,
\ee
assuming $\delta_I''(\bar k_v) \Delta k_v^2 \ll 1$. 
Although, mathematically, the amplitude is peaked at $\Delta v =2 \delta_I'(\bar k_v)$, in practice, the amplitude is order unity over a range of width $1/\Delta k$ --- this is the resolvability criterion coming from the uncertainty principle. 
Thus a more precise statement is that the time delay is
\be
\Delta v_I = \frac{2 \partial \delta_I(\bar k_v)}{\partial \bar k_v} \pm {\cal O}(\Delta k_v^{-1})\, .
\ee
Given that $\Delta k_v \ll \bar k_v$ in order to trust the saddle point approximation, the strongest statement of causality we can infer from this argument is
\be
\Delta v_I \gtrsim -\bar k_v^{-1} \, .
\ee

\subsection{Uncertainty of time delay}

\label{sec:uncertainty}

A generic incoming state is not an eigenstate of the $S$-matrix or time delay operator. 
Indeed, normalisable scattering states are always wave packets. 
Since na\"ively $[\Delta  \hat {\mathbb{V}}, \hat P_v^0 ]=i$, there is an inevitable uncertainty relation between the time delay and the asymptotic null energy. 
We may define this uncertainty in the usual way
\be
\Delta_{\rm un} v^2 = \langle g,{\rm in}| {\widehat{\Delta {\mathbb{V}}}} {\,}^2| g, {\rm in} \rangle- \( \langle g,{\rm in}|{\widehat{\Delta {\mathbb{V}}}} | g, {\rm in} \rangle\)^2 \, .
\ee
Evaluating the expectation value on the single-particle wave packets \eqref{wave packet} gives
\be
\langle g,{\rm in} |    {\widehat{\Delta {\mathbb{V}}}} {\,}^2 | g,{\rm in} \rangle = \int_0^{\infty}\frac{\d k_v}{2\pi}  \sum_I  \left(  |g_I(k_v)|^2  \[ \frac{2 \partial \delta_I(k_v)}{\partial k_v}\]^2 + \bigg|\frac{\partial g_I(k_v)}{\partial k_v} \bigg|^2 \right)\, ,
\ee
and so
\be
\Delta_{\rm un} v^2 = \left\llangle \(\frac{2 \partial \delta_I}{\partial k_v}\)^2 \right\rrangle-\left\llangle \frac{2 \partial \delta_I}{\partial k_v} \right\rrangle^2 +  \int_0^{\infty}\frac{\d k_v}{2\pi} \sum_I \Big|\frac{\partial g_I(k_v)}{\partial k_v} \Big|^2 \, ,
\ee
where $\llangle  X_I \rrangle =  \int_0^{\infty}\frac{\d k_v}{2\pi}  \sum_I   |g_I(k_v)|^2 X_I(k_v) $. 
The last term is the usual uncertainty due to the fact that we are considering wave packets in $k_v$, while the first two terms are contributions to the uncertainty which come directly from the scattering and vanish in the limit of no scattering. 
We thus have the bound
\be
\Delta_{\rm un} v^2 \ge \int_0^{\infty}\frac{\d k_v}{2\pi}  \sum_I \Big|\frac{\partial g_I(k_v)}{\partial k_v} \Big|^2 \, .
\ee
Although $k_v\ge 0$, we may view $f(k_v)$ as a wavefunction that happens to vanish for $k_v<0$, and so by the usual reasoning we have
\be
\Delta_{\rm un} v \, \Delta  k_v \ge 1/2 \, ,
\ee
with
\be
( \Delta  k_v)^2 =  \llangle k_v^2 \rrangle -  \llangle k_v \rrangle^2 \, .  \ee
This was explicit in the wave packets considered above.
The contribution to the uncertainty from scattering at fixed $k_v$ can be expressed in terms of the single-particle Hilbert space time delay as
\be
\delta  v = \sqrt{\bra{\rm in}\widehat{\Delta v}{\,}^2\ket{\rm in} -\bra{\rm in}\widehat{\Delta v}\ket{\rm in} ^2} \, ,
 \ee
 with
 \be
\bra{\rm in} \widehat{\Delta v}{\,}^2\ket{\rm in} =\bra{\rm in} \frac{\p \hat{S}^{\dagger}}{\p k_v} \frac{\p \hat{S}}{\p k_v}\ket{\rm in} \,.
 \ee
We shall make crucial use of this fact later. 

\subsection{Asymptotic time delay in Schr\"odinger picture }
\label{sec:Schrodinger1}

Consider the Schr\"odinger-like equation of the form
\be
i \frac{\partial}{\partial u} \psi(u,\mathbf{x})  =- \frac{1}{2 k_v} \nabla^2 \psi(u,\mathbf{x}) + V(u,{\bf x}) \psi(u,\mathbf{x}) \, ,
\ee
or more abstractly, in first-quantised language,
\be
i \frac{\partial}{\partial u} |\psi \rangle =\hat H  |\psi \rangle \, ,
\ee
where $\hat H= \hat H_0+\hat H_1$ with $\hat H_0 = \hat p^2/(2 k_v)$ and $\hat H_1= V(u,\hat x)$. Let us assume without loss of generality that interactions vanish for $u<u_i$ and $u>u_f$, with the usual case recovered in the limit $|u_i|,|u_f| \rightarrow \infty$. 
Then, the $S$-matrix is the time-evolution operator in interacting picture
\be
\hat S =\hat {\cal U}(u_f,u_i)={\cal T} e^{-i \int_{u_i}^{u_f} \d u \hat H_1^{\rm int}(u) } = e^{i \hat H_0 u_f} \hat U(u_f,u_i) e^{-i \hat H_0 u_i}   \, ,
\ee
where $\hat H_1^{\rm int}(u) =e^{i \hat H_0 u} \hat H_1 e^{-i \hat H_0 u}$ is the perturbation in interacting picture and $\hat U(u_f,u_i) = {\cal T} e^{-i \int_{u_i}^{u_f} \d u \hat H(u) } $
 is the Schr\"odinger-picture evolution operator. 
 To determine the time delay we need
\ba
 \frac{\partial}{\partial k_v} \hat S &=&   \frac{\partial}{\partial k_v}{\cal T} e^{-i  \int_{u_i}^{u_f}\d u \hat H_1^{\rm int}(u) } =  - i  \int_{u_i}^{u_f}\d u' {\cal T} \left[  \frac{\partial \hat H_1^{\rm int}(u') }{\partial k_v} e^{-i  \int_{u_i}^{u_f} \d u \hat H_1^{\rm int}(u) }  \right] \nn \\
  &=&  - i  \int_{u_i}^{u_f} \d u'  \hat {\cal U}(u_f,u') \frac{\partial \hat H_1^{\rm int}(u') }{\partial k_v}  \hat {\cal U}(u',u_i)  \, ,
\ea
from which we can infer
\ba
 \widehat{\Delta v} &=&  -i \hat{S}^{\dagger}\frac{\p \hat{S}}{\p k_v} =-  \int_{u_i}^{u_f}  \d u' \, \hat {\cal U}(u_f,u_i)^{\dagger} \hat {\cal U}(u_f,u') \frac{\partial \hat H_1^{\rm int}(u') }{\partial k_v}  \hat {\cal U}(u',u_i) \nn \\
&=& - \int_{u_i}^{u_f}  \d u' \, \hat {\cal U}(u',u_i)^{\dagger} \frac{\partial \hat H_1^{\rm int}(u') }{\partial k_v}  \hat {\cal U}(u',u_i) \, .
\ea
Now let us convert this into Schr\"odinger picture. 
First note since $\hat H_0$ commutes with $ \frac{\partial \hat H_0 }{\partial k_v}$, from the definition of the interacting picture Hamiltonian, we have
\ba
\frac{\partial \hat H_1^{\rm int}(u) }{\partial k_v}  &=& e^{i \hat H_0 u} \frac{\partial \hat H_1(u) }{\partial k_v}  e^{-i \hat H_0 u} + i u \left[ \frac{\partial \hat H_0 }{\partial k_v}, \hat H_1^{\rm int}(u) \right] \, \\
&=& e^{i \hat H_0 u} \left( \frac{\partial \hat H_1(u) }{\partial k_v} + i u \left[ \frac{\partial \hat H_0 }{\partial k_v}, \hat H_1(u) \right]  \right) e^{-i \hat H_0 u} \, .
\ea
Denoting the time delay operator with finite initial and final times as
\be
 \widehat{\Delta v}= - \int_{u_i}^{u_f}  \d u' \, \hat {\cal U}(u',u_i)^{\dagger} \frac{\partial \hat H_1^{\rm int}(u') }{\partial k_v}  \hat {\cal U}(u',u_u) \, ,
\ee
this becomes, in terms of Schr\"odinger picture quantities,
\be
 \widehat{\Delta v}= - \int_{u_i}^{u_f}  \d u \, e^{i \hat H_0 u_i} \hat U(u,u_i)^{\dagger}  \left( \frac{\partial \hat H_1(u) }{\partial k_v} + i u \left[ \frac{\partial \hat H_0 }{\partial k_v}, \hat H_1(u) \right]  \right)   \hat U(u,u_i) e^{-i \hat H_0 u_i} \, .
\ee
At the level of the expectation value this is
\ba\label{eq:Time delay exact}
\Delta v  = \langle {\rm in} | \widehat{\Delta v} |{ \rm in} \rangle
= - \int_{u_i}^{u_f} \d u  \langle \psi(u)| \hat Q(u)   | \psi(u) \rangle \, ,
\ea
where
\be\label{eq:Time delay exact2}
\hat Q(u) =  \frac{\partial \hat H_1(u) }{\partial k_v} + i u \left[ \frac{\partial \hat H_0 }{\partial k_v}, \hat H_1(u) \right]   \, ,
\ee
and
\be
|\psi(u) \rangle = \hat U(u,u_i) e^{-i \hat H_0 u_i}  |{\rm in} \rangle \, ,
\ee
which we recognise as the solution to the Schr\"odinger equation with initial condition
\be
|\psi(u_i)\rangle = e^{-i \hat H_0 u_i}  |{\rm in} \rangle \, .
\ee
The $u$ in the second term in \eqref{eq:Time delay exact2} is specific to the convention that states in the Heisenberg picture are defined as Schr\"odinger states evaluated at $u=0$, and interacting picture states are referred to similarly. 
A different choice $u \rightarrow u-u_0$ can be absorbed by a unitary transformation in $|\psi \rangle$, so \eqref{eq:Time delay exact} is universal when considering the set of all possible states $|\psi \rangle$.

In the situation in which all interactions vanish for $u<0$, \textit{i.e.} $u_i=0$, the time delay can be written simply in terms of Heisenberg operators and the Heisenberg state $|\psi(0) \rangle=|\rm in \rangle$,
\be\label{Timedelay2}
\Delta v  = - \int_{0}^{u_f} \d u  \langle \psi(0) | \hat Q_H(u)   | \psi(0) \rangle \, ,
\ee
with
\be
\hat Q_H(u) =\hat U(u,0)^{\dagger}  \hat Q(u)  \hat U(u,0) \, .
\ee
We may view $\hat Q_H(u)$ as the operator version of the classical quantity that determines the rate of increase in the $v$-time delay per unit $u$-time
\be
Q(u) = - \frac{\d \Delta v(u)}{\d u }  \, .
\ee

\subsection{EFT time delay in Schr\"odinger picture }
\label{EFTSmatrix}

In the previous section, we defined the total time delay.
This is typically a combination of the GR time delay and corrections that come from interactions which, in the EFT, are captured by contributions suppressed by inverse powers of the cutoff. 
More precisely, the form of the Hamiltonian can be split as
\be
\hat H(u) = \hat H_{\rm GR}(u) + \hat H_{\rm EFT}(u) \, ,
\ee
for which
\be
\lim_{\Lambda \rightarrow \infty } \hat H_{\rm EFT} =0 \, .
\ee
In many situations, the cutoff $\Lambda$ must be necessarily below the Planck scale, so one should also take the limit $\mpl \to \infty$. Note that, since shockwaves are exact solutions, we can always appropriately scale $f$ in that limit so as to maintain a non-trivial background. 

Alternatively, a more pragmatic (and generally applicable) definition of the split is that $\hat H_{\rm GR}$ denotes the terms in the Hamiltonian that come from the two- (and lower) derivative part of the action, while $\hat H_{\rm EFT}$ are those terms that arise from higher-derivative (irrelevant) operators. 
Given this split, we may give a more local definition of the time delay which encodes the effect of the EFT interactions relative to the GR background. 
To do this, we choose the map from Schr\"odinger to interacting picture by
\be
|\psi(u) \rangle = {\cal T} e^{-i \int_0^{u} \d u' \, \hat H_{\rm GR}(u')} |\psi_I(u) \rangle \, .
\ee
Then we may define the EFT $S$-matrix via
\be
\hat S =\hat {\cal W}(u_f,u_i)={\cal T} e^{-i \int_{u_i}^{u_f} \d u \hat H_{\rm EFT}^{\rm int}(u) }   \, ,
\ee
with now
\be
\hat H_{\rm EFT}^{\rm int}(u) =\hat {\cal K}(u,0)^{\dagger} \hat H_{\rm EFT}(u) \hat {\cal K}(u,0) \, ,
\ee
and
\be
\hat {\cal K}(u,u_i)={\cal T} e^{-i \int_{u_i}^{u} \d u' \, \hat H_{\rm GR}(u')}  \, .
\ee
The EFT time delay is then
\be
 \widehat{\Delta v}_{\rm EFT}= - \int_{u_i}^{u_f}  \d u' \, \hat{\cal K}(u',u_i)^{\dagger} \frac{\partial \hat H_{\rm EFT}^{\rm int}(u') }{\partial k_v}  \hat {\cal K}(u',u_u) \, .
\ee
Using
\ba
\frac{\partial}{\partial k_v} \hat{\mathcal K}(u,0)&=& - i \int_0^{u} \d u'\, \hat {\cal K}(u,u')   \frac{\partial  \hat H_{\rm GR}(u')}{\partial k_v} \hat {\cal K}(u',0) \nn \\
&=&- i \int_0^{u} \d u' \hat {\cal K}(u,u')   \frac{\partial  \hat H_{\rm GR}(u')}{\partial k_v}\hat {\cal K}(u,u')^{\dagger}   \hat {\cal K}(u,0) \, ,
\ea
this then becomes
\be\label{Timedelay3}
\Delta v_{\rm EFT}  = - \int_{u_i}^{u_f} \d u  \langle \psi(u)| \hat Q_{\rm EFT}(u)| \psi(u) \rangle \, ,
\ee
where now
\be
\hat Q_{\rm EFT}(u)=  \frac{\partial \hat H_{\rm EFT}(u) }{\partial k_v} + i \int_0^u \d u' \left[ \hat {\cal K}(u,u')   \frac{\partial \hat H_{\rm GR}(u')}{\partial k_v} \hat {\cal K}(u,u')^{\dagger}  , \hat H_{\rm EFT}(u) \right]    \, ,
\ee
and
\be
| \psi(u) \rangle =\hat {\cal K}(u,0)  \hat {\cal W}(u,u_i) | \rm in \rangle \, .
\ee
For most purposes, it is sufficient to determine the EFT time delay to first order in $\hat H_{\rm EFT}$ and so we just directly compute
\be
\Delta v_{\rm EFT}  \approx  -\frac{\partial}{\partial k_v} \int_{u_i}^{u_f} \d u  \,  \langle{ \rm in} | \hat H^{\rm int}_{\rm EFT}(u) | {\rm in }\rangle \, .
\ee

\subsection{Time delay from Wigner functions}

In practice, the exact formula for the scattering time delay \eqref{eq:Time delay exact} is difficult to determine other than by means of an approximation or by numerical evolution. 
It is also relatively straightforward to infer the time delay in perturbation theory (Born series), and we shall do so in Section \ref{sec: quantum calcs}. 
Our main tool will however be the eikonal and semi-classical approximations. 
It proves useful to rewrite \eqref{eq:Time delay exact} in terms of Wigner functions since they are most closely tied to a (nearly) classical interpretation of the scattering. 
Denoting the density operator as
\be
\hat \rho(u) = | \psi(u) \rangle \langle \psi(u) | \, ,
\ee
then the total time delay is
\be
\Delta v = -  \int_{u_i}^{u_f} \d u\, {\rm Tr}[\hat \rho(u) \hat Q(u) ]\, .
\ee
Denoting the Wigner phase-space function by
\be
W({\bf x},{\bf p},u) = \int \d^d {\bf a} \,  e^{- i {\bf a} \cdot {\bf p}} \langle {\bf x} + {\bf a}/2 | \hat \rho(u) | {\bf x} - {\bf a}/2 \rangle \, ,
\ee
and the associated phase space representation of $\hat Q(u)$ by
\be
Q({\bf x},{\bf p},u) = \int \d^d {\bf a} \,  e^{- i {\bf a} \cdot {\bf p}} \langle {\bf x} + {\bf a}/2 | \hat Q(u) | {\bf x} - {\bf a}/2 \rangle \, ,
\ee
then the asymptotic/total time delay can be rewritten as
\be
\Delta v= -  \int_{u_I}^{u_f} \d u \int \d^d {\bf x} \int \frac{\d^d {\bf p}}{(2 \pi)^d} W({\bf x},{\bf p},u)  Q({\bf x},{\bf p},u)  \, .
\ee
A similar expression holds for the EFT time delay
\be
\Delta v_{\rm EFT}= -  \int_{u_i}^{u_f} \d u \int \d^d {\bf x} \int \frac{\d^d {\bf p}}{(2 \pi)^d} W({\bf x},{\bf p},u) Q_{\rm EFT}({\bf x},{\bf p},u)  \, .
\ee
In both cases, the operator $\hat Q$ as a function of its constituent operators can be replaced with the phase-space function as a function of the phase-space versions of the same, with operator products replaced by the Moyal product, denoted by $\star$. 
Similarly, the equation for the density operator in Schr\"odinger picture 
\be
i  \frac{\partial \hat \rho(u) }{\partial u}  =-\[ \hat \rho(u), \hat H \] \, ,
\ee
is replaced by the quantum Liouville equation
\be
i  \frac{\partial  W({\bf x},{\bf p},u)}{\partial u}  =-\[ W({\bf x},{\bf p},u) \star H({\bf x},{\bf p},u) - H({\bf x},{\bf p},u) \star W({\bf x},{\bf p},u) \] \, .
\ee
These expressions may appear unwieldy, but they prove to be the most useful for deriving the semi-classical approximation.

\subsection{Semi-classical approximation}
\label{subsec: semiclassical}

In the semi-classical approximation, we intuitively expect the dynamics to be dominated by that of the classical particle trajectories. 
This is most straightforwardly derived by reintroducing $\hbar$ in the conventional places within the Schr\"odinger equation, and taking the limit $\hbar \rightarrow 0$. 
In this limit, the Moyal product reduces to the ordinary product
\be
\lim_{\hbar \rightarrow 0} A({\bf x},{\bf p},u) \star B({\bf x},{\bf p},u) \rightarrow  A({\bf x},{\bf p},u) B({\bf x},{\bf p},u) \, ,
\ee
and the expression for $Q=-\d \Delta v/\d u$ is given by its classical version with commutators replaced by Poisson brackets
\ba
Q({\bf x},{\bf p},u)  &=& \frac{\partial H_1({\bf x},{\bf p},u)}{\partial k_v} -u \left\{\frac{\partial H_0({\bf x},{\bf p},u)}{\partial k_v}, H_1({\bf x},{\bf p},u)  \right\}  \\
&=& \frac{\partial}{\partial k_v } V(u,{\bf x})- u \frac{\bf p}{k_v^2}\cdot \boldsymbol{\nabla} V(u,{\bf x}) \, .\label{eq:Q} 
\ea
Finally, at leading order in the semi-classical approximation, the dynamical equation for the Wigner function reduces to the Liouville equation
\be
\frac{\partial}{\partial u} W({\bf x},{\bf p},u) =-\{W({\bf x},{\bf p},u), H({\bf x},{\bf p},u)  \} \, .
\ee
The solution of this linear equation is straightforward
\be
 W({\bf x},{\bf p},u) = \int \d^d {\bf x_0} \int \frac{\d^d {\bf p_0}}{(2 \pi)^d}   W_0({\bf x}_0,{\bf p}_0) \delta^{(d)}\big({\bf x} - {\bf x}_{\rm cl}({\bf x}_0,{\bf p}_0,u)\big) (2 \pi)^d  \delta^{(d)}\big({\bf p} - {\bf p}_{\rm cl}({\bf x}_0,{\bf p}_0,u)\big) \, ,\nn
\ee
where $  W_0({\bf x}_0,{\bf p}_0) $ is the Wigner function at time $u=0$, and $\big(\bf{x}_{\rm cl}({\bf x}_0,{\bf p}_0,u) \, , {\bf p}_{\rm cl}({\bf x}_0,{\bf p}_0,u)\big) $ denotes the solution of the classical equations of motion in phase space, given the initial data $({\bf x}_{\rm cl},{\bf p}_{\rm cl})=({\bf x}_0,{\bf p}_0)$ at $u=0$.

Substituting into the expression for the time delay, we finally obtain at leading order in the semi-classical approximation, 
\be\label{eq:Time delay semiclassical}
(\Delta v)_{\rm semi-classical}  = -  \int \d^d {\bf x}_0 \int \frac{\d^d {\bf p}_0}{(2 \pi)^d} W_0({\bf x}_0,{\bf p}_0)   \int_{u_i}^{u_f} \d u \,  Q({\bf x}_{\rm cl}({\bf x}_0,{\bf p}_0,u),{\bf p}_{\rm cl}({\bf x}_0,{\bf p}_0,u),u)  \, .
\ee
The classical approximation would further neglect the effect of uncertainty and quantum diffusion:
\be\label{eq:Time delay classical}
(\Delta v)_{\rm classical}  = -  \int_{u_i}^{u_f} \d u \,  Q({\bf x}_{\rm cl}({\bf x}_0,{\bf p}_0,u),{\bf p}_{\rm cl}({\bf x}_0,{\bf p}_0,u),u) \, .
\ee
Stated in words: In order to determine the semi-classical scattering time delay, we first infer the classical time delay \eqref{eq:Time delay classical} associated with an arbitrary classical initial condition $({\bf x}_0,{\bf p}_0)$ in phase space. 
We then perform the average of those classical initial conditions according to whatever the initial wavefunction and hence initial Wigner distribution is.
The latter step automatically incorporates the effects of the uncertainty principle and quantum diffusion into the calculation of the time delay. 
It is clear that the classical approximation can only be trusted when most of the time delay is built up well before the diffusion time, for which we can no longer neglect the quantum induced spread.

\subsection{Quantum diffusion}\label{subsec: becoming quantum}
A crucial fact which is embedded in the above discussion so far, is that a generic initial wavefunction $\psi_{0}$ can never be perfectly localised both in position and momenta by virtue of the uncertainty principle $\Delta \mathbf{x} \cdot \Delta \mathbf{k} \ge 1/2$. 
Thus, generic initial wave packets have some spread in transverse space.
Viewed as a field in a low-energy EFT, its spatial extent $\sigma \sim \abs*{\Delta \mathbf{x}}$ is additionally bounded below by the EFT cut-off as $\sigma \gg \Lambda^{-1}$ (see appendix~\ref{app: bound on wavefunction spread}).
Even in the absence of a potential, this initial uncertainty grows with time due to quantum diffusion and limits the amount of time delay that can be built up.
Consider a spherically symmetric Gaussian initial profile
\begin{equation}\label{eq: Gaussian wave packet}
  \psi_{0}(\mathbf{x})  =\frac{1}{(2 \pi \sigma^2)^{d/4}}\text{exp}\left[- \frac{|\mathbf{x}-\mathbf{b}|^2}{4 \sigma^2}\right] \, ,
\end{equation}
centred at ${\bf b}$ with width $\sigma$ much smaller than the distance to the source(s) $\sigma \ll b$.
The associated Wigner function at the initial time is
\be
W_0({\bf x_0},{\bf p_0}) =2^d \text{exp}\left[- \frac{|\mathbf{x_0}-\mathbf{b}|^2}{2 \sigma^2}\right] \text{exp}\left[- \frac{\mathbf{p_0}^2}{2 \Delta {p}^2}\right] \, ,
\ee
with
\be
\sigma \Delta p = \frac{1}{2} \, .
\ee
After a time $u$ under free evolution, the Wigner function will be peaked around the straight-line trajectory
\be
W({\bf x},{\bf p},u) =2^d \text{exp}\left[- \frac{|\mathbf{x}-{\bf p} u/k_v-\mathbf{b}|^2}{2 \sigma^2}\right] \text{exp}\left[- \frac{\mathbf{p}^2}{2 \Delta {p}^2}\right] \, ,
\ee
and the wave packet will diffuse to one with an effective width of $\sigma_{\text{eff}}^2(u) = \sigma^2 +  (u/2 k_v \sigma)^2$.

This diffusion inevitably provides a cutoff on the scattering process and the build-up of the time delay/advance. 
For example, if we consider the scattering of two localised objects of impact parameter $b$, we no longer expect a significant contribution to the time delay once the diffusion of the wave packet is comparable to the impact parameter. 
This occurs at the ``diffusion time" for which
\begin{equation}\label{eq: diffusion timescale}
    \sigma_{\text{eff}}(u_{\text{diff}}) \sim  b \quad \implies \quad u_{\text{diff}} \sim  2 k_v \sigma \sqrt{b^2 - \sigma^2}.
\end{equation}
The maximum diffusion time is then seen to be
\be
u_{\rm diff} \sim k_v b^2 \, .
\ee
However, if the initial wave packet is well-localised, it may be even smaller $u_{\rm diff} \sim k_v  \sigma b$. 
Either way, the wavefunction $\psi(u, \mathbf{x})$ can no longer be considered localised after a finite time $u_{\text{diff}}$. 
We shall confirm this explicitly by more detailed calculations below.

\subsection{Eikonal approximation}

Having written everything in terms of the Schr\"odinger picture makes it easier to compare with the standard eikonal approximation for the Schr\"odinger equation. 
Let us first do this for the $S$-matrix describing scattering from states of momenta ${\bf k}$ to ${\bf k'}$. 
In order to maintain symmetry between the incoming and outgoing state, we define the average momenta ${\bf n = (k'+k)}/2$. 
The eikonal approximation is derived by first factorising out the free evolution at the average momenta
\be
\psi = e^{i {\bf n} \cdot {\bf x}}e^{-i {\bf n}^2 u/2k_v}  \chi \, ,
\ee
so that the Schr\"odinger equation becomes
\be
i \frac{\partial}{\partial u} \chi({\bf x},u) =- i \frac{\bf n}{k_v} \cdot \boldsymbol{\nabla} \chi({\bf x},u)-\frac{1}{2 k_v} \nabla_{d}^2 \chi({\bf x},u) + V(u,{\bf x}) \chi({\bf x},u) \, .
\ee
At leading order in the approximation, we neglect the Laplacian term, meaning that one of the conditions for the approximation to be valid is that the momentum transfer is small, which is achieved for small scattering angles.
Then, changing variables to ${\bf  x} = {\bf y}+ {\bf n}u/k_v$, the resulting equation
\be
i \frac{\partial}{\partial u} \chi({\bf y},u) \approx  V(u,{\bf y}+{\bf n} u/k_v) \chi({\bf y},u) \, ,
\ee
can be exactly integrated
\be
\chi({\bf y},u_f) \approx e^{-i \int^{u_f}_{u_i} \d u\, V(u,{\bf y}+{\bf n} u/k_v)  } \chi({\bf y},u_i) \, .
\ee
The $S$-matrix between initial state $\chi_i = e^{i {\bf (k-n)\cdot y}}$ and final state $\chi_f = e^{i {\bf (k'-n)\cdot y}}$ is then
\be\label{eikonal1}
\langle {\bf k'}| \hat S | {\bf k} \rangle = \int \d^d {\bf y } e^{-i {\bf q \cdot y}} e^{-i \int^{u_f}_{u_i} \d u\, V(u,{\bf y}+{\bf n} u/k_{v})  }  \, ,
\ee
with momentum transfer ${\bf q=k'-k}$. 
In the familiar case, for which $V$ is independent of $u$ with $|u_{i,f}| \rightarrow \infty$, we may further define ${\bf y = b + \hat n }z$, where ${\bf n \cdot b}=0$. 
Then, by shifting $u \rightarrow u-|{\bf n}|z$, the $S$-matrix becomes
\be
\langle {\bf k'}| \hat S | {\bf k} \rangle = 2 \pi\delta({\bf q \cdot \hat n}) \int \d^{d-1}{\bf b } \, e^{-i {\bf q \cdot b}} e^{-i \int^{\infty}_{-\infty} \d u\, V({\bf b}+{\bf n} u/k_v)  }  \, ,
\ee
which we recognise to be the familiar eikonal approximation for scattering off a time-independent potential. 

To determine the scattering time delay for an incoming state which is peaked at incoming momentum ${\bf p}_0$, it is convenient to further approximate \eqref{eikonal1} by replacing ${\bf n}$ by ${\bf p_0}$
\be\label{eikonal2}
\langle {\bf k'}| \hat S | {\bf k} \rangle \approx \int \d^d {\bf y } e^{-i {\bf q \cdot y}} e^{-i \int^{u_f}_{u_i} \d u\,  V(u,{\bf y}+{\bf p_0} u/k_v)  }  \, ,
\ee
so that
\be
\langle {\bf k'}| \hat S | \psi \rangle \approx \int \d^d {\bf y } e^{-i {\bf k'.y}} \psi(y) e^{-i \int^{u_f}_{u_i} \d u \, V(u,{\bf y}+{\bf p_0} u/k_v)  }  \, .
\ee
This allows the integrals to be performed, giving
\begin{equation}
    \begin{aligned}
    & \Delta v = -i \langle \psi | \hat S^{\dagger} \frac{\partial}{\partial k_v} \hat S | \psi \rangle \\
    & =  -i  \int \d^d {\bf k'} \langle \psi | \hat S^{\dagger} |{\bf k'} \rangle \frac{\partial}{\partial k_v} \langle {\bf k'}| \hat S | \psi \rangle \\
    &\approx -\int^{u_f}_{u_i} \d u \int \d^d {\bf y} |\psi(y)|^2 \[\frac{\partial}{\partial k_v} V\(u,{\bf y}+\frac{{\bf p_0} u}{k_v}\)  - u  \frac{{\bf p_0}}{k_v^2} \cdot \boldsymbol{\nabla}  V\(u,{\bf y}+\frac{{\bf p_0} u}{k_v}\) \] \, .
    \end{aligned}
\end{equation}
Given the Wigner function will be similarly peaked at $p_0$, we can rewrite the eikonal scattering time delay as
\ba
(\Delta v)_{\rm eikonal} &\approx&-\int^{u_f}_{u_i} \d u \int \d^d {\bf y} \int \d^d {\bf p_0}  \, W_0({\bf y,p_0}) \[\frac{\partial}{\partial k_v} V\(u,{\bf y}+\frac{{\bf p_0} u}{k_v}\)  - u  \frac{{\bf p_0}}{k_v^2} \cdot \boldsymbol{\nabla}  V\(u,{\bf y}+\frac{{\bf p_0} u}{k_v}\) \] \,  \nn \\
&\approx&-\frac{\partial }{\partial k_v}  \int^{u_f}_{u_i} \d u \int \d^d {\bf y} \int \d^d {\bf p_0}  \, W_0({\bf y,p_0})  V\(u,{\bf y}+\frac{{\bf p_0} u}{k_v}\)  \, .
\ea
Identifying $\bf y = x_0$ as the classical position at $u=0$, we recognise the leading-order eikonal approximation to simply be the semi-classical approximation evaluated such that the trajectories of the particles are taken to be straight lines, which would be expected to be valid for small angle scattering. 
As such, it neglects certain higher-order effects of scattering, which cause the mean trajectory to depart from a straight line.

If we further neglect the effect of quantum uncertainty/diffusion then this reduces to the ``classical" eikonal form
\be\label{eq:classicaleikonal}
(\Delta v)_{\rm eikonal}  \approx-\frac{\partial }{\partial k_v}  \int^{u_f}_{u_i} \d u\, V\(u,{\bf x_0}+\frac{{\bf p_0} u}{k_v}\)  \, .
\ee
In realistic situations, the potential vanishes at large distances, so this integral will be maximised for ${\bf p_0} $ as small as possible. 
At the impact parameter $\bf x_0 = b$, the eikonal time delay reduces to
\ba
(\Delta v)_{\rm eikonal} &\approx&- \frac{\partial }{\partial k_v}  \int^{u_f}_{u_i} \d u\,  V(u,{\bf b})  \, ,
\ea
which is the expression considered in \cite{Camanho:2014apa}. 
This result may be obtained more directly by neglecting scattering in the transverse directions from the outset.
This amounts to ignoring the Laplacian-term in \eqref{eq: master eq as Schr eq} so that the particle deviates from its initial configuration only by a phase.
Setting up our initial conditions at time $u = 0$ with initial particle displacement $\mathbf{x} = \mathbf{b} = b\mathbf{\hat{b}}$, then the approximate solution at a later time is
\be
    \psi(u,\mathbf{b}) = \psi_{0}\, \text{exp}\left[2 i \delta(u,\mathbf{b})\right] \, , \quad 2\delta(u,\mathbf{x}) = -\int_0^u \text{d}u' V(u',\mathbf{x}) \,,\label{eq: 0th order phase}
\ee
from which the time delay is easily obtained
\be\label{impact}
(\Delta v)_{\rm eikonal}=2\frac{\partial \delta(u,{\bf b})}{\partial k_v}=-\frac{\partial }{\partial k_v}  \int_0^u \text{d}u' V(u',\mathbf{b}) \, .
\ee
It is clear from our previous derivations that \eqref{impact} can only be trusted when the integral is taken over a range of $u$ for which the effects of diffusion and scattering are neglected.

\section{Shockwave scattering in the eikonal approximation}\label{subsec: pt source scattering}

As outlined above, there are two physical effects which limit the maximum time delay/advance that can be built up in a given experiment: (i) Quantum diffusion and (ii) Scattering. 
In this section we shall consider the maximum time delay/advance that can be obtained by a sequence (or continuum) of shockwaves, each of which may be viewed as a point source in the transverse direction. 
In this case it will turn out that the main limitation on an accumulation of time delay is from (ii) the scattering, and so we shall neglect the quantum diffusion.

\subsection{Warm-up: Scattering due to a single point source}

Before proceeding to the case of multiple (or a continuum of) shockwaves, let us consider the simplest scenario of a particle scattering off a single point shockwave. 
An idealised shockwave describing a boosted black hole/particle of null momentum $P_u^{(2)}$ takes the form
\be
\d s^2 = 2 \d u\, \d v + \alpha \delta(u) \d u^2 + \d {\bf x}^2 \, ,
\ee
with $\alpha =4\Gamma((d-2)/2)/\pi^{(d-2)/2}\, G P_u^{(2)} $. 
As discussed in Section \ref{regulatedshockwave}, this solution is singular from the point of view of the EFT, and to be treated within the EFT it must be smoothed out over a null distance $L \sim k_v/\Lambda^2$ to, for example,
\be
    \d s^2 = 2 \d u\, \d v + \alpha \frac{1}{\sqrt{2\pi L^2}}\exp\left[-\frac{u^2}{2L^2}\right] \d u^2 + \d {\bf x}^2 \, .
\ee
If $L$ is taken to be too large then we need to account for the evolution of the trajectory through the Gaussian which we do in the next section. 
For now we choose it to be the smallest possible consistent with the EFT. 
Since the potential lasts only for a short $u$-time, the eikonal approximation is good, and including the EFT correction, this is
\be \label{eq:singleshocktimedelay}
\Delta v = \frac{\alpha}{2 b^{d-2}} - A   (d-2) \frac{\cGB}{\Lambda^2}\frac{\alpha}{2 b^{d}} \, 
\ee
to a good approximation.
It is easy to see that the GR/Shapiro contribution can be made arbitrarily large and resolvable without contradicting the validity of the EFT. 
This is of course as it should be, since the GR effects are experimentally observable. 
By contrast, in this situation, the EFT contribution to the time delay is bounded. 
The EFT bound on the Riemann curvature applied at the impact parameter amounts \eqref{eq: EFT RoV strength bound} to
\be
\frac{k_v^2 \alpha}{L b^d} = \frac{\Lambda^2 k_v^2 \alpha}{k_v b^d}  \ll \Lambda^4 \, ,
\ee
which simplifies to
\be
\frac{ k_v \alpha}{ b^d}  \ll \Lambda^2 \, ,
\ee
from which we see
\be
k_v |\Delta v_{\rm EFT}| \lesssim |\cGB| \, .
\ee
Thus, the EFT contribution to the time delay, meaning the contribution from the GB-term, is only resolvable if $|\cGB|$ is taken to be larger than unity.
This is outside of the expectations of the EFT, and more precisely the expectations of positivity/bootstrap bounds. 
The same argument does not apply to the GR/Shapiro contribution because it is larger by a factor of $(\Lambda b)^2$ and there is no constraint on how large $b$ can be.

\subsection{Scattering due to a set of point sources}

We now want to extend the argument in the previous section to multiple shockwaves or extended (not localised) pp-wave configurations, including a continuum. 
For now, we continue to consider just the point source \eqref{eq: point source metric function} with arbitrary $u$-profile for the function $f(u)$ and hence $H$. 
This captures the main physics at play except for the possibility of an equilibrium point in the potential, which will be addressed in Section~\ref{subsec: balancing sources}. 
The point source potential is
\begin{equation}
    V(u,r) = -\frac{k_v}{2} H(u,r) + A k_v \frac{\cGB}{\Lambda^2}\frac{\partial_{r}H(u,r)}{r}.
\end{equation}
The eikonal time delay, with diffusion and scattering neglected, accumulated up until time $u$ is
\begin{align}\label{eq: EW time delay single source}
\begin{split}
    (\Delta v)_{\rm eikonal}(u)
    &=-\frac{\partial }{\partial k_v}  \int_0^u \text{d}u' V(u',\mathbf{b}) \\
    &= \left.\left(\frac{1}{2}\int_0^u \text{d}u' H(u',r) -  A \frac{\cGB}{\Lambda^2} \int_0^u \text{d}u' \frac{\partial_r H(u',r)}{r}\right)\right|_{r=b}.
\end{split}
\end{align}
If we consider $N$ copies of the previous shockwave solution, stacked together and sufficiently spaced in null time to be distinguishable, it would at first sight appear that we can generate a contribution arbitrarily large in magnitude from the GB-term and still remain consistent with the EFT validity constraints \eqref{eqs: collated EFT RoV bounds}. 
This is however not the case, because in such a stacked shockwave configuration, the approximations used to obtain \eqref{eq: EW time delay single source} do not hold for all time. 
As we have discussed, there are two effects that limit this: One is diffusion and the second is scattering in the transverse directions. 
In the present case scattering in the transverse directions is the main constraint. As such, we need only concern ourselves with the magnitude of the time delay accumulated by some time $u_{\text{max}}$ when the eikonal approximation breaks down.

\subsubsection{Scattering time estimate}

A simple estimate for this time can be obtained by considering the classical trajectory a particle of mass $k_v$ experiencing a force due to the potential $V(u,\mathbf{x})$ would follow:
\begin{equation}\label{eq: Newton force eq}
    k_v \frac{\d^2 \mathbf{x}}{\d u^2} = - \boldsymbol{\nabla} V(u,\mathbf{x}).
\end{equation}
Since the main contribution to the potential is from the GR term, we will neglect the GB-term to get an estimate on $u_{\text{max}}$.
The amount scattered in the transverse direction is thus approximately
\begin{equation}
    \Delta r(u) \sim \left.-\frac{1}{k_v}\int_0^{u} \sd u' \int_0^{u'} \sd u'' \partial_r V(u'',r)\right|_{r=b} \sim \left.-\int_0^{u} \sd u' \int_0^{u'} \sd u'' \partial_r H(u'',r)\right|_{r=b}.
\end{equation}
When the amount scattered becomes comparable to the impact parameter $b$, we can no longer trust the eikonal approximation, and so $u_{\text{max}}$ is defined by the condition
\begin{equation}\label{eq: Newtonian scattering time}
    \Delta r(u_{\text{max}}) \sim b \quad \implies \quad \int_0^{u_{\text{max}}} \sd u \int_0^u \sd u' \frac{f(u')}{b^d} \sim \mathcal{O}(1).
\end{equation}
The above form of the time delay \eqref{eq: EW time delay single source} can only be trusted for null times less than this scattering time. 
Imposing this upper bound, the maximum EFT time delay generated by time $u_{\text{max}}$ is
\begin{equation}\label{eq: GB time delay at umax}
    \abs*{\Delta v_{\text{EFT}}(u_{\text{max}})} \sim \frac{\abs*{\cGB}}{\Lambda^2}\int_0^{u_{\text{max}}} \sd u \frac{f(u)}{b^d},
\end{equation}
and making the judicious choice to express $f(u)$ as the integral of its derivative,
\begin{equation}
    \abs*{\Delta v_{\text{EFT}}(u_{\text{max}})} \sim \frac{\abs*{\cGB}}{\Lambda^2}\int_0^{u_{\text{max}}} \sd u \int_0^u \sd u' \frac{f'(u')}{b^d} \, .
\end{equation}
If we now apply the EFT validity bound \eqref{eq: EFT RoV u-derivative bound}, or rather its averaged version \eqref{eq:averagefprime}, we find that
\begin{equation}
    k_v \abs*{\Delta v_{\text{EFT}}(u_{\text{max}})} \ll \abs*{\cGB} \int_0^{u_{\text{max}}} \sd u \int_0^u \sd u' \frac{f(u')}{b^d} \ll |\cGB|.
\end{equation}
That is, the EFT time delay is unresolvable for any $|\cGB| \lesssim \mathcal{O}(1)$. 
We stress that this argument not only applies to a single extended shockwave, but any number of shocks stacked together.

\subsection*{Special situation: Sequence of shockwaves}
\begin{figure}
    \centering
    \includegraphics[width=0.40\linewidth]{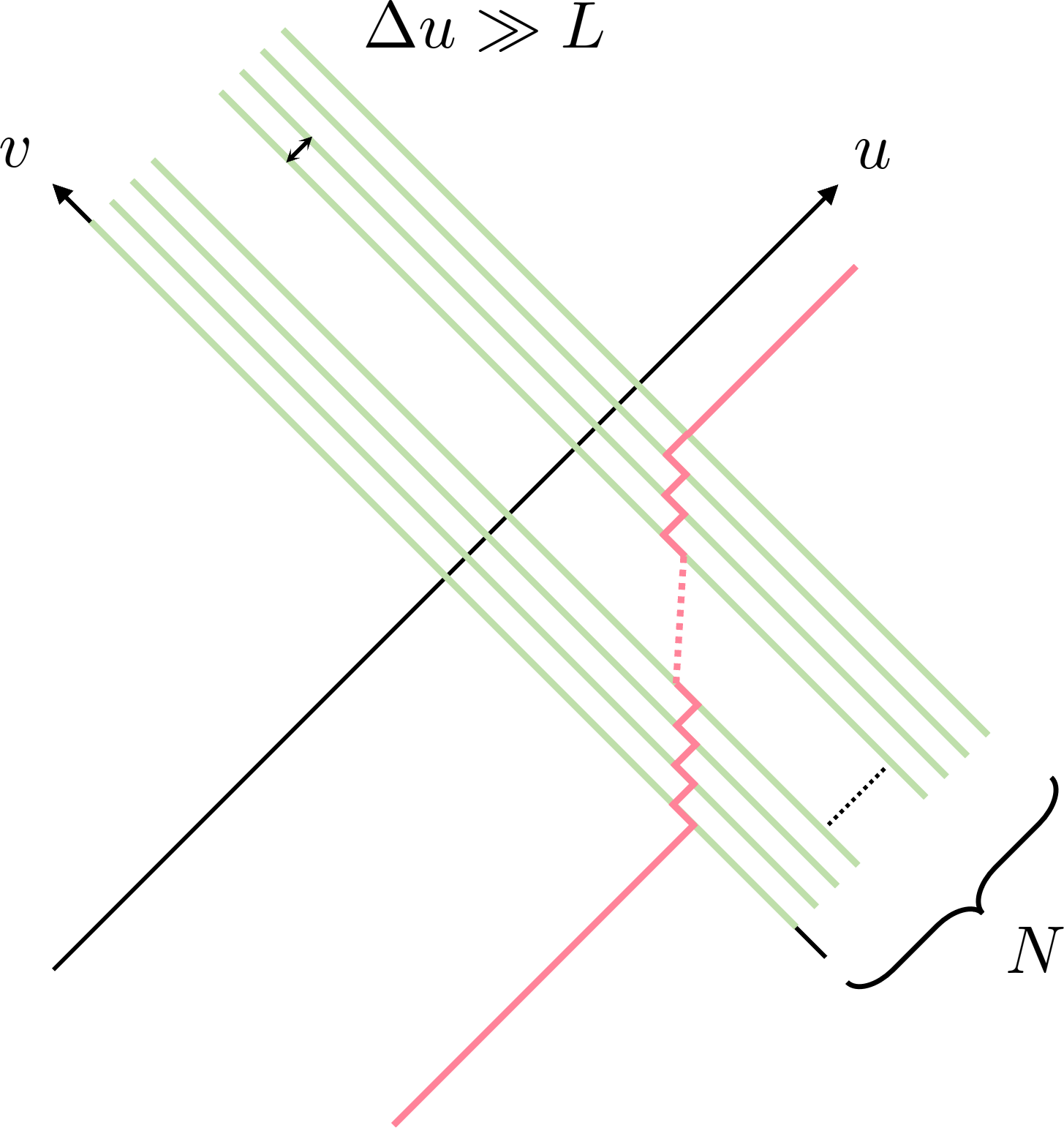}
    \caption{Illustration of the time delay accumulated by scattering off a pp-wave with $N$ sources.}
    \label{fig: stacked shocks}
\end{figure}
Let us run this argument in more detail for the sequence of $N$ shockwaves, illustrated in figure \ref{fig: stacked shocks} and described by the metric
function \be
f_{ssw}(u) = \sum_{n=0}^{N-1} \alpha \delta(u- n \Delta u) \, .
\ee
The time delay is now $N$ times larger than \eqref{eq:singleshocktimedelay} and so, returning to the argument of the introduction, we may think that by making $N$ large enough, we can make the EFT contribution resolvable. 
The total EFT time delay is
\be \label{eq:mulitpleNdelay}
\Delta v_{\rm EFT} \sim -N A (d-2) \frac{\cGB}{\Lambda^2}\frac{\alpha}{2 b^{d}} \, .
\ee
The key observation is that the total time $N \Delta u$, over which we can trust the eikonal approximation, is bounded by the scattering time. 
Identifying $u_{\rm max}=N \Delta u$, then \eqref{eq: Newtonian scattering time} amounts to
\be
N^2 \Delta u \frac{\alpha}{b^d} = u_{max}^2 \frac{\alpha}{\Delta u b^d}  \sim 1 \, . 
\ee
The scattering time is
\be
u_{\rm max}=u_s = \sqrt{\frac{b^d}{f_0}} \, ,
\ee
with $f_0=\alpha/\Delta u$ being the mean value of $f(u)$ during the period of the shocks.
Substituting in \eqref{eq:mulitpleNdelay} we obtain
\be
k_v |\Delta v_{\rm EFT}| \approx   |\cGB| \frac{k_v}{\Lambda^2 u_s} \, .
\ee
However, to trust the EFT, the spacing between the shocks must satisfy \eqref{eq: EFT RoV u-derivative bound} and \eqref{eq: EFT RoV strength bound} both of which amount to
\be
u_s \gg \frac{k_v}{\Lambda^2} \, ,
\ee
which is just the statement that we cannot meaningfully consider null times smaller than the EFT cutoff $k_v/\Lambda^2$. 
Putting this together 
we obtain
\be
k_v |\Delta v_{\rm EFT}| \ll |\cGB| \, ,
\ee
regardless of $N$. 
Thus, the superficial gain from multiple shocks of the same amplitude is more than compensated by the increased scattering they induce. 
We may, of course, think to compensate this by repositioning the location of each source to track the scattered state. 
A configuration which effectively does this will be considered in Section~\ref{subsec: balancing sources}.

\subsubsection{Breakdown of eikonal}

For a more sophisticated argument, we can account for the leading corrections to the eikonal approximation. 
This may be done by directly computing the corrections in \eqref{eq:Time delay classical} which depart from \eqref{eq:classicaleikonal}. 
A more immediate way is to follow  \cite{Camanho:2014apa} and write the wavefunction in WKB form
\begin{equation}\label{eq: WKB ansatz}
    \psi(u,\mathbf{x})  = \psi_{0} \exp \left[2 i \delta(u,\mathbf{x})\right] \, ,
\end{equation}
for which the WKB phase must obey
\begin{equation}\label{eq: phase equation}
    -2 \frac{\partial \delta\left(u,\mathbf{x}\right)}{\partial u} = \frac{2}{k_v} \left|\boldsymbol{\nabla} \delta\left(u,\mathbf{x}\right)\right|^2 - \frac{i}{ k_v} \lp\delta\left(u,\mathbf{x}\right) + V\left(u,\mathbf{x}\right).
\end{equation}
The leading-order solution $\delta^{(0)}\left(u,\mathbf{x}\right)$, obtained by neglecting gradient terms, is the eikonal expression already given by \eqref{eq: 0th order phase}.
The first correction $\delta^{(1)}\left(u,\mathbf{x}\right)$ is obtained by substituting the leading term back into \eqref{eq: phase equation},
\begin{equation}
    -2 \frac{\partial \delta^{(1)}\left(u,\mathbf{x}\right)}{\partial u} = \frac{2}{ k_v} \left|\boldsymbol{\nabla} \delta^{(0)}\left(u,\mathbf{x}\right)\right|^2 - \frac{i}{ k_v} \lp \delta^{(0)}\left(u,\mathbf{x}\right).
\end{equation}
Given that $V$ is harmonic, the solution is simply
\begin{equation}\label{eq:delta1}
    \delta^{(1)}\left(u,\mathbf{x}\right) = -\frac{1}{4 k_v} \int_0^u \sd u'\left|\int_0^{u'} \sd u'' \boldsymbol{\nabla} V\left(u'',\mathbf{x}\right)\right|^2 \, ,
\end{equation}
and so the leading correction to the eikonal time delay is
\be
    (\Delta v)^{(1)}_{\rm eikonal}(u) = -\frac{\partial}{\partial k_v } \left(  \frac{1}{2 k_v} \int_0^u \sd u'\left|\int_0^{u'} \sd u'' \boldsymbol{\nabla} V\left(u'',\mathbf{b}\right)\right|^2  \right)\, .
\ee
This result can be obtained directly from \eqref{eq:classicaleikonal}, or more precisely \eqref{eq:Time delay classical}, by accounting for the leading correction to the straight line trajectory. 
When $ (\Delta v)^{(1)}_{\rm eikonal}(u) $ becomes comparable to $ (\Delta v)^{(0)}_{\rm eikonal}(u) $, we can clearly no longer trust the leading-order eikonal result. 

In a situation with repeated shocks, the integral in \eqref{eq:delta1} continues to grow, meaning that there is inevitably some time $u_{\text{max}}$ at which the eikonal approximation breaks down. 
At the level of the phase equation \eqref{eq: phase equation}, we may ask when their partial-$u$ derivatives become comparable, \textit{i.e.}
\begin{equation}\label{eq: comparing 1st and 0th order WKB solns}
    \frac{1}{2 k_v}\left|\int_0^{u_{\text{max}}} \sd u \boldsymbol{\nabla} V\left(u,\mathbf{b}\right)\right|^2 \sim V\left(u_{\text{max}},\mathbf{b}\right).
\end{equation}
For our spherically symmetric pp-wave this condition for $u_{\text{max}}$ is
\begin{equation}\label{eq: WKB scattering time}
    \int_0^{u_{\text{max}}} \sd u \frac{f(u)}{b^d} \sim \sqrt{\frac{f(u_{\text{max}})}{b^d}},
\end{equation}
see eq. \eqref{eq: Newtonian scattering time}.
The left-hand side of \eqref{eq: WKB scattering time} can be directly replaced by the right-hand side in the expression for the EFT time delay \eqref{eq: GB time delay at umax} to give
\begin{equation}
    \abs*{\Delta v_{\text{EFT}}(u_{\text{max}})} \sim \frac{\abs*{\cGB}}{\Lambda^2}\sqrt{\frac{f(u_{\text{max}})}{b^d}}.
\end{equation}
We can now deploy a second EFT validity bound \eqref{eq: EFT RoV strength bound}, in the form
\be
\sqrt{\frac{f(u)}{b^d}}\ll \frac{\Lambda^2}{k_v} \, ,
\ee
to arrive again at the key result:
\be
k_v \abs*{\Delta v_{\text{EFT}}(u_{\text{max}})} \ll \abs*{\cGB} \, .
\ee
We thus see that, provided $|\cGB| \lesssim {\cal O}(1)$, the EFT contribution to the time delay computed in the eikonal regime is necessarily unresolvable. 

Neither of the above arguments required a specific choice for the time-dependence of the background spacetime, or in particular $f(u)$, and are confirmed by the explicit choice of constant the $f(u)=f_0$ used in Section~\ref{semiclassical1}, in both the eikonal and semi-classical approximations. 

\section{Scattering bypass by balancing shockwaves}\label{subsec: balancing sources}

It should now be clear that the main obstruction to generating an observably large EFT time advance within the regime of validity of the EFT is scattering in the transverse directions.
This scattering is the result of an attractive gravitational potential pulling waves propagating in the $v$-direction towards the source.

Na\"ively, it may seem like this fate could be avoided by engineering an equilibrium point in the potential at which the $v$-moving-waves could sit without deviation.
However, point-particles should be understood as limits of highly, but crucially not perfectly, localised wave packets --- these do feel the gravitational pull towards the sources.

\subsection{Surfing on an extremum}

In order to see why it is not possible to indefinitely accumulate time delay even when the potential possesses an equilibrium point, we need to understand this set-up within the semi-classical approximation, as discussed in Section \ref{subsec: semiclassical}.

The prototypical example of such a configuration with an (unstable) equilibrium is associated to the balanced metric function \eqref{eq: balancing source metric function}, which sets up a symmetric potential sourced by two pp-waves located at $\mathbf{x}=\pm \mathbf{b}$.
The GR-term in the potential is
\begin{equation}
    V(u,\mathbf{x}) = -\frac{k_v f(u)}{2} \left(\frac{1}{\abs*{\mathbf{x}-\mathbf{b}}^{d-2}}+\frac{1}{\abs*{\mathbf{x}+\mathbf{b}}^{d-2}}\right).
\end{equation}
We refer to this scenario as ``balancing" pp-waves, because there is an unstable equilibrium at $\mathbf{x}=\mathbf{0}$ at which the forces are perfectly balanced $\boldsymbol{\nabla} V = 0$ (see also \cite{Goon:2016une}).
An ideal point-particle, perfectly localised at the origin in transverse space, could in principle move in the $v$-direction without fear of scattering into a gravitational well.

At first glance, since a classical trajectory initially at the equilibrium can remain there indefinitely, it would seem reasonable to suppose that the only cutoff on the time delay is the quantum diffusion which occurs by $u_{\rm diff} = k_v b^2$. 
Indeed, if this were the case, it is easy to see that it would be possible to generate a resolvable EFT time delay within the regime of validity of the EFT
\be
k_v |\Delta v_{\rm EFT}| \sim \frac{k_v}{\Lambda^2 b^d} f \times u_{\rm diff} \sim  (\Lambda b)^2 \frac{k_v^2 f}{ \Lambda^4 b^d} \ll (\Lambda b)^2 \, .
\ee
However, with the potential being unstable along the $\mathbf{b}$-direction, any small perturbation will lead to an instability on a timescale associated to the second derivative of the potential in that direction.
More precisely, if we consider a wave packet initially localised near the equilibrium point ${\bf x}=\mathbf{0}$, then the potential may be approximated as a time-dependent unstable harmonic oscillator
\ba
\label{harmonicapprox}
  V(u,\mathbf{x}) \approx -\frac{k_v f(u)}{b^{d-2}} +\frac{1}{2} \frac{k_v f(u)}{b^d} (d-2) \( {\bf x}^2-d \frac{(\bf x \cdot \bf b)^2}{b^2} \) \,  \\
  \approx -\frac{k_v f(u)}{b^{d-2}} +\frac{1}{2} \frac{k_v f(u)}{b^d} (d-2) \( {\bf x}_{\perp}^2-(d-1) {\bf x}_{\parallel}^2 \) \, ,\label{harmonicapprox2}
\ea
where we split the position vector in the orthogonal and parallel directions $\mathbf{x}=\mathbf{x}_{\perp}+\mathbf{x}_{\parallel}$, with $\mathbf{x}_{\perp} \cdot \mathbf{b}=\mathbf{x}_{\perp} \cdot\mathbf{x}_{\parallel}=0$. 
In order to determine the semi-classical time delay, we need to solve the classical equations of motion for arbitrary initial data. 
These are unstable in the direction parallel to ${\bf b}$ and stable in the direction perpendicular, and take the form
\be
\frac{\d^2  {\bf x}_{\parallel} }{\d u^2}= \Omega_{\parallel} ^2(u){\bf x}_{\parallel} \, , \quad \frac{\d^2 {\bf x}_{\perp} }{\d u^2} = -\omega_{\perp}^2(u){\bf x}_{\perp} \, ,
\ee
with
\be
 \Omega_{\parallel} ^2(u) =  \frac{f(u)}{b^d}(d-2)(d-1) \, , \quad \omega_{\perp}^2(u) =  \frac{f(u)}{b^d}(d-2) \, .
\ee
Since $ {\bf x}_{\parallel}={ x}_{\parallel} {\bf b}/b$ is the unstable direction, it sets the maximum null time over which the time delay can be built up. 
In particular, once the expectation value of ${ x}_{\parallel}^2$ becomes comparable to $b^2$, we can no longer trust the harmonic approximation to the potential and hence rely on the symmetry of the balancing sources to neglect scattering.

Assuming, for simplicity, that the source varies slowly so that we may use the WKB approximation (or more precisely that, averaged over several shockwaves, the source may be taken to vary slowly), then for the solution in the parallel direction we have
\be
 { x}_{\parallel}(u) \approx  { x}_{\parallel}(0) \sqrt{\frac{\Omega_{\parallel}(0) }{\Omega_{\parallel}(u) }} \cosh\( \int_0^u \d u' \Omega_{\parallel}(u') \)+ \frac{{ p}_{\parallel}(0)}{k_v}\frac{1}{\sqrt{\Omega_{\parallel}(u)\Omega_{\parallel}(0) }} \sinh\( \int_0^u \d u' \Omega_{\parallel}(u') \) \, .\nn
\ee
Classically, it is possible to tune ${ p}_{\parallel}(0)=0$ and ${ x}_{\parallel}(0)=0 $, so that the particle remains at $ { x}_{\parallel}(u)=0$ --- quantum mechanically, this is not possible due to the inevitable spread in phase space. 
This can be captured by the parallel part of the Wigner distribution, which, assuming the initial Gaussian state considered previously, is
\be
W_{\parallel,0}({x}_{\parallel}(0),{ p}_{\parallel}(0)) =2 \,  \text{exp}\left[- \frac{{ x}^2_{\parallel}(0)}{2 \sigma^2}\right] \text{exp}\left[- 2 \sigma^2 { p}^2_{\parallel}(0)\right] \, .
\ee
Using this Wigner distribution to compute the expectation value of $ {x}_{\parallel}(u)^2$, and that $\cosh(x) \sim \sinh(x) \sim 2^{-1}e^x$ once the instability has kicked in, we find 
\be
\langle \psi | \hat {x}^2_{\parallel}(u)| \psi  \rangle \approx \frac{1}{4 \Omega_{\parallel}(u) } \( \sigma^2 \Omega_{\parallel}(0) +\frac{1}{4 k^2_v \sigma^2 \Omega_{\parallel}(0) } \) \exp\(2 \int_0^u \d u' \Omega_{\parallel}(u') \) \, .
\ee
This expectation value is minimised for
\be
\sigma^2 = \frac{1}{2 k_v \Omega_{\parallel}(0) } \, ,
\ee
so that
\be
\langle \psi | \hat { x}^2_{\parallel}(u)| \psi \rangle  \gtrsim \frac{1}{4 k_v \Omega_{\parallel}(u) }\exp\(2 \int_0^u \d u' \Omega_{\parallel}(u') \) \, .
\ee
Clearly, we can no longer trust the harmonic treatment of the potential when this is comparable to $b^2$, and so the maximum null time over which the balancing time delay can be accumulated is estimated by
\be
\frac{1}{ {k_v \Omega_{\parallel}(u_{\rm max})} }\exp\(2 \int_0^{u_{\rm max}} \d u' \Omega_{\parallel}(u') \) \le b^2 \, ,
\ee
so that
\be \label{eq: instability timescale}
 \int_0^{u_{\rm max}} \d u' \Omega_{\parallel}(u') \le \ln \( b^2 k_v \Omega_{\parallel}(u_{\rm max})\) \, .
\ee
We can again use this to bound the maximum achievable EFT time delay. 
The master equations in the balancing potential are very similar to the point source potential and can be found in appendix~\ref{app: master variables in balancing background}, so the parameter dependence of the EFT-induced time delay is still given by \eqref{eq: GB time delay at umax}.
Therefore,
\ba
   k_v  \abs*{\Delta v_{\text{EFT}}(u_{\text{max}})}& \sim& k_v \frac{\abs*{\cGB}}{\Lambda^2}\int_0^{u_{\text{max}}} \sd u \frac{f(u)}{b^d} \\
    & \le & \frac{k_v}{\Lambda^2} \abs*{\cGB} \text{max}  \left\{ \sqrt{\frac{f(u)}{b^d}}\right\} \int_0^{u_{\text{max}}} \sd u \sqrt{\frac{f(u)}{b^d}} \, .
\ea
Then, using the expression for the instability timescale \eqref{eq: instability timescale}, we arrive at:
\ba \label{Deltaequation101}
   k_v  \abs*{\Delta v_{\text{EFT}}(u_{\text{max}})} &\le&  \abs*{\cGB} {\rm max}\left\{ \sqrt{\frac{k_v^2 f(u)}{\Lambda^4 b^d}}\right\} \ln\( b^2 k_v  \Omega_{||}^{\rm max}\) \, .
\ea
The combination $b^2 k_v  \Omega_{||}^{\rm max}$ is parametrically of order $u_{\rm diff}/u_s$, where $u_s$ is the scattering time defined later \eqref{scatteringtime} and may be regarded as the (inverse of the) scale which controls quantum corrections  $\hh ^{-1}=u_{\rm diff}/u_s$ --- see 
equations ~\eqref{eq:Schrodinger_h} and \eqref{eq:h_loop_counting} and discussions in
Section~\ref{semiclassical1}. 
Using the EFT validity bound \eqref{eq: EFT RoV strength bound} $f(u)/b^d \ll \Lambda^4/k_v^2$, we infer
\be \label{balancingtimedelay}
 k_v  \abs*{\Delta v_{\text{EFT}}(u_{\text{max}})} \ll   \abs*{\cGB}     \ln\(\hh ^{-1} \) \, .
\ee
Provided $\hh $ is tuned to be exponentially small to minimise the quantum diffusion as much as possible, it appears to be possible in the balancing case to generate a resolvable EFT time delay. 
In fact this is not the case, since we have tacitly assumed the time delay has a small intrinsic uncertainty, which is not always the case.

\subsection{Time delay uncertainty}
\label{balancinguncertainty}

The Gaussian localised on the equilibrium point chosen to generate a large time delay/advance is not an eigenstate of the $S$-matrix, and there is an intrinsic uncertainty in the time delay associated with this, as discussed in Section~\ref{sec:uncertainty}. 
To determine this, we consider
\be
\widehat {\Delta v}{\,}^2 = \frac{\partial \hat S^{\dagger}}{\partial k_v}\frac{\partial \hat S}{\partial k_v} \, .
\ee
Following the discussion of Section~\ref{sec:Schrodinger1}, we can write its expectation value in Schr\"odinger picture as
\be
\langle {\rm in} | \widehat {\Delta v}{\,}^2 |{ \rm in} \rangle = \int_{u_i}^{u_f} \d u  \int_{u_i}^{u_f} \d u' \langle \psi(u) | \hat Q(u) \hat U(u,u') \hat Q(u') | \psi(u') \rangle \, ,
\ee
or equivalently in Heisenberg picture as
\be
\langle {\rm in} | \widehat {\Delta v}{\,}^2 |{ \rm in} \rangle = \int_{u_i}^{u_f} \d u  \int_{u_i}^{u_f} \d u' \langle \psi(0) | \hat Q_H(u)  \hat Q_H(u') | \psi(0) \rangle \, ,
\ee
so that the uncertainty is
\ba
\delta v^2&=&\langle {\rm in} | \widehat {\Delta v}{\,}^2 |{ \rm in} \rangle-\langle {\rm in} | \widehat {\Delta v} |{ \rm in} \rangle^2\\
&=&\int_{u_i}^{u_f} \d u  \int_{u_i}^{u_f} \d u' \Big[  \langle \psi(0) | \hat Q_H(u)  \hat Q_H(u') | \psi(0) \rangle \\
&&\phantom{\int_{u_i}^{u_f} \d u  \int_{u_i}^{u_f} \d}
- \langle \psi(0) | \hat Q_H(u) |\psi(0) \rangle \langle \psi(0)|\hat Q_H(u') | \psi(0) \rangle   \Big] \, .\nn
\ea
The main contribution to the uncertainty will come from the unstable direction along $x_\parallel$, so we neglect motion in the perpendicular direction $\mathbf{x}_\perp$. 
In the near-equilibrium approximation \eqref{harmonicapprox2}, the expression for $\hat{Q}$ is then
\be
\hat Q \approx  -\frac{f(u)}{b^{d-2}} -\frac{1}{2} \frac{f(u)}{b^d} (d-2)(d-1) \left( \hat x_{\parallel}^2-\frac{u}{k_v} (\hat p_{\parallel} \hat x_{\parallel} + \hat x_{\parallel} \hat p_{\parallel} )\right) \, ,
\ee
which may easily be translated into Heisenberg picture. 
The first term, which is the main contribution to the Shapiro delay, will drop out of the uncertainty. 
By the time the instability has kicked in, we may use the approximation 
\be
\hat p^H_{\parallel} \approx k_v \Omega_{\parallel}(u) \hat x^H_{\parallel} \, ,
\ee
so that
\be
\hat Q_H \approx  -\frac{f(u)}{b^{d-2}} -\frac{1}{2} \frac{f(u)}{b^d} (d-2)(d-1) \left( 1-2 u \Omega_{\parallel}(u)  \)  ({\hat x^H_{\parallel}}(u))^2 \, ,
\ee
where
\be
{\hat x^H_{\parallel}}(u) \approx  \frac{1}{2} \( { x}_{\parallel}(0) \sqrt{\frac{\Omega_{\parallel}(0) }{\Omega_{\parallel}(u) }}  +\frac{{ p}_{\parallel}(0)}{k_v}\frac{1}{\sqrt{\Omega_{\parallel}(u)\Omega_{\parallel}(0) }}  \)\exp\( \int_0^u \d u' \Omega_{\parallel}(u') \) \, .
\ee
Assuming  $u_i=0$ without loss of generality, a straightforward calculation gives
\ba
\delta v 
&\approx&  
\left|  \int_{0}^{u_f} \d u   \left( 1-2 u \Omega_{\parallel}(u)  \) \Omega_{\parallel}(u) \exp\( 2\int_0^u \d u' \Omega_{\parallel}(u') \) \right| \nn \\
&& \times 2^{-3/2}\Omega_{\parallel}(0) \( \sigma^2+\frac{1}{4 k_v^2 \Omega^2_{\parallel}(0) \sigma^2 } \)
\, ,
\ea
and from extremising over $\sigma$ we infer the bound 
\be \label{Deltavscatt1}
k_v \delta v \gtrsim 2^{-3/2} \left| \int_{0}^{u_f} \d u   \left( 1-2 u \Omega_{\parallel}(u)  \right)  \Omega_{\parallel}(u) \exp\(2 \int_0^u \d u' \Omega_{\parallel}(u') \) \right|   \, .
\ee
If $u_f$ is taken to be such that $ \exp \(2 \int_0^{u_f} \d u' \Omega_{\parallel}(u') \) \lesssim 1$, then the right-hand side is of order unity or less, and in this situation $ k_v  \abs*{\Delta v_{\text{EFT}}(u_{\text{max}})} \lesssim \abs*{\cGB}  $. 
In order to make the EFT contribution to the time delay large, we must take $u_f$ such that $\exp\(2 \int_0^{u_f} \d u' \Omega_{\parallel}(u') \) \gg  1$. But then the integral on the RHS of \eqref{Deltavscatt1} is exponentially large, and grows faster than $k_v  \abs*{\Delta v_{\text{EFT}}}$. 
Thus, precisely at the point where it appears we can generate a resolvable EFT time delay, the uncertainty in the GR contribution to the time delay swamps it.

\section{Shockwave scattering in the semi-classical approximation}

\label{semiclassical1}

The considerations of Section~\ref{subsec: pt source scattering} showed that, from a single or a stack of point source shockwaves, it was impossible to generate a resolvable EFT time delay within the regime in which the eikonal approximation can be trusted. 
The limitation comes from scattering, which necessarily takes us outside of the eikonal limit, where it is assumed that the deflection angle is small. 
However, we already know that the semi-classical approximation automatically incorporates the effects of scattering, and so we should further address the question of what happens when we take this into account.

Let us consider a sequence of closely spaced shockwaves of separation $\Delta u$
\be
f_{ssw}(u) = \sum_{n=0}^{\infty} f_0 \Delta u \delta(u-n \Delta u) \, .
\ee
Na\"ively, the maximal time delay/advance will be obtained by having the largest number of shocks over the diffusion time scale, which would amount to $\Delta u$ as small as possible. 
Thus a reasonable estimate of the scattering of the classical trajectories is obtained by replacing $f(u)$ by a constant
\be
f(u) \approx f_0 \theta(u) \, .
\ee
In fact, it is easier to consider a constant continuum $f(u)=f_0$ for all $u$ so that the system regains a Killing vector in the $u$-direction. 
Then, the $S$-matrix commutes with the asymptotic null Hamiltonian $\hat P_u^0$ and it is sufficient to analyse the system in the same manner as a time-independent Schr\"odinger equation. 
We shall further restrict to the case of spherically symmetric shockwaves, so that angular momentum squared $\ell (\ell+(d-2))$ is additionally conserved. 
Since the classical time delay is uniquely determined by the energy and angular momentum, there is no uncertainty for any partial wave ${\rm in}$-state which has definite energy, so the semi-classical time delay \eqref{eq:Time delay semiclassical} reduces to the classical one \eqref{eq:Time delay classical}.

Consider the redefinition of variables ${\bf x} = r_* {\bf z}$ and $u = u_s \rho$, with
\be \label{scatteringtime}
u_s = \sqrt{\frac{r_* ^d }{f_0}} \, ,
\ee
which is the scattering time scale. 
Here, $r_*$ is the turning point, which is generically similar to the impact parameter $b$ defined via
\be
b =\frac{ (\ell+(d-2)/2)}{\sqrt{2 k_v P_u}} \, .
\ee
Under this change of variables and with $f(u)=f_0$, the Schr\"odinger equation (ignoring EFT contributions) becomes 
\be
\label{eq:Schrodinger_h}
i { \hh } \frac{\partial \psi(\rho,{\bf z})}{\partial \rho} = - \frac{\hh ^2}{2} \nabla^2_z \psi(\rho,{\bf z}) - \frac{1}{2 |{\bf z}|^{d-2}} \psi(\rho,{\bf z})\,,
\ee
where we have defined an effective $\hbar$ or loop counting parameter
\be
\label{eq:h_loop_counting}
\hh  = \frac{u_s}{u_{\rm diff}}= \frac{u_s}{k_v r_*^2} =\frac{r_*^{d/2-2}}{k_v \sqrt{f_0}} \, ,
\ee
and we have now identified the diffusion time as $u_{\rm diff}=k_v r_*^2$. 
The semi-classical approximation is expected to be valid when $\hh  \ll 1$, which requires that the diffusion time is longer than the scattering time. 
We see that this leads to a very different condition on the impact parameter for $d<4$ and $d>4$. 
In fact, it is well-known that the non-relativistic Schr\"odinger equation has very different properties depending on whether $\lim_{r \rightarrow 0}r^2V(r)$ vanishes, is finite, or diverges. 
The situation $d<4$ corresponds to a regular potential, $d=4$ to a transition potential, and $d>4$ to a singular potential. 
Singular potentials and strongly attractive transition potentials can have undetermined $S$-matrices due to an ambiguity in the solutions near $r=0$, and without further conditions may have bound state energies unbounded below~\cite{Frank:1971xx}. 
However, any reasonable solution to these issues is expected to match the semi-classical phase shift in the appropriate region. 

The classical equation of motion takes the dimensionless form
\be\label{dimensionless}
\frac{\d^2 {\bf z}}{\d \rho^2}= -\frac{(d-2)}{2} \frac{{\bf z}}{|{\bf z}|^{d}} \, ,
\ee
with dimensionless initial conditions. On dimensional grounds, the GR contribution to the time delay is of the form
\be
\Delta v_{\rm GR} = \frac{u_s f_0}{r_*^{d-2}} \Delta \tilde v_{\rm GR} \, ,
\ee
where $\Delta \tilde v $ is the time delay computed with dimensionless initial conditions with what amounts to the dimensionless equation \eqref{dimensionless}.

The EFT corrections are further suppressed by $(\Lambda r_*)^{-2}$, so schematically
\be \label{EFT600}
\Delta v_{\rm EFT} \approx \frac{u_s f_0}{\Lambda^2 r_*^d} c_{\rm GB}\Delta \tilde v _{\rm EFT} \, ,
\ee
where, typically, $\Delta \tilde v _{\rm EFT} \sim (\Delta \tilde v )^{\rm GR}_{\rm semi-classical}$ is the dimensionless form of the EFT corrections.
Whence
\be \label{semiclassical estimate}
k_v \Delta v_{\rm EFT} \approx \sqrt{\frac{r_*^d }{f_0}}  f_0\frac{k_v}{\Lambda^2 r_*^d} c_{\rm GB}\Delta \tilde v _{\rm EFT} \approx \frac{k_v}{\Lambda^2} \sqrt{\frac{f_0}{r_*^d}} c_{\rm GB} \Delta \tilde v _{\rm EFT}  \, .
\ee
Thus, applying the EFT bounds \eqref{eq: EFT RoV strength bound} at the turning point $r_*$, we infer that
\be
|k_v \Delta v_{\rm EFT}| \ll |c_{\rm GB} | \,| \Delta \tilde v _{\rm EFT}  | \, .
\ee
The question now is whether it is possible to generate a large dimensionless time delay $\Delta \tilde v _{\rm EFT} $ for classical trajectories dominated by  \eqref{dimensionless}.

Given the assumed spherical symmetry, the time delay is best determined for partial waves and states of definite $\hat P_u$, \textit{i.e.} classical trajectories of definite angular momenta and energy, for which the $S$-matrix is diagonal.
Rewriting the equation \eqref{dimensionless} in radial form we have
\be
\frac{\d^2 z_r}{\d \rho^2} = - \frac{\partial V_{\rm eff}(z_r) }{\partial z_r} \, ,
\ee
where the effective potential includes the centrifugal term
\be
V_{\rm eff}(z_r) = \frac{L^2}{2 z_r^2}- \frac{1}{2}\frac{1}{z_r^{d-2}}  \, .
\ee
Here, $L$ is the dimensionless form of the angular momentum, which is related to the partial wave $\ell$ (including the Langer correction \cite{Langer:1937qr,deRham:2020zyh,Chen:2021bvg}) via
\be
L^2 =\hh ^2 \(\ell+\frac{1}{2}(d-2)\)^2 \, .
\ee
The classical trajectory is determined by the energy conservation equation
\be
\(\frac{\d z_r}{\d \rho}\)^2 + \frac{L^2}{ z_r^2}-\frac{1}{z_r^{d-2}} =2\hh  u_s P_u=L^2-1 \, ,
\ee
with the turning point $z_r=1$.

Following the standard WKB treatment of equation \eqref{eq:Schrodinger_h}, the semi-classical scattering phase shift, including the Langer correction and ignoring EFT corrections, is
\be
\delta_{\ell}^{\rm GR} = \frac{1}{\hh } \int_1^{\infty} \d z_r \left[  \sqrt{L^2-1-\(\frac{L^2}{z_r^2} -\frac{1}{z_r^{d-2} }\)}  - \sqrt{L^2-1} \right]   -\frac{\sqrt{L^2-1}}{\hh } + \frac{\pi}{2} \(\ell+\frac{1}{2}(d-2) \)\, .
\ee
The EFT corrections to the phase shift are
\be \label{EFTphaseshift101}
\delta_{\ell}^{\rm EFT} = -\frac{A c_{\rm GB} (d-2) }{\hh  \Lambda^2 r_*^2} \int_1^{\infty} \d z_r  \frac{1}{z_r^{d}}\frac{1}{\sqrt{L^2-1-\(\frac{L^2}{z_r^2} -\frac{1}{z_r^{d-2} }\)}}  \, .
\ee
In deriving these expressions, we use the WKB matching formula. 
To do so, we assume that the forbidden region barrier is sufficiently large, that the solution to the left of the turning point may be well-approximated by the WKB mode which decays exponentially as $z$ decreases. 
A more precise treatment includes both modes leading to an additional exponentially suppressed imaginary contribution to the phase shift, associated with tunnelling across the barrier. 
This term plays no significant role in determining the phase shift. 
From these scattering phase shifts we may directly infer the time delays.

\subsection{Recovery of eikonal}

It is helpful at this point to compare the above result with the eikonal expectation. 
In eikonal scattering, the trajectory is assumed to not depart significantly from a straight line, which will be true for large $L \gg 1$. 
The GR phase shift can then be approximated as
\be
\delta_{\ell}^{\rm GR} \approx -\frac{1}{2\hh } \int_1^{\infty} \d z_r  \frac{1}{z_r^{d-2}} \frac{1}{ \sqrt{L^2-\frac{L^2}{z_r^2}}}  \, .
\ee
To translate back into more familiar language, we may perform the change of variables $z_r^2 =1+p_0^2 u^2/(k_v b)^2$ and use $r_* \sim b$ for $L \gg 1$ to get
\be
\delta_{\ell}^{\rm GR} \approx -\frac{1}{2} \int_{-\infty}^{\infty} \d u  \frac{1}{2} \frac{k_v f_0}{(b^2+p_0^2 u^2/k_v^2)^{(d-2)/2}}  \, ,  \,
\ee
with $p_0=\sqrt{2k_v P_u}$. We recognise this to be precisely the eikonal phase shift that leads to \eqref{eq:classicaleikonal} with the assumption that ${\bf x_0}={\bf b}$ and ${\bf p_0}$ is orthogonal to ${\bf b}$. In the same limit the EFT time delay reduces to
\be
\delta_{\ell}^{\rm EFT} = -\frac{A c_{\rm GB} (d-2) }{ 2 \Lambda^2 } \int_{-\infty}^{\infty} \d u  \frac{k_v f_0}{(b^2+p_0^2 u^2/k_v^2)^{d/2}}  \, ,
\ee
which is again consistent with the EFT correction to  \eqref{eq:classicaleikonal}. \\

\subsection{Beyond eikonal}

Returning to the semi-classical WKB form, it is clear that the magnitude of the EFT phase shift and hence the time delay is determined by the properties of the dimensionless integral
\be \label{eq:integral}
I_d(L)= \int_1^{\infty} \d z_r  \frac{1}{z_r^{d}}\frac{1}{\sqrt{L^2-1-\(\frac{L^2}{z_r^2} -\frac{1}{z_r^{d-2} }\)}} \, .
\ee
Furthermore, the time delay is also sensitive to the integral
\be\label{eq:integral2}
\tilde I_d(L)= -L \frac{ \d I_d(L)}{\d L} =\int_1^{\infty} \d z_r  \frac{L^2}{z_r^{d}} \(1-\frac{1}{z_r^2} \)\frac{1}{\( L^2-1-\(\frac{L^2}{z_r^2} -\frac{1}{z_r^{d-2} }\) \)^{3/2}} \, .
\ee
The properties of these integrals are dimension-dependent, being strongly sensitive to whether $d$ is less than, equal to, or greater than $d=4$ ($D=6$).
We therefore consider the 3 different cases separately below.

\subsubsection{Case $D<6$}

The particular choice $d=2$ corresponds to $D=4$, for which the GB-term makes no contribution to the equations of motion and does not need to be considered. 
For $d=3$ ($D=5$), the integrals \eqref{eq:integral} and \eqref{eq:integral2} converge at $z_r=1$ and at $z_r = \infty$, and direct evaluation confirms that both $I_3(L)$ and $\tilde I_3(L)$ are at most of order unity for all $L^2 \ge 1$ (which is necessary for scattering states) and fall of as $1/L$ at large $L$. Specifically
\be
I_{3}(L)=\frac{2L\sqrt{L^{2}-1}+\pi-2 \text{arccot}\left(\frac{L}{\sqrt{L^{2}-1}}\right)}{2L^{3}} \, ,
\ee
and so $I_3(1) = \pi/2$ and $\tilde I_3(1)=3 \pi/2$. Thus, generically $ \Delta \tilde v _{\rm EFT} \sim {\cal O}(1)$, implying that
\ba
k_v |\Delta v_{\rm EFT}| &\ll&  |c_{\rm GB}|\, \, .
\ea

\subsubsection{Case $D=6$}

The case $d=4$ requires more care. The integrals are easily evaluated to give
\ba
&& I_4(L) = \frac{\pi}{4} \frac{1}{\sqrt{L^2-1}} \, , \\
&& \tilde I_4(L) = \frac{\pi L^2}{4} \frac{1}{(L^2-1)^{3/2}} \, .
\ea
Generically, these integrals are of order unity or less.
However, it appears they can be made arbitrarily large by tuning $L$ close to $1$. 
This reflects the cancellation between the centrifugal repulsion and the gravitational attraction. 

In fact, for $d=4$, the Schr\"odinger equation may be solved exactly for the pure GR potential, giving the exact GR phase shift
\be
\delta_{\ell}^{\rm GR} = \frac{\pi}{2}(\ell+1) -\frac{\pi}{2} \sqrt{(\ell+1)^2- k_v^2 f_0}\, ,
\ee
which happens to be the exact WKB result (with Langer correction). 
This is associated with the partial wave scattering state
\be
\psi_{\ell} \propto \frac{1}{\rho} J_{\nu}\(\sqrt{L^2-1} \rho\)  \propto \frac{1}{r} J_{\nu}\(\sqrt{2 k_v P_u} r\)  \, ,
\ee
with $\nu =  \sqrt{(\ell+1)^2- k_v^2 f_0}$. The GR time delay is then
\be
\Delta v_{\rm GR} = \frac{\pi k_v f_0}{ \sqrt{(\ell+1)^2- k_v^2 f_0}}=\frac{\pi}{ k_v}\frac{1}{\sqrt{L^2-1}} \, ,
\ee
which, as anticipated, can be made arbitrarily large by sending $L \rightarrow 1$.

However, crucially, this only makes sense if $ k_v^2 f_0<1$. 
If $ k_v^2 f_0>1$, then this phase shift is not well-defined for small $\ell$, and this is a familiar result for transitionary potentials \cite{Frank:1971xx}. 
It arises because the wavefunction of both modes are finite as $r \rightarrow 0$ and so the phase shift is ambiguous. 
An additional consequence is that the system admits an infinity of bound states with arbitrarily negative energies \cite{Frank:1971xx} suggesting it is unstable. 
Either way, this implies that the phase shift is strongly sensitive to the UV physics that regulates the $1/r^2$ potential. 
Indeed, this aspect will occur for essentially any field with energy $k_v$ in this given pp-wave background. 
Since the monopole ($\ell=0$) of essentially all fields are explicitly UV sensitive, we cannot trust the spherically symmetric background solution we started with, and while it may not be unstable, it cannot be described within the remit of the EFT. 
Consequently we regard the case $k_v^2 f_0>1$ outside of the regime of validity of the EFT at least as long as the source is point-like. 
We shall consider extended sources below.

\paragraph{Case $k_v^2 f_0<1$ ---} The phase shift of the EFT $S$-matrix defined in Section \ref{EFTSmatrix} can be computed in the Born approximation using the known exact GR partial wave states, and takes the form
\ba
\delta_{\ell}^{\rm EFT} &=& -\frac{1}{2}\langle P_u,\ell| \hat H_{\rm EFT} | P_u,\ell\rangle \nn \\
&\sim& c_{\rm GB} \int_0^{\infty} \d r \, r^3 \, \frac{k_v^2 f_0}{\Lambda^2 r^4} \frac{1}{r^2}\(J_{\nu}(\sqrt{2 k_v P_u} r) \)^2 \, ,
\ea
which should in general be more accurate than \eqref{EFTphaseshift101}. 
Since $J_{\nu}(\rho) \propto \rho^{\nu}$ as $\rho \rightarrow 0$, this perturbative computation of the EFT time delay is only valid for $\nu \ge 1$, and for $0<\nu \le 1$ we must remember that the EFT only makes sense for $r \ge \Lambda^{-1}$ --- this will regulate the divergence.

Let us first consider $\nu \ge 1$, for which the integral is
\be
\delta_{\ell}^{\rm EFT} \sim c_{\rm GB} \frac{k_v^2 f_0}{\Lambda^2} \frac{2 k_v P_u}{4 (\nu^3-\nu)} \sim c_{\rm GB} \frac{k_v^2 f_0}{\Lambda^2 r_*^2} \frac{\nu}{4 (\nu^2-1)} \, .
\ee
For generic $\nu \gg 1$, this scales as
\be \label{delatEFT500}
\delta_{\ell}^{\rm EFT} \sim  c_{\rm GB} \frac{k_v^2 f_0}{\Lambda^2 r_*^2} \frac{1}{\nu} \sim c_{\rm GB} \frac{k_v \sqrt{f_0}}{\Lambda^2 r_*^2} \frac{1}{\sqrt{L^2-1}}  \, ,
\ee
which is consistent with our estimate \eqref{semiclassical estimate}. 
Restricting to the physical region $ k_v^2 f_0 < 1$ and imposing the EFT condition $\Lambda r_* \gg 1$ is sufficient to render this effect unresolvable
\be
k_v |\Delta v_{\rm EFT}|   \lesssim|c_{\rm GB}|  \frac{\nu}{4 (\nu^2-1)} \, 
\ee
for generic $\nu$. 
However, it still appears that we can enhance this effect arbitrarily by sending $\nu \rightarrow 1$.
As we tune $L \rightarrow 1$, or equivalently $\nu \rightarrow 0$, we are sending the GR effective potential to zero. 
But at some point the approximation that the EFT potential can be treated perturbatively breaks down. 
By comparing terms in the effective potential, we require the scattering state to be valid in EFT perturbation theory
\be\label{6Dbound1}
\langle \psi  | \frac{\nu^2}{k_v \hat r^2} | \psi \rangle \gg  \langle \psi  |  \frac{k_v f_0}{\Lambda^2 \hat r^4} | \psi \rangle  \,  .
\ee
Imposing this condition on the EFT time delay we have
\ba
k_v |\Delta v_{\rm EFT}| &\ll& |c_{\rm GB}| \int_0^{\infty} \d r \, r^3 \,  \frac{\nu^2}{ r^2} \frac{1}{r^2}\(J_{\nu}(\sqrt{2 k_v P_u} r) \)^2 \,  \\
 &\ll& |c_{\rm GB}| \nu^2  \int_0^{\infty} \d x \,   \frac{1}{x} J_{\nu}(x)^2  = |c_{\rm GB}| \frac{1}{2} \nu\, ,\label{6Dbound2}
\ea
When \eqref{6Dbound2} is combined with \eqref{6Dbound1} we find
\ba
k_v |\Delta v_{\rm EFT}| \ll  |c_{\rm GB}|\, \, .
\ea

\paragraph{Extended sources: Case $k_v^2 f_0>1$ ---}
As already noted, for $k_v^2 f_0>1$ the singular nature of the potential leads to an ill-defined $S$-matrix for low multipoles. However, if the source is extended of width $w \gg \Lambda^{-1}$ then we may replace the GR potential by something of the form
\be
-\frac{k_v f_0}{2r^2} \rightarrow  -\frac{k_v f_0}{2(r^2 + w^2)} \, .
\ee
In this situation, the centifugal term will win out at small $r$, leading to a well-defined $S$-matrix. 
The monopole will scatter at $r=r_0 \ll w$ for which
\be
P_u+\frac{k_v f_0}{2w^2} \approx \frac{1}{2k_v r_0^2} \, .
\ee
which implies, given $P_u>0$, $f_0>0$
\be
r_0 \lesssim \frac{w}{\sqrt{k_v^2 f_0}} \, .
\ee
Using \eqref{eq: EFT RoV strength bound} applied at $r_0$ implies 
\be
\frac{\sqrt{k_v^2 f_0}}{\Lambda^2} \ll \frac{w^2}{k_v^2 f_0} \, .
\ee
Then, returning to high multipoles, to trust our previous expression \eqref{delatEFT500} we require $w \ll r_*$ and so
\be
k_v |\Delta v_{\rm EFT}| \sim  |c_{\rm GB}| \frac{k_v^2 f_0}{\Lambda^2 r_*^2} \frac{1}{\nu} \ll |\cGB|\frac{k_v^2 f_0}{\Lambda^2 w^2} \frac{1}{\nu}\ll |\cGB|\frac{1}{\sqrt{k_v^2 f_0}} \frac{1}{\nu} \ll |\cGB| \, .
\ee
Thus for fixed $k_v$ and $P_u$, demanding that the monopole scattering lies in the regime of validity of the EFT ensures unresolvability of the high multipole scattering. 

\subsubsection{Case $D>6$}

For the general case $d > 4$ ($D>6$), the attractive gravitational potential dominates over the centrifugal repulsion at small $z_r$, so that the effective potential has a maximum. 
Scattering states will be those whose energy is less than the maximum, which means there is a minimum impact parameter. 
States with energy above this spiral into the centre and do not reach infinity. 
For the scattering states, in general, the integral on the RHS of \eqref{eq:integral} is of order unity, consistent with expectations. 
However it is possible to make it arbitrarily large for $d \ge 5$ by tuning the energy close to the maximum of the potential. 
This maximum occurs when $2 L^2 z_r^{d-4} = (d-2)$, \textit{i.e.} for
\be
L^2 = \frac{(d-2)}{2} \, .
\ee
For this choice, by construction, the function $h(z_r)$ entering the square root in \eqref{eq:integral} satisfies $h(1)=h'(1)=1$, implying that the integral is logarithmically divergent near $z_r=1$. 
If we tune $L$ to be exponentially close to the maximum
\be
L^2 = \frac{(d-2)}{2} +e^{-N} \, ,
\ee
then we expect\be \label{EFT601}
 \Delta \tilde v_{\rm EFT}  \sim  N \, ,\ee
or that
\be
k_v |\Delta v_{\rm EFT}| \ll |c_{\rm GB} | N \, .
\ee
Thus, it na\"ively appears that, by making $N$ large enough, we can generate an arbitrarily large EFT contribution to the time delay. 
Once again, there is a caveat to this argument. 
In this case, the validity of the WKB approximation for the phase shift requires there to be sufficient exponential damping in the forbidden region that we may ignore the solution on the left of the barrier. 
This requires
\be
\frac{1}{\hh } \int_{z_*}^1\sqrt{\frac{L^2}{z_r^2} -\frac{1}{z_r^{d-2}}-(L^2-1) }   \d z_r  \gg 1 \, ,
\ee
where $z_*$ is the turning point on the left-hand side of the maximum. 
This sets the maximum $N$ in terms of $\hh $ such that $N \sim \ln (\hh ^{-1})$. 
The same condition can be obtained by demanding the WKB approximation to be valid at the maximum of the effective potential $V'_{\rm eff}(z_{\rm max})=0$
\be
\Big|\frac{\partial_z \omega}{\omega^2}\Big|_{z=z_{\rm max}} \ll 1 \, ,
\ee
with
\be
\omega^2=-\frac{1}{\hh ^2} \(\frac{L^2}{z_r^2} -\frac{1}{z_r^{d-2}}-(L^2-1) \) \, .
\ee
Putting this together, the maximum time delay generated in the regime of validity of the EFT is
\be
k_v |\Delta v_{\rm EFT}| \ll |c_{\rm GB}| {\rm ln}(\hh ^{-1})\, .
\ee
Thus, at best, we are forced to tune the quantum hierarchy $\hh $ to be exponentially small, \textit{e.g.} $\hh  \sim e^{-100} \sim 10^{-44}$, to have an observable EFT contribution to the time delay. 
This result is closely analogous to the balancing shockwave result in \eqref{balancingtimedelay}. 
This is no surprise since, in both cases, the time delay is being maximised by engineering an equilibrium point in the effective potential --- in the present case due to the balancing of the centrifugal force and the gravitational attraction. 
In the case of the balancing shockwaves, we argued in Section~\ref{balancinguncertainty} that the uncertainty in the time delay necessarily becomes larger at precisely the point that the logarithm becomes large. 
In the present case, since we are considering $P_u$ eigenstates of definite partial wave, there is no scattering uncertainty $\delta v=0$ and so we cannot rely on that argument.

The issue in the present case is closely analogous to the $D=6$ case. 
To tune $N$ to be large, say $100$, we need $\hh  \sim e^{-100}$ in order to trust the WKB approximation. 
At the same time, we are tuning $L^2=(d-2)/2$ which means $\ell \sim e^{100}$. 
However, for all lower multipoles of the same energy, the centrifugal barrier is too low, meaning such states fall into the origin --- more precisely, they have an underdetermined $S$-matrix due to the ambiguity in the wavefunctions at $r=0$, which can only be resolved by the actual UV completion. 
This situation is in fact worse than $D=6$, since it is now an exponentially large number of low multipoles which are ill-defined. 
To avoid this pathology, and to have the background under control within the regime of validity of the EFT, we need the monopole $\ell=0$ to be scattered. To do that within the above framework requires $\hh  \sim 1$ and so we obtain
\be
k_v |\Delta v_{\rm EFT}| \ll |c_{\rm GB}| \, .
\ee

\paragraph*{Extended sources ---}
As previously discussed, the situation is improved if the source is extended, for example in the manner
\be
-\frac{k_v f_0}{2r^{d-2}} \rightarrow  -\frac{k_v f_0}{2(r^2 + w^2)^{(d-2)/2}} \, ,
\ee
so that the $S$-matrix is well-defined for low multipoles. Assuming we have tuned $L^2 \approx (d-2)/2$, the monopole will pass over the barrier and scatter at the point $r=r_0$ for which the centrifugal term dominates
\be
P_u + \frac{k_v f_0}{2w^{(d-2)}} \approx \frac{((d-2)/2)^2}{2 k_v r_0^2} \, .
\ee
Given $P_u>0$ and $f_0>0$, this implies $r_0 \lesssim \frac{w^{(d-2)/2}}{\sqrt{k_v^2 f_0}} $.
Imposing  \eqref{eq: EFT RoV strength bound} at $r=r_0$ implies
\be
\(\frac{k_v^2 f_0}{\Lambda^4} \)^{1/d} \ll\frac{w^{(d-2)/2}}{\sqrt{k_v^2 f_0}} \, .
\ee
Returning to the high multipoles, to trust our previous formulae for the phase shift we require $r_* \gg w$ and so
\be \label{EFT603}
\(\frac{k_v^2 f_0}{\Lambda^4} \)^{1/d} \ll\frac{r_*^{(d-2)/2}}{\sqrt{k_v^2 f_0}}  \quad \quad \Rightarrow \quad \quad \frac{k_v^2 f_0}{r_*^{d-4}} \ll (r_* \Lambda)^{\frac{8}{(d+2)}}\, .
\ee
From \eqref{EFT600} and \eqref{EFT601} we estimate the time delay to be
\ba
k_v |\Delta v_{\rm EFT}| &\sim&  |c_{\rm GB}| \frac{1}{(\Lambda r_*)^2} \sqrt{\frac{k_v^2 f_0}{r_*^{d-4}} } \ln \(\frac{k_v^2 f_0}{r_*^{d-4}} \)    \\
& \ll &  |c_{\rm GB}| \frac{1}{\(\Lambda r_*\)^{\frac{2(d-2)}{(d+2)}}} \ln (r_* \Lambda) \, ,
\ea
which combined with \eqref{eq: EFT RoV distance bound}, \textit{i.e.}, $\Lambda r_* \gg 1$ implies $k_v |\Delta v_{\rm EFT}| \ll |c_{\rm GB}| $.


\section{Shockwave scattering in the Born approximation}\label{sec: quantum calcs}

In this section we compute the scattering time delay in perturbation theory, identifying the first few terms in the Born series. 
In general, the Born series has a smaller regime of validity than either the eikonal or semi-classical approximations, and so we do not draw any major conclusions from its results. 
However, we include it because the calculation may be performed more precisely and it is informative in terms of demonstrating the distinction between the single source and balancing shockwaves. 
It is also worth emphasising that almost all treatments of the eikonal approximation in the literature rely on resumming the Born amplitude, and the leading Born approximation to the time delay is identical to the eikonal result. 
The exponential eikonal resummation is implicit in the definition of the time delay operator. 

\begin{figure}
    \centering
    \includegraphics[width=\textwidth]{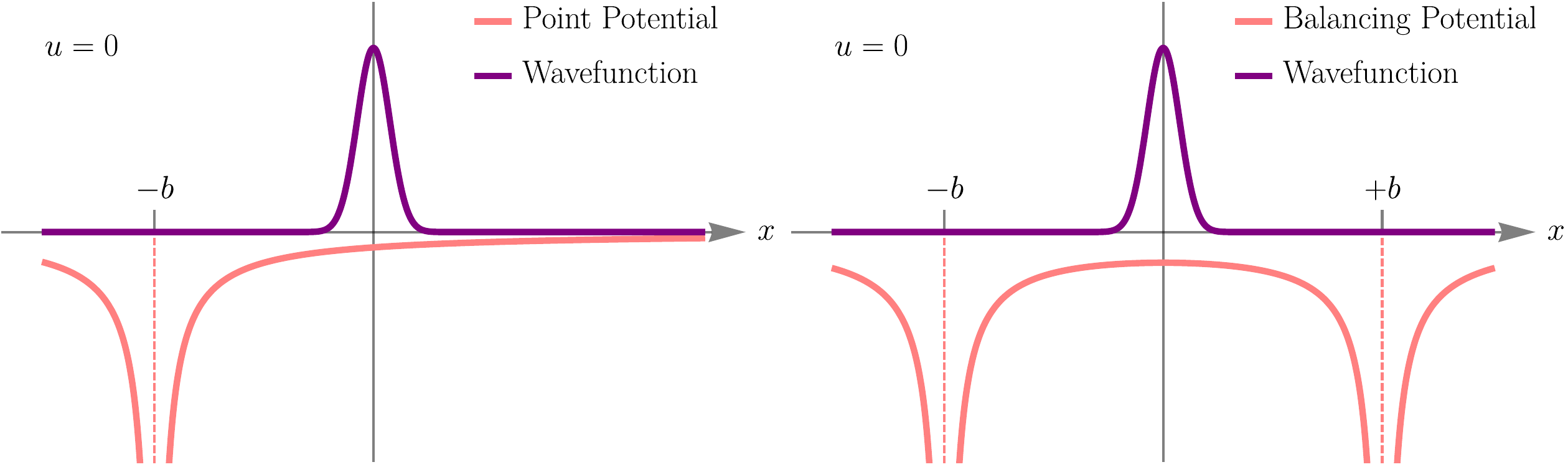}
    \caption{Snapshot at time $u=0$ of the point-source and balancing potentials and the wavefunction which probes the spacetime.}
    \label{fig: wavefn on potentials}
\end{figure}

The physical scenario for the two types of potential we have in mind is illustrated in Fig.~\ref{fig: wavefn on potentials}.
Then the potential is generated by either a single point source located at $\mathbf{x} = -\mathbf{b}$ or balancing sources at $\mathbf{x} = \pm \mathbf{b}$:
\begin{equation}
    V(u,\mathbf{x}) =  -\frac{k_v f(u)}{2} \left(\frac{\Theta_{\text{bal}}}{\abs*{\mathbf{x}-\mathbf{b}}^{d-2}} + \frac{1}{\abs*{\mathbf{x}+\mathbf{b}}^{d-2}}\right) \, ,
\end{equation}
where we have introduced the notation
\begin{equation*}
	\Theta_{\text{bal}} =
	\begin{cases}
		0,\quad \text{point source} \\
		1,\quad \text{balancing sources} \, .
	\end{cases}
\end{equation*}
Note that, in order to accommodate the point source case and balancing source case simultaneously, the location of the point source is shifted relative to the discussion of Section~\ref{sec: pp-waves}. 
The interaction picture potential is
\begin{equation}
   \hat H_1^{\rm int}(u) = -\frac{2 \pi^{\frac{d}{2}}}{\Gamma\left(\frac{d-2}{2}\right)} k_v f(u) \int \frac{\sd^d q}{(2\pi)^d}\frac{e^{i \mathbf{q}\cdot\mathbf{b}} + \Theta_{\text{bal}}e^{-i \mathbf{q}\cdot\mathbf{b}}}{\mathbf{q}^2}e^{i \mathbf{q}\cdot\left(\mathbf{\hat{x}} + \frac{u}{k_v}\mathbf{\hat{k}}\right)}.
\end{equation}
which is conveniently rewritten as
\begin{equation}
\begin{split}
    \hat H_1^{\rm int}(u) = -\frac{2 \pi^{\frac{d}{2}}}{\Gamma\left(\frac{d-2}{2}\right)} k_v f(u) \int &\frac{\sd^d q}{(2\pi)^d} \int \frac{\sd^d k}{(2\pi)^d} \int \sd^d x \frac{e^{i \mathbf{q}\cdot\mathbf{b}} + \Theta_{\text{bal}}e^{-i \mathbf{q}\cdot\mathbf{b}}}{\mathbf{q}^2}\\
     &\exp\left[i \mathbf{x}\cdot\left(\mathbf{q}+\mathbf{k}\right)+ \frac{iu}{k_v}\mathbf{q}\cdot\left(\frac{\mathbf{q}}{2} + \mathbf{k}\right)\right] ||\ket{\mathbf{x}}\bra{\mathbf{k}}.
\end{split}
\end{equation}
Denoting the terms in the Dyson series as
\begin{equation}
    \hat{S} = 1+\underbrace{(-i)\int_{-\infty}^{+\infty}\sd u\, \hat H_1^{\rm int}(u)}_{\hat{S}^{(1)}}+\underbrace{(-i)^{2}\int_{-\infty}^{+\infty}\sd u \int_{-\infty}^{u}\sd u'\, \hat H_1^{\rm int}(u)\hat H_1^{\rm int}(u')}_{\hat{S}^{(2)}}+\dots,
\end{equation}
the first three terms in the series expansion for the time delay are
\begin{align}\label{eq: perturbative time delay op in terms of perturbative $S$-matrix}
    \widehat{\Delta v}{}^{(1)}=&-i\frac{\partial\hat{S}^{(1)}}{\partial k_{v}}, \\
    \widehat{\Delta v}{}^{(2)}=&-i\left(\frac{\partial\hat{S}^{(2)}}{\partial k_{v}}+\hat{S}^{(1)\dagger}\frac{\partial\hat{S}^{(1)}}{\partial k_{v}}\right), \\
    \widehat{\Delta v}{}^{(3)}=&-i\left(\frac{\partial\hat{S}^{(3)}}{\partial k_{v}}+\hat{S}^{(1)\dag}\frac{\partial\hat{S}^{(2)}}{\partial k_{v}}+\hat{S}^{(2)\dag}\frac{\partial\hat{S}^{(1)}}{\partial k_{v}}\right).
\end{align}
Without ambiguity we denote the corresponding $n$-th order term in the expectation value for the in-state \eqref{eq: Gaussian wave packet} simply by
\begin{equation}
    \Delta v^{(n)} = \bra{\rm in}\widehat{\Delta v}{}^{(n)}\ket{\rm in} \, .
\end{equation}
We consider interactions over a finite time $0 < u < u_{\text{max}}$. 
To first order, the GR time delay is
\ba\label{eq: first order in perturbative time delay}
        \Delta v^{(1)}_{\text{GR}}(u) = \int_{0}^{u}\sd u_1\,f(u_{1}) \left(1-u_{1}\frac{\partial}{\partial u_{1}}\right)K^{(1)}(u_{1})\,.
\ea\label{eq: second order in perturbative time delay}
The second order GR time delay is given by 
\ba
        \Delta v^{(2)}_{\text{GR}}(u) &= &k_{v} \left(\prod_{i=1}^{2} \int_{0}^{u}\sd u_{i}\,f(u_{i})\right)\times\\
        &&\bigg[\theta(u_1-u_2)\left(1-u_{1}\frac{\partial}{\partial u_{1}}\right)-\theta(u_2-u_1)\left(1-u_{2}\frac{\partial}{\partial u_{2}}\right)\bigg]K^{(2)}(u_{1},u_{2})\,,\nn
\ea\label{eq: third order in perturbative time delay}
and the third order one by
\ba
        \Delta v^{(3)}_{\text{GR}}(u) &=&k_{v}^{2} \left(\prod_{i=1}^{3} \int_{0}^{u}\sd u_{i}\,f(u_{i})\right)\times\\
        &\bigg[&\theta(u_1-u_2)\theta(u_2-u_3)\left(3-u_{1}\frac{\partial}{\partial u_{1}}-u_{2}\frac{\partial}{\partial u_{2}}-u_{3}\frac{\partial}{\partial u_{3}}\right)\nn \\
        &-&\theta(u_2-u_3)\left(2-u_{2}\frac{\partial}{\partial u_{2}}-u_{3}\frac{\partial}{\partial u_{3}}\right) + \theta(u_2-u_1)\left(1-u_{3}\frac{\partial}{\partial u_{3}}\right)\bigg]K^{(3)}(u_{1},u_{2},u_{3})\,,\nn
\ea
where $\theta(u)$ is the Heaviside step function and
\ba\label{eq: general n K expression}
    K^{(n)}(u_1,\dots,u_n) &=& (-i)^{n+1}\left(-\frac{2 \pi^{\frac{d}{2}}}{\Gamma\left(\frac{d-2}{2}\right)}\right)^{n} \left(\prod_{i=1}^{n}\int\frac{\sd^{d}q_{i}}{(2\pi)^{d}}\frac{e^{i\mathbf{q}_{i}\cdot\mathbf{b}}+\Theta_{\text{bal}}e^{-i\mathbf{q}_{i}\cdot\mathbf{b}}}{\mathbf{q}_{i}^{2}}\right)\\
   &\times& \exp\left[-\frac{\sigma^{2}}{2}\left\{\left|\sum_{i=1}^{n}\mathbf{q}_{i}\right|^{2}+\left|\frac{1}{2k_{v}\sigma^{2}}\sum_{i=1}^{n}u_{i}\mathbf{q}_{i}\right|^{2}\right\}
   +\frac{i}{2k_{v}}\sum_{i<j=1}^n(u_i-u_j)\mathbf{q}_i\cdot\mathbf{q}_j\right] \nonumber \, .
\ea

\subsection{The classical limit: Identifying diffusion and scattering effects}\label{subsec: classical limit}
The effects of uncertainty and diffusion are captured, at every order in $\Delta v$, by the term in curly brackets in the exponential of each $K^{(n)}$ in \eqref{eq: general n K expression}. 
In the classical limit, where this term is negligible, the first-order term becomes
\begin{equation}\label{eq: classical limit of first order time delay}
    \lim_{\rm classical}  \Delta v_{\text{GR}}^{(1)}(u) = -\frac{1}{k_v}\int_0^u \sd u_1 V(u_1,\mathbf{b})(1 + \Theta_{\text{bal}}),
\end{equation}
which is exactly the eikonal expression of \eqref{eq: EW time delay single source}. 
This is to be expected since the leading-order eikonal phase can be identified with the leading-order Born approximation. 

After some manipulation, the classical limit of the second-order term in the time delay expansion can be written as
\begin{align}\label{eq: classical limit of second order time delay}
    \lim_{\rm classical} \Delta v^{(2)}_{\text{GR}}(u) = &-\left(-\frac{2 \pi^{\frac{d}{2}}}{\Gamma\left(\frac{d-2}{2}\right)}\right)^2 k_v \int_0^u \sd u_1 f(u_1) \int_0^{u_1} \sd u_2 f(u_2) \int \frac{\sd^d q_1}{(2\pi)^2} \frac{\sd^d q_2}{(2\pi)^2}\nonumber\\
    &\frac{e^{i\mathbf{q}_{1}\cdot\mathbf{b}}+\Theta_{\text{bal}}e^{-i\mathbf{q}_{1}\cdot\mathbf{b}}}{\mathbf{q}_{1}^{2}} \frac{e^{i\mathbf{q}_{2}\cdot\mathbf{b}}+\Theta_{\text{bal}}e^{-i\mathbf{q}_{2}\cdot\mathbf{b}}}{\mathbf{q}_{2}^{2}}\\
    &\left(\sin\left[\frac{1}{2k_v}(u_1-u_2)\mathbf{q}_1\cdot\mathbf{q}_2\right] - \frac{u_1}{2k_v}\mathbf{q}_1\cdot\mathbf{q}_2 \cos\left[\frac{1}{2k_v}(u_1-u_2)\mathbf{q}_1\cdot\mathbf{q}_2\right]\right) . \nonumber
\end{align}
In the balancing case, $\Theta_{\text{bal}} = 1$, it is easy to see that the integrand of \eqref{eq: classical limit of second order time delay} is odd under $\mathbf{q}_1 \rightarrow - \mathbf{q}_1$ (or $\mathbf{q}_2 \rightarrow - \mathbf{q}_2$) and thus the integral vanishes.
In the point source case, there is no such anti-symmetry so the integral does not vanish.

The classical limit of the third-order term in the time delay expansion can be written as
\begin{align}\label{eq: classical limit of third order time delay}
    \lim_{\rm classical} \Delta v^{(3)}_{\text{GR}}(u) = &\left(-\frac{2 \pi^{\frac{d}{2}}}{\Gamma\left(\frac{d-2}{2}\right)}\right)^3 k_v^2 \int_0^u \sd u_1 f(u_1) \int_0^{u} \sd u_2 f(u_2) \int_0^{u} \sd u_3 f(u_3) \int \frac{\sd^d q_1}{(2\pi)^2} \frac{\sd^d q_2}{(2\pi)^2} \frac{\sd^d q_3}{(2\pi)^2}\nonumber\\
    &\frac{e^{i\mathbf{q}_{1}\cdot\mathbf{b}}+\Theta_{\text{bal}}e^{-i\mathbf{q}_{1}\cdot\mathbf{b}}}{\mathbf{q}_{1}^{2}} \frac{e^{i\mathbf{q}_{2}\cdot\mathbf{b}}+\Theta_{\text{bal}}e^{-i\mathbf{q}_{2}\cdot\mathbf{b}}}{\mathbf{q}_{2}^{2}} \frac{e^{i\mathbf{q}_{3}\cdot\mathbf{b}}+\Theta_{\text{bal}}e^{-i\mathbf{q}_{3}\cdot\mathbf{b}}}{\mathbf{q}_{3}^{2}}\nonumber\\
    &\left[\theta(u_1-u_2)\theta(u_2-u_3)\left(1-u_1\frac{\partial}{\partial u_1}\right)\right.\nonumber\\
    &- \theta(u_2-u_1)\theta(u_2-u_3)\left(1-u_2\frac{\partial}{\partial u_2}\right)\\\
    &\left.+ \theta(u_2-u_1)\theta(u_3-u_2)\left(1-u_3\frac{\partial}{\partial u_3}\right)\right]\nonumber\\
    &\exp\left[\frac{i}{2k_v}\left((u_1 - u_2)\mathbf{q}_1\cdot \mathbf{q}_2+(u_1 - u_3)\mathbf{q}_1\cdot \mathbf{q}_3+(u_2 - u_3)\mathbf{q}_2\cdot \mathbf{q}_3\right)\right].\nonumber
\end{align}
The vanishing of this in the balancing case is less straightforward. 
First note that, as the eigenvalue of an Hermitian operator, the expression \eqref{eq: classical limit of third order time delay} should be real.
We can verify that it is real by simultaneously exchanging
\begin{equation}
    u_1 \longleftrightarrow u_3 \quad \text{and} \quad \mathbf{q}_1 \longleftrightarrow \mathbf{q}_3,
\end{equation}
in the integrand to retrieve its complex conjugate.
Thus, the exponential factor at the bottom of \eqref{eq: classical limit of third order time delay} may be replaced by its real part, a cosine.
Then, in the balancing case only, we have the following situation
\begin{align}
    \lim_{\rm classical} \Delta v^{(3)}_{\text{GR}}(u) = \int_0^u \sd u_1 \sd u_2 \sd u_3 &f(u_1) f(u_2)f(u_3) \left[\theta(u_1-u_2)\theta(u_2-u_3)\left(1-u_1\frac{\partial}{\partial u_1}\right)\right.\nonumber\\
    &\left.- \theta(u_2-u_1)\theta(u_2-u_3)\left(1-u_2\frac{\partial}{\partial u_2}\right)\right.\\
    &\left.+ \theta(u_2-u_1)\theta(u_3-u_2)\left(1-u_3\frac{\partial}{\partial u_3}\right)\right] G(u_1,u_2,u_3)\,,\nonumber
\end{align}
where $G(u_1,u_2,u_3)$ is a function that is invariant under any permutation of the $u_i$'s.
For example, the exchange $u_1 \leftrightarrow u_2$ in $G$, may be offset by the exchange of the integration variables $\mathbf{q}_1 \leftrightarrow \mathbf{q}_2$ and the change of variables $\mathbf{q}_3 \rightarrow -\mathbf{q}_3$ to leave it invariant.
With this in mind, the first and third terms in the square brackets give the same contribution overall (by exchange of $u_1 \leftrightarrow u_3$).
Merging these terms and additionally exchanging $u_1 \leftrightarrow u_2$ in the second term, we obtain:
\begin{align}\label{eq: classical limit of third order time delay written with explicit integration limits}
\begin{split}
    \lim_{\rm classical} \Delta v^{(3)}_{\text{GR}}(u) = \int_0^u &\sd u_1 f(u_1) \int_0^{u_1} \sd u_2 f(u_2) \left(2\int_0^{u_2} \sd u_3 - \int_0^{u_1} \sd u_3\right)f(u_3)\\
    &\left(1-u_1\frac{\partial}{\partial u_1}\right)G(u_1,u_2,u_3).
\end{split}
\end{align}
Now, note that
\begin{align}
\begin{split}
    \int_0^{u_1}\sd u_2\left(2\int_0^{u_2} \sd u_3 - \int_0^{u_1} \sd u_3\right) &= \int_0^{u_1}\sd u_2\left(\int_0^{u_2} \sd u_3 - \int_{u_2}^{u_1} \sd u_3\right)\\
    &= \int_0^{u_1}\sd u_2 \int_0^{u_2} \sd u_3 - \int_0^{u_1} \sd u_3 \int_0^{u_3} \sd u_2 \, ,
\end{split}
\end{align}
and since the integrand is invariant under the exchange of $u_2 \leftrightarrow u_3$, the whole integral in \eqref{eq: classical limit of third order time delay written with explicit integration limits} is zero.

It is expected that all higher-order corrections to the leading term \eqref{eq: classical limit of first order time delay} in the balancing-source time delay vanish in the classical limit.
Again, let it be stressed that this is a reflection of the fact that there is, classically, no scattering from the unstable equilibrium point in the balancing potential.
Thus, outside the classical limit, the higher-order terms $\Delta v^{(n>1)}$ may be regarded as the result of diffusion-induced scattering in the quantum problem.

\subsection{Perturbative regime of validity}\label{subsec: scattering bounds}

The requirement that the higher-order scattering-induced terms $\Delta v_{\text{GR}}^{(n>1)}(u)$ are under control compared to the leading-order GR term $\Delta v^{(1)}_{\text{GR}}(u)$ imposes a time cut-off $u_{\text{max}}$ beyond which we may no longer trust \textit{perturbation theory}. 
Below, we will determine the parameter ``$\epsilon$" which controls the series expansion of $\Delta v_{\text{GR}}(u)$.
As we will see, this parameter will grow with time $u$ itself --- at early times, the series expansion is under better control and $\Delta v^{(1)}_{\text{GR}}(u)$ represents a good approximation for $\Delta v_{\text{GR}}(u)$, but at late times that approximation begins to fail.
The scattering timescale $u_{\text{pert}}$ will be defined as the time at which it becomes large, $\epsilon(u_{\text{pert}}) \sim 1$.

As a concession to simplicity, consider the limit of a large number of closely spaced shockwaves for which we may approximate $f(u) = f_0$ as a constant.
With this simplification, the $u$-derivative-terms in expressions \eqref{eq: first order in perturbative time delay}-\eqref{eq: third order in perturbative time delay} may be integrated by parts. For $u < u_{\text{max}}$, we therefore get, explicitly, at first order
\ba
        \Delta v^{(1)}_{\text{GR}}(u)  = -2\left(-\frac{2 \pi^{\frac{d}{2}}}{\Gamma\left(\frac{d-2}{2}\right)}\right) f_0 \int_0^u \text{d} u_1 \int \frac{\text{d}^d q_1}{(2 \pi)^d}
        \frac{e^{i \mathbf{q}_1\cdot\mathbf{b}}+\Theta_{\text{bal}}e^{-i\mathbf{q}_1\cdot \mathbf{b}}}{\mathbf{q}_1^2} \text{exp}\left[-\frac{\sigma^2 \mathbf{q}_1^2}{2}-\frac{u_1^2 \mathbf{q}_1^2}{8 k_v^2 \sigma^2}\right]\,,\nn
\ea
at second order,
\ba
        \Delta v^{(2)}_{\text{GR}}(u) &= &-2\left(-\frac{2 \pi^{\frac{d}{2}}}{\Gamma\left(\frac{d-2}{2}\right)}\right)^2 k_v f_0^2
        \int_0^u \text{d} u_1 \text{d} u_2 \int \frac{\text{d}^d q_1}{(2 \pi)^d} \frac{\text{d}^d q_2}{(2 \pi)^d}\\
        &&\frac{e^{i \mathbf{q}_1\cdot\mathbf{b}}+\Theta_{\text{bal}}e^{-i\mathbf{q}_1\cdot \mathbf{b}}}{\mathbf{q}_1^2} \frac{e^{i \mathbf{q}_2\cdot\mathbf{b}}+\Theta_{\text{bal}}e^{-i\mathbf{q}_2\cdot \mathbf{b}}}{\mathbf{q}_2^2}\nn \\
        &&\text{exp}\left[-\frac{\sigma^2}{2}\left|\mathbf{q}_1+\mathbf{q}_2\right|^2-\frac{1}{8 k_v^2 \sigma^2}\left|u_1 \mathbf{q}_1 + u_2 \mathbf{q}_2\right|^2\right]\nn \\
        && \text{sin}\left[\frac{1}{2k_v} (u_1 - u_2)\mathbf{q}_1\cdot \mathbf{q}_2\right]\,,\nn
\ea
and at third order
\ba
        \Delta v^{(3)}_{\text{GR}}(u) &= &2\left(-\frac{2 \pi^{\frac{d}{2}}}{\Gamma\left(\frac{d-2}{2}\right)}\right)^3 k_v^2 f_0^3
        \int_0^u \text{d} u_1 \text{d} u_2 \text{d} u_3 \int \frac{\text{d}^d q_1}{(2 \pi)^d} \frac{\text{d}^d q_2}{(2 \pi)^d} \frac{\text{d}^d q_3}{(2 \pi)^d}\\
        &&\frac{e^{i \mathbf{q}_1\cdot\mathbf{b}}+\Theta_{\text{bal}}e^{-i\mathbf{q}_1\cdot \mathbf{b}}}{\mathbf{q}_1^2} \frac{e^{i \mathbf{q}_2\cdot\mathbf{b}}+\Theta_{\text{bal}}e^{-i\mathbf{q}_2\cdot \mathbf{b}}}{\mathbf{q}_2^2} \frac{e^{i \mathbf{q}_3\cdot\mathbf{b}}+\Theta_{\text{bal}}e^{-i\mathbf{q}_3\cdot \mathbf{b}}}{\mathbf{q}_3^2}\nn \\
        &&\text{exp}\left(-\frac{\sigma^2}{2}\left|\mathbf{q}_1+\mathbf{q}_2+\mathbf{q_3}\right|^2-\frac{1}{8 k_v^2 \sigma^2}\left|u_1 \mathbf{q}_1 + u_2 \mathbf{q}_2 + u_3 \mathbf{q}_3\right|^2\right)\nn \\
        &&\text{cos}\left(\frac{1}{2k_v}\left[(u_1 - u_2)\mathbf{q}_1\cdot \mathbf{q}_2+(u_1 - u_3)\mathbf{q}_1\cdot \mathbf{q}_3+(u_2 - u_3)\mathbf{q}_2\cdot \mathbf{q}_3\right]\right)\nn\\
        &&\left[2 \Theta(u_1-u_2)\Theta(u_2-u_3) - \Theta(u_2-u_1)\Theta(u_2-u_3)\right]\,.\nn
\ea
While the cross terms mean these expressions are not readily integrated, we can estimate the parameter dependence of each $\Delta v^{(n)}_{\text{GR}}(u)$ as long as we make the following assumptions:
\begin{align}
    &\sigma \ll b, \label{eq: sigma much smaller than b}\\
    &u \ll k_v \sigma b. \label{eq: u much smaller than udiff}
\end{align}
Physically, \eqref{eq: sigma much smaller than b} corresponds to the requirement that the width of the initial Gaussian wave packet $\sigma$ is much smaller than the distance to the source $b$.
The second assumption \eqref{eq: u much smaller than udiff} is that the timescale under consideration $u$ is smaller than the diffusion time $u_{\text{diff}} \sim k_v \sigma b$ (see \eqref{eq: diffusion timescale}).
We also need to distinguish between the two possible cases where scattering is before diffusion and \textit{vice versa}.

\subsubsection{Scattering before diffusion}

By switching to dimensionless integration variables $w_i = u_i/u$ and $\mathbf{p}_i = b \mathbf{q}_i$, the parameter dependence of $\Delta v^{(n)}_{\text{GR}}(u)$ can be extracted to leading order in $\sigma/b$ and $u/k_v \sigma b$:
\begin{subequations}\label{eq: parameter dependence of first three orders in time delay}
\begin{align}
    &\begin{aligned}\label{eq: parameter dep of first order time delay}
        \Delta v^{(1)}_{\text{GR}}(u) \sim \frac{f u}{b^{d-2}},
    \end{aligned}\\[\baselineskip]
    &\begin{aligned}
        \Delta v^{(2)}_{\text{GR}}(u) \sim \frac{f^2 u^3}{b^{2d-2}}\left[1+\Theta_{\text{bal}}\left(-1 + a^{(2)}\frac{\sigma^2}{b^2} + \tilde{a}^{(2)}\frac{u^2}{k_v^2 b^2 \sigma^2}\right)\right],
    \end{aligned}\\[\baselineskip]
    &\begin{aligned}
        \Delta v^{(3)}_{\text{GR}}(u) \sim \frac{k_v^2 f^3 u^3}{b^{3d-6}}\left[1+\Theta_{\text{bal}}\left(-1 + a^{(3)}\frac{\sigma^2}{b^2} + \tilde{a}^{(3)}\frac{u^2}{k_v^2 b^2 \sigma^2}\right)\right],
    \end{aligned}
\end{align}
\end{subequations}
where $a^{(n)}$ and $\tilde{a}^{(n)}$ are numbers expressed in terms of $n (d+1)$ dimensionless integrals (given in appendix~\ref{app: expressions for a's}).

The ratio of one term $\Delta v^{(n)}_{\text{GR}}(u)$ to the next $\Delta v^{(n+1)}_{\text{GR}}(u)$ is not the same for all $n$.
The even terms $\Delta v^{(2n)}_{\text{GR}}(u)$ are artificially suppressed by the presence of an odd sine in their integrands, compared to the even cosine in odd terms $\Delta v^{(2n+1)}_{\text{GR}}(u)$.
In other words, the third-order term $\Delta v^{(3)}_{\text{GR}}(u)$ is not necessarily subdominant to the second-order term $\Delta v^{(2)}_{\text{GR}}(u)$, and so on.
This means that the ratio which truly controls the expansion is not the ratio of one term to the next $\Delta v^{(n+1)}_{\text{GR}}(u)/\Delta v^{(n)}_{\text{GR}}(u)$ but instead it is of even (or equally, odd) terms $\Delta v^{(n+2)}_{\text{GR}}(u)/\Delta v^{(n)}_{\text{GR}}(u)$.

Although the explicit parameter dependence of the generic $n$-th term is not given here, it is apparent from the general expression \eqref{eq: general n K expression} and the examples \eqref{eq: parameter dependence of first three orders in time delay} that the series expansion of the time delay is controlled by
\begin{equation}
    \epsilon(u) \sim \frac{\Delta v^{(n)}_{\text{GR}}(u)}{\Delta v^{(n+2)}_{\text{GR}}(u)} \sim \left(\frac{k_v f u}{b^{d-2}}\right)^2.
\end{equation}
The $S$-matrix expansion is under control when $\epsilon(u) \ll 1$ and so we may identify the time at which perturbation theory breaks down as
\begin{equation}\label{eq: perturbation theory scattering timescale}
    \epsilon(u_{\text{pert}}) \sim 1 \quad \implies \quad u_{\text{pert}} \sim \frac{1}{k_v}\frac{b^{d-2}}{f}.
\end{equation}
Unlike the estimates \eqref{eq: Newtonian scattering time}, \eqref{eq: WKB scattering time} and \eqref{eq: instability timescale}, this estimate for the cut-off timescale depends on the $v$-momentum $k_v$, which is a property of the test wave itself and not of the background. 
In fact 
\be
\frac{u_{\rm pert}}{u_s} \sim \frac{u_s}{u_{\rm diff}} \, ,
\ee
and so for $u_s<u_{\rm diff}$ the Born series breaks down before the eikonal approximation, which is why we regard it as less informative.

Using the expressions for the two timescales, \eqref{eq: diffusion timescale} and \eqref{eq: perturbation theory scattering timescale}, and the approximation for the leading-order time delay \eqref{eq: parameter dep of first order time delay}, we find
\begin{equation}
    u_{\text{pert}} \ll u_{\text{diff}} \quad \iff \quad k_v \abs*{\Delta v^{(1)}_{\text{GR}}(u_{\text{diff}})} \gg 1.
\end{equation}
In other words, the assumption that perturbation theory breaks down before the diffusion time is consistent, as long as the GR time delay generated by $u_{\text{diff}}$ is resolvable.

In this formalism, the EFT contribution to the time delay can generally be estimated as
\begin{equation}
    \Delta v^{(1)}_{\text{EFT}}(u) \sim \frac{\cGB}{b^2 \Lambda^2} \Delta v^{(1)}_{\text{GR}}(u) \sim  \cGB \frac{f u}{b^d \Lambda^2} \, .
\end{equation}
The total EFT time delay by the time at which perturbation theory breaks down is
\begin{equation}
     \Delta v^{(1)}_{\text{EFT}}(u_{\text{pert}}) \sim \frac{\cGB}{b^2 \Lambda^2} \frac{1}{k_v} \, ,
\end{equation}
and, given the EFT-validity bound $b \Lambda \gg 1$, satisfies
\be
k_v |  \Delta v^{(1)}_{\text{EFT}}(u_{\text{pert}})| \ll |\cGB| \, .
\ee

\subsubsection{Diffusion before scattering}

We have already seen that the assumption $u_{\text{scatt}} < u_{\text{diff}}$ is self-consistent, and we will now show that it is actually the only regime of interest.
Assuming that we can trust the expressions for $\Delta v^{(n)}_{\text{GR}}(u)$ given in \eqref{eq: parameter dependence of first three orders in time delay} all the way up until $u \sim u_{\text{diff}}$, we can write
\begin{equation}
    \Delta v^{(3)}_{\text{GR}}(u_{\text{diff}}) \sim k_v^2\left(\Delta v^{(1)}_{\text{GR}}(u_{\text{diff}})\right)^3 \left[1+\Theta_{\text{bal}}\left(-1 + a^{(3)}\frac{\sigma^2}{b^2} + \tilde{a}^{(3)}\right)\right].
\end{equation}
Of the parts that are exclusive to the balancing case, $\tilde{a}^{(3)}$ will dominate over $a^{(3)} \left(\sigma/b\right)^2$, so we can write-off the totality of the square brackets as an $\mathcal{O}(1)$ number plus suppressed corrections.
If the scattering time comes after the diffusion time then, by definition, the first-order term $\Delta v^{(1)}_{\text{GR}}(u)$ should still dominate the series expansion by $u = u_{\text{diff}}$.
However, asking for the third-order term to be sub-dominant, $\abs*{\Delta v^{(3)}_{\text{GR}}(u_{\text{diff}})} < \abs*{\Delta v^{(1)}_{\text{GR}}(u_{\text{diff}})}$ leads to the conclusion
\begin{equation}
   \abs*{ k_v^2\left(\Delta v^{(1)}_{\text{GR}}(u_{\text{diff}})\right)^3} < \abs*{\Delta v^{(1)}_{\text{GR}}(u_{\text{diff}})} \quad \implies \quad k_v \abs*{\Delta v^{(1)}_{\text{GR}}(u_{\text{diff}})} < 1,
\end{equation}
\textit{i.e.} the first-order GR time delay itself is unresolvable.
As hinted at previously, we see that if we try to slow down the effects of scattering for long enough such that diffusion takes hold first, then the price to pay is a potential so weak (\textit{i.e.} $f$ so small) that even the leading-order GR time delay cannot be resolved.
Since the EFT contribution is further suppressed by $1/b^2 \Lambda^2 \ll 1$ compared to the GR contribution, it will certainly not be resolvable.

\section{Conclusions}\label{sec:Conclusion}

In this paper, we reviewed the formal definition of the scattering time delay appropriate to spacetimes with a null Killing vector (namely pp-wave geometries), showed how to compute the time delay for arbitrary initial states in the eikonal, semi-classical, and Born approximations, and discussed their validity and connections between them. 
We considered the generic EFT of gravity in arbitrary dimensions $D>4$, focusing on the Gauss-Bonnet term as the leading correction. 
Assuming a generic weakly coupled ($\Lambda \ll \mpl$) Wilsonian EFT expansion \eqref{eq: generic EFT action}, we carefully computed the regime of validity necessary to determine scattering amplitudes in a general pp-wave background. 
We saw that, while it is straightforward to engineer background configurations for which there is an arbitrarily large gravitational (Shapiro) time delay, the corrections due to the GB-term that can be consistently inferred in the regime of validity of the low-energy EFT are bounded in the manner
\be
k_v |\Delta v_{\rm EFT}| \lesssim |\cGB| \, .
\ee
In particular, this remains true for the rather special balancing shockwave solutions considered in Section \ref{subsec: balancing sources}, for which scattering is minimised. 
To this end, one must choose highly-localised wave packets which are far from $S$-matrix eigenstates. 
As such, the quantum mechanical uncertainty in the time delay grows, swamping any potentially observable contribution.

These conclusions may be read in two ways: (1) Assuming that $|\cGB| \lesssim {\cal O}(1)$, which is expected for gravitational EFTs with consistent UV completions based on positivity/bootstrap bounds \cite{Caron-Huot:2022jli} as well as by EFT power counting arguments, the time delay induced by the GB-term is generically unresolvable, or (2) Demanding infrared causality (namely causality with respect to the locally inertial metric) $\Delta v_{\rm EFT}  \gtrsim -k_v^{-1}$ and accounting for the fact that the Gauss-Bonnet term can lead to time advances for scattering of specific polarisations, then we infer $|\cGB| \lesssim {\cal O}(1)$. 
Finally, we note that asymptotic causality in the form $\Delta T>0$ is automatically satisfied, since the magnitude of the EFT correction is $|\cGB|/(\Lambda b)^2$ suppressed relative to the GR/Shapiro delay and the validity of the EFT requires $\Lambda b \gg 1$.
While satisfied, it does not provide much insight. 

Either way, these results indicate a close connection between causality via time delays and positivity/bootstrap bounds previously observed in \cite{Adams:2006sv,deRham:2020zyh,Chen:2021bvg,CarrilloGonzalez:2022fwg,CarrilloGonzalez:2023cbf}. 
Further, they show that, while asymptotic causality is automatically satisfied, it is generically too weak a condition.
This confirms the findings of \cite{deRham:2021bll} and that, in EFTs with a consistent UV completion, causality is maintained with respect to the local metric, which defines the local inertial frame and to which at least one field couples to (taking care of field redefinition ambiguity, see \cite{deRham:2019ctd} for more details on that point).

\section*{Acknowledgements}

This work is supported by STFC Consolidated Grant ST/T000791/1. CdR is also supported by a Simons Investigator award 690508. CC and AM are funded by a Presidents Scholarship.

\newpage

\appendix

\section{EFT field equations for metric perturbations}\label{app: full field eqs}
In lightcone gauge, where $h_{v\alpha} = 0$, the EFT field equations for the metric perturbations
\begin{equation}
    \delta \mathcal{E}_{\alpha \beta} = \delta G_{\alpha \beta} +\frac{2\cGB}{\Lambda^2}\delta B_{\alpha \beta} = 0,
\end{equation}
are explicitly given by
\begin{subequations}
\begin{align}
    &-2\delta\mathcal{E}_{vv} = \partial_v^2 h_{ii},\\[\baselineskip]
    &-2\delta\mathcal{E}_{vu} = -\tilde{\Box} h_{ii} + \partial_v\partial_u h_{ii} + \partial_i (\partial_v h_{iu} + \partial_j h_{ij}),\\[\baselineskip]
    &-2\delta\mathcal{E}_{vi} = \partial_v \partial_i h_{jj} - \partial_v(\partial_v h_{iu} + \partial_j h_{ij}),\\[\baselineskip]
    &\,\,\begin{aligned}
    -2\delta\mathcal{E}_{uu} = &\,\,\tilde{\Box}h_{uu} - 2\partial_u(\partial_v h_{uu} + \partial_i h_{iu}) + \partial_i H \partial_v h_{iu} - \partial_j(\partial_i H h_{ij}) + \partial_u^2 h_{ii} - H\tilde{\Box}h_{ii}\\
    &+ \frac{1}{2}\partial_j H \partial_j h_{ii} - \frac{1}{2}\partial_u H \partial_v h_{ii} + H\partial_i(\partial_v h_{iu} + \partial_j h_{ij}) + H\partial_v(\partial_v h_{uu} + \partial_i h_{iu})\\
    &+ \frac{4\cGB}{\Lambda^2} \partial_i \partial_j H(\partial_v\partial_i h_{ju} - \partial_v\partial_j h_{iu} + \partial_k\partial_kh_{ij} - \partial_i \partial_k h_{jk} - \partial_j\partial_k h_{ik} + \partial_i\partial_j h_{kk}),
    \end{aligned}\\[\baselineskip]
    &\,\,\begin{aligned}
    -2\delta\mathcal{E}_{ui} = &\,\,\tilde{\Box}h_{ui} + \partial_j H \partial_v h_{ij} - \partial_i(\partial_v h_{uu} + \partial_j h_{ju}) -\partial_u(\partial_v h_{iu} + \partial_jh_{ij})\\
    &+\partial_i\partial_uh_{jj} - \frac{1}{2}\partial_i H \partial_v h_{jj}\\
    &-\frac{4\cGB}{\Lambda^2}\partial_v\left[\partial_i\partial_j H (\partial_j h_{kk} - \partial_k h_{jk})+\partial_j\partial_k H(\partial_i h_{jk} - \partial_j h_{ik})\right],
    \end{aligned}\\[\baselineskip]
    &\,\,\begin{aligned}
    -2\delta\mathcal{E}_{ij} = &\,\,\tilde{\Box}h_{ij} - \partial_i(\partial_v h_{ju} + \partial_k h_{jk}) - \partial_j(\partial_v h_{iu} + \partial_k h_{ik}) + \partial_i\partial_j h_{kk} - \delta_{ij}\tilde{\Box}h_{kk}\\
    &+\delta_{ij}\partial_k(\partial_vh_{ku} + \partial_l h_{kl}) + \delta_{ij}\partial_v(\partial_v h_{uu} + \partial_k h_{ku})\\
    &+\frac{4\cGB}{\Lambda^2}\partial_v^2(\partial_i\partial_j H h_{kk} - \partial_i\partial_k H h_{jk} - \partial_j\partial_k H h_{ik} + \delta_{ij}\partial_k\partial_l H  h_{kl}).
    \end{aligned}
\end{align}
\end{subequations}
These equations are satisfied with the constraints \eqref{eq: constraint 1: traceless} -- \eqref{eq: constraint 3: huu} and the equation of motion for $h_{ij}$ \eqref{eq: eom for hij}.

\section{Master variables in balancing source background}\label{app: master variables in balancing background}
Identifying the master variables for the metric perturbations on the point-source background was a straightforward procedure described above equation \eqref{eq: master equation for point source}.
When a second source is introduced (to ``balance" the first, as discussed in Section \ref{subsec: balancing sources}), spherical symmetry is lost and the previous decomposition is no longer suitable.
Without loss of generality, assume the two sources are aligned along the $z$-axis.
Then, we can write the transverse-space metric in cylindrical coordinates as:
\begin{equation}
    \delta_{ij} \d x^i \d x^j = \d z^2 + \d r^2 + r^2 \d\Omega_{d-2}^2.
\end{equation}
For this appendix only, $\gamma_{ab}$ will refer to the metric on the $(d-2)$-dimensional sphere.
In these coordinates, the harmonic condition on the metric function is
\begin{equation}
    \frac{\partial^2}{\partial z^2} H(u,r,z) = -\frac{\partial^2}{\partial r^2} H(u,r,z) - (d-2)\frac{1}{r}\frac{\partial}{\partial r} H(u,r,z).
\end{equation}
While most of the components of the metric perturbations --- $h_{rr}$, $h_{rz}$, $h_{zz}$, $h_{ra}$, $h_{za}$ and the diagonal components of $h_{ab}$ --- are coupled in a complicated way via \eqref{eq: eom for hij}, it happens that the off-diagonal components of $h_{ab}$ are completely decoupled from the rest and evolve independently.
These account for $(D-5)(D-4)/2$ of the total $D(D-3)/2$ propagating modes.
Their equation of motion is
\begin{equation}
    \tilde{\Box}h_{ab} - 8 \frac{\cGB}{\Lambda^2}\frac{\partial_r H}{r}\partial_v^2 h_{ab} = 0, \quad a\neq b \, ,
\end{equation}
which has the same format as the equations for the modes in the spherically-symmetric background \eqref{eq: master equation for point source}.

\section{Determining the EFT regime of validity}\label{app: determining the EFT regime of validity}

In Section \ref{subsec: EFT regime of validity}, we argue that the EFT regime of validity is given schematically by
\begin{equation}
     \left(\frac{\nabla}{\Lambda}\right)^{m} \left(\frac{\bar{R}}{\Lambda^2}\right)^n\left(\frac{k}{\Lambda}\right)^p \ll 1.
\end{equation}
In this appendix, we will explain why only the particular contractions given in \eqref{eqs: non-trivial EFT bounds together} lead to non-trivial bounds on the energy of the perturbations and the background parameters.

Firstly, contracting two momentum vectors together will lead to a weaker bound than if they had been contracted with other components.
The reason for this is that the gravitational wave momentum is a null vector in GR, so its norm vanishes $k_{\alpha}k^{\alpha} \approx 0$ up to EFT corrections.

Secondly, any contractions of indices on the same Riemann tensor produces a Ricci tensor (or scalar) which is zero in vacuum.
Therefore, each of the four indices on a Riemann tensor need to be contracted with either a momentum vector $k_{\alpha}$, a covariant derivative $\nabla_{\alpha}$ or a different Riemann tensor.
Also, it can be checked that the divergence of the Riemann tensor vanishes in vacuum
\begin{equation}
    \nabla^{\alpha}\bar{R}_{\alpha\beta\gamma\delta} = 0 \quad \text{when}\quad \lp H = 0.
\end{equation}
Finally, by symmetry, the Riemann tensor may only enter via contractions with at most two $k$'s, at most two $\nabla$'s and at most two other Riemann tensors.

The expressions in \eqref{eqs: non-trivial EFT bounds together} represent the only cases where no covariant derivatives have been contracted with an index on a Riemann tensor.
The remaining options are to contract one covariant derivative with a Riemann tensor index,
\begin{equation}
    (\bar{R}\cdot \nabla)^{2p} k^{4p} \ll \Lambda^{10p},
\end{equation}
or two covariant derivatives with Riemann tensor indices,
\begin{equation}
    (\bar{R}\cdot \nabla \nabla)^{2q} k^{4q} \ll \Lambda^{12q}.
\end{equation}
Both cases produce weaker bounds, which follow from those already given in \eqref{eqs: collated EFT RoV bounds}.

\section{Higher-dimension EFT field equations}
\label{app: higher-dim EFT}

In Section \ref{subsec: RoV by example}, we show that control over the higher-dimension EFT action \eqref{eq: EFT action with higher-dim ops} amounts to exactly the regime of validity found by generic arguments in Section \ref{subsec: EFT regime of validity}.
In this appendix, we provide the field equations for that action and also identify the tensor-type master variable and its associated master equation.

The background equation is
\begin{equation}\label{eq: higher dim vacuum eq}
    \mathcal{E}_{\alpha\beta} \coloneqq G_{\alpha\beta} + 2\frac{\cGB}{\Lambda^{2}}B_{\alpha\beta} + 2\frac{\cRt}{\Lambda^{4}}C_{\alpha\beta} + 2\frac{\cRf}{\Lambda^{6}}D_{\alpha\beta} = 0,
\end{equation}
where
\begin{align}
    B_{\alpha \beta} = & 4 R_{\sigma \alpha \beta \rho} R^{\sigma \rho} + 2 R\indices{_{\alpha}^{\sigma \rho \kappa}}R_{\beta \sigma \rho \kappa}- 4 R_{\alpha \sigma}R\indices{_{\beta}^{\sigma}}+ 2 R R_{\alpha \beta} -\frac{1}{2}  \RGB^2\, g_{\alpha \beta}, \\
    C_{\alpha\beta} = & -\frac{1}{2} g_{\alpha \beta} (R^{3}) + 3(R^{3})\indices{_{\alpha\sigma\beta}^{\sigma}} + 6\nabla_{\sigma}\nabla_{\rho}(R^{2})\indices{_{\alpha}^{\sigma}_{\beta}^{\rho}}, \\
    D_{\alpha\beta} = & -\frac{1}{2}g_{\alpha \beta} (R^{4})+4(R^{4}) \indices{_{\alpha\sigma\beta}^{\sigma}} + 8\nabla_{\sigma}\nabla_{\rho} (R^{3})\indices{_{\alpha}^{\sigma}_{\beta}^{\rho}},
\end{align}
and
\begin{align}
    &(R^n)_{\alpha \beta \sigma \rho} = R_{\alpha \beta \gamma \delta} R^{\gamma \delta \bullet \bullet} \dots R^{\bullet \bullet \kappa \lambda} R_{\kappa \lambda \sigma \rho},\\
    &(R^n) = (R^n)\indices{_{\alpha \beta}^{\alpha \beta}}.
\end{align}
The vacuum pp-wave metric is again a solution to \eqref{eq: higher dim vacuum eq}.
The perturbations of these tensors are
\begin{align}
    &\begin{aligned}
        \delta B_{\mu\nu} = & -4R\indices{_{(\mu}^{\alpha}_{\nu)}^{\beta}} \nabla_{\lambda} \nabla_{\beta} h\indices{_{\alpha}^{\lambda}} + 2R\indices{_{\mu}^{\alpha}_{\nu}^{\beta}} \left(\Box h_{\alpha\beta}+\nabla_{\beta}\nabla_{\alpha}h\right) \\
        & + 4R_{(\mu|\alpha\beta\rho} \nabla^{\rho} \nabla_{|\nu)} h^{\alpha\beta} - 4R_{(\mu|\rho\alpha\beta} \nabla^{\rho} \nabla^{\beta} h\indices{_{|\nu)}^{\alpha}}	\\
        & + 2g_{\mu\nu} R_{\alpha\rho\beta\sigma} \nabla^{\sigma} \nabla^{\rho}h^{\alpha\beta}
    \end{aligned}  \\
    &\begin{aligned}
        \delta C_{\mu\nu} = & 12R\indices{_{(\mu|\rho\alpha\beta}} \nabla^{\rho} \Box\nabla^{\beta} h\indices{_{|\nu)}^{\alpha}} - 12R\indices{_{(\mu|\rho\alpha\beta}} \nabla^{\rho} \nabla^{\beta} \nabla_{\sigma} \nabla_{|\nu)} h^{\sigma\alpha} \\
        & -12 \nabla_{\sigma} R\indices{_{(\mu|\rho\alpha\beta}} \nabla^{\rho} \nabla^{\beta} \nabla_{|\nu)} h^{\alpha\sigma} + 12\nabla_{\sigma} R\indices{_{(\mu|\rho\alpha\beta}} \nabla^{\rho} \nabla^{\sigma} \nabla^{\beta} h\indices{_{|\nu)}^{\alpha}} \\
        & +3 \left(R\indices{_{\mu}^{\rho\lambda\alpha}} R\indices{_{\nu\lambda}^{\sigma\beta}} + R\indices{_{\mu}^{\lambda\rho\alpha}} R\indices{_{\nu}^{\sigma}_{\lambda}^{\beta}}\right) \nabla_{\rho} \nabla_{\sigma} h_{\alpha\beta}
    \end{aligned} \\
    &\begin{aligned}
        \delta D_{\mu\nu} = & -16R_{(\mu|\rho\beta\lambda} R_{|\nu)\sigma\alpha\kappa} \nabla^{\sigma} \nabla^{\rho} \nabla^{\lambda} \nabla^{\kappa} h^{\alpha\beta} - 32\nabla_{\sigma} R\indices{_{(\mu|\rho\beta\lambda}} R\indices{_{|\nu)}^{\sigma}_{\alpha\kappa}} \nabla^{\rho} \nabla^{(\kappa} \nabla^{\lambda)} h^{\alpha\beta} \\
        & -16\nabla^{\sigma} R\indices{_{(\mu|}^{\rho}_{\alpha\kappa}} \nabla_{\rho} R\indices{_{|\nu)\sigma\beta\lambda}} \nabla^{\lambda} \nabla^{\kappa} h^{\alpha\beta}.
    \end{aligned}
\end{align}
In lightcone gauge, the presence of the $R^3$-operator modifies the $v$-component perturbation equations
\begin{subequations}
\begin{align}
    &\begin{aligned}
    -2 \delta \mathcal{E}_{vv} = \partial_v^2 h_{ii},
    \end{aligned}\\[\baselineskip]
    &\begin{aligned}
    -2 \delta \mathcal{E}_{vu} = &-\tilde{\Box}_v^2 h_{ii} + \partial_v \partial_u h_{ii} + \partial_i(\partial_v h_{iu} + \partial_j h_{ij})\\
    &- 12 \frac{\cRt}{\Lambda^4}\partial_v^2 \partial_i\left[\partial_i\partial_j H(\partial_v h_{ju} + \partial_k h_{jk}) + \partial_i\partial_j\partial_k H h_{jk}\right],
    \end{aligned}\\[\baselineskip]
    &\begin{aligned}
    -2\delta\mathcal{E}_{vi} = &\,\,\partial_v \partial_i h_{jj} - \partial_v(\partial_v h_{iu} + \partial_j h_{ij})\\
    &+ 12 \frac{\cRt}{\Lambda^4}\partial_v^3\left[\partial_i\partial_j H(\partial_v h_{ju} + \partial_k h_{jk}) + \partial_i\partial_j\partial_k H h_{jk}\right],
    \end{aligned}
\end{align}
\end{subequations}
such that, while the traceless condition $h_{ii} = 0$ is unchanged, the second constraint equation \eqref{eq: constraint 2: hui} becomes
\begin{equation}\label{eq: higher dim modified second constraint eq}
    \partial_v h_{iu} + \partial_j h_{ij} = 12 \frac{\cRt}{\Lambda^4}\partial_v^2\left[\partial_i\partial_j H(\partial_v h_{ju} + \partial_k h_{jk}) + \partial_i\partial_j\partial_k H h_{jk}\right].
\end{equation}
The $\delta \mathcal{E}_{ij}$ is also modified by both new operators
\begin{align}
\begin{split}
    -2\delta\mathcal{E}_{ij} = \,\,&\tilde{\Box}h_{ij} - \partial_i(\partial_v h_{ju} + \partial_k h_{jk}) - \partial_j(\partial_v h_{iu} + \partial_k h_{ik}) + \partial_i\partial_j h_{kk} - \delta_{ij}\tilde{\Box}h_{kk}\\
    &+\delta_{ij}\partial_k(\partial_vh_{ku} + \partial_l h_{kl}) + \delta_{ij}\partial_v(\partial_v h_{uu} + \partial_k h_{ku})\\
    &+\frac{4\cGB}{\Lambda^2}\partial_v^2(\partial_i\partial_j H h_{kk} - \partial_i\partial_k H h_{jk} - \partial_j\partial_k H h_{ik} + \delta_{ij}\partial_k\partial_l H  h_{kl})\\
    &\begin{aligned}-12 \frac{\cRt}{\Lambda^4}\partial_v^2&\left[\partial_i \partial_k H \tilde{\Box}h_{jk} + \partial_i \partial_k H \tilde{\Box}h_{jk} + \partial_u\partial_i \partial_k H \partial_v h_{jk} + \partial_u\partial_j \partial_k H \partial_v h_{ik}\right.\\
    &-\partial_i\partial_k H\partial_j(\partial_v h_{ku} + \partial_l h_{kl}) - \partial_j\partial_k H\partial_i(\partial_v h_{ku} + \partial_l h_{kl})\\
    &\left.-\partial_i\partial_k\partial_l H(\partial_j h_{kl} - \partial_k h_{jl}) -\partial_j\partial_k\partial_l H(\partial_i h_{kl} - \partial_k h_{il}) \right]
    \end{aligned}\\
    &+16 \frac{\cRf}{\Lambda^6} \partial_i \partial_k H \partial_j \partial_l H \partial_v^4 h_{kl}\,.
\end{split}
\end{align}
The final constraint will again come from the trace of this equation.
However, the process is complicated by the presence of the $R^3$-operator.
In EGB theory, the second constraint \eqref{eq: constraint 2: hui} allowed us to exactly remove the $h_{iu}$ components, which appeared in the $ij$-equation.
But now, the $h_{iu}$ components also appear on the right hand side of \eqref{eq: higher dim modified second constraint eq}.
As a result, they may only be replaced \textsl{perturbatively} in inverse powers of the cut-off inside $\delta \mathcal{E}_{ij}$.
Up to corrections of $\mathcal{O}(\Lambda^{-8})$, the third constraint equation is
\begin{align}
\begin{split}
    \partial_v h_{uu} + \partial_i h_{iu} = \,\,&-4\frac{d-2}{d}\frac{\cGB}{\Lambda^2}\partial_i\partial_j H \partial_v h_{ij}\\
    &+\frac{24}{d}\frac{\cRt}{\Lambda^4}\partial_v\left[\partial_i\partial_j H \tilde{\Box}h_{ij} + \partial_u \partial_i \partial_j H \partial_v h_{ij} - \frac{d-2}{2} \partial_i \partial_j \partial_k H \partial_i h_{jk}\right]\\
    &-\frac{16}{d}\frac{\cRf}{\Lambda^6} \partial_i \partial_j H \partial_i \partial_k H \partial_v^3 h_{jk}.
\end{split}
\end{align}
As before, we see that the dynamical degrees of freedom are contained within the metric perturbations in the transverse directions $h_{ij}$.
Applying the three constraint equations, their equation of motion becomes (up to corrections of $\mathcal{O}(\Lambda^{-8})$):
\begin{align}\label{eq: higher dim ij eom}
    \tilde{\Box}h_{ij} - 8\frac{\cGB}{\Lambda^2}\partial_v^2 X_{ij} - 24 \frac{\cRt}{\Lambda^4}\partial_v^2 Y_{ij} - 16 \frac{\cRf}{\Lambda^6}\partial_v^4 Z_{ij} = 0,
\end{align}
where
\begin{align}
    &X_{ij} = \frac{1}{2}\left(h_{ik}\partial_j\partial_k H + h_{jk}\partial_i\partial_k H\right) - \frac{1}{d}\delta_{ij} h_{kl}\partial_k\partial_l H,\\
    &\begin{aligned}
        \,Y_{ij} = \,\,&\frac{1}{2}\left(\tilde{\Box}h_{ik}\partial_j \partial_k H +\tilde{\Box}h_{jk}\partial_i \partial_k H \right) - \frac{1}{d} \delta_{ij} \tilde{\Box}h_{kl}\partial_k \partial_l H\\
        &+\frac{1}{2}\left(\partial_v h_{ik} \partial_u \partial_j \partial_k H +\partial_v h_{jk} \partial_u \partial_i \partial_k H\right) - \frac{1}{d}\delta_{ij} \partial_v h_{kl} \partial_u \partial_k \partial_l H\\
        &+\frac{1}{2}\left(\partial_l h_{ik} \partial_j \partial_k \partial_l H + \partial_l h_{jk} \partial_i \partial_k \partial_l H\right) - \frac{1}{d}\delta_{ij} \partial_m h_{kl} \partial_k \partial_l \partial_m H\\
        &+ h_{kl} \partial_i \partial_j \partial_k \partial_l H,
    \end{aligned}\\
    &Z_{ij} = -h_{kl}\partial_i\partial_k H \partial_j \partial_l H + \frac{1}{d}\delta_{ij} h_{lm}\partial_k \partial_l H \partial_k \partial_m H.
\end{align}

\subsection{Tensor modes}

The perturbation equations \eqref{eq: higher dim ij eom} couple all components of $h_{ij}$ in a non-trivial fashion when $\cRt \neq 0$.
Even in the point source case, the identification of master variable used in Section \ref{subsec: metric perturbations} for EGB theory no longer works when the $R^3$-operator is included.
For illustrative purposes, we consider just the spherically symmetric point source case.
Due to the symmetry, we may perform a scalar-vector-tensor (SVT) decomposition with each type of perturbation (scalar, vector or tensor) evolving independently of the other types.
The simplest of them are the tensor modes, which are parameterised by
\begin{align}
    h_{rr} = h_{ra} = 0, \quad h_{ab} = r^2 \Phi_T \mathbb{T}_{ab},
\end{align}
where $\mathbb{T}_{ab}$ is a tensor spherical harmonic on the $(d-1)$-sphere and satisfies
\begin{subequations}
    \begin{align}
        &\left(\hat{\Delta}_{d-1} + \kappa_T^2\right)\mathbb{T}_{ab} = 0,\\
        &\mathbb{T}\indices{^a_a} = 0, \quad \hat{D}^a \mathbb{T}_{ab} = 0.
    \end{align}
\end{subequations}
We used $\kappa_T^2$ to represent the eigenvalue of $\mathbb{T}_{ab}$ instead of the traditional $k^2$ to avoid any possible confusion with the momentum of the perturbations $(k_u,k_v,\mathbf{k})$ used elsewhere.
There is one tensor mode $\Phi_T$ for each tensor spherical harmonic on the $(d-1)$-sphere.
They all share a master equation, which can be derived from the $(a,b)$-components (with $a \neq b$) of \eqref{eq: higher dim ij eom}
\begin{align}
\begin{split}
        0 = \,\, &\tilde{\Box} \Phi_{T} - \frac{\kappa_T^2+2}{r^2}\Phi_T - 8\frac{\cGB}{\Lambda^{2}} \frac{\partial_{r}H}{r} \partial_{v}^{2} \Phi_{T} + 24 \frac{\cRt}{\Lambda^{4}}\partial_v^2 \left[d\frac{\partial_r H}{r^2}\left(\partial_r \Phi_T + \frac{\Phi_T}{r}\right)\right.\\
        &\left.-\frac{\partial_{u}\partial_{r}H}{r} \partial_{v} \Phi_{T}\right] + 16\frac{\cRf-12\cGB\cRt}{\Lambda^{6}} \left(\frac{\partial_{r}H}{r}\right)^{2} \partial_{v}^{4} \Phi_{T}.
\end{split}
\end{align}
In arriving at this equation, the $\tilde{\Box}h_{ab}$ terms present in $Y_{ab}$ have been replaced perturbatively using the lower-order equation of motion, giving rise to the cross term $\propto \cGB \cRt$.
Once again, this equation is valid up to corrections of order $\mathcal{O}(\Lambda^{-8})$.

\section{EFT validity bound on wavefunction spread}\label{app: bound on wavefunction spread}

In Section \ref{subsec: becoming quantum}, we discuss how a generic initial quantum state has a non-zero spread $\sigma$ in transverse space and thus cannot be perfectly balanced at an unstable equilibrium point.
As a state in a low-energy EFT, this spread is in fact also bounded below by $\Lambda^{-1}$.
To demonstrate this, we must bound a Lorentz-invariant quantity dependent on $\sigma$ (as we did for $H(u,r)$, $k_v$ and $b$ in Section \ref{subsec: EFT regime of validity}).
The spread in momentum space is inversely related to the spread in real space, $\abs*{\Delta \mathbf{k}} \sim \sigma^{-1}$, so we would like to bound the inner product of two opposing wave vectors $\mathbf{k^{\pm}} = \mathbf{k^0} \pm \Delta \mathbf{k}$, where $\mathbf{k^0}$ is some central momentum.
However, to construct a scalar quantity, they must first be embedded in $D$-vectors on the full $D$-dimensional pp-wave spacetime
\begin{equation}
    \left(k^{\pm}_{\alpha}\right) = \left(k^{\pm}_u, k^{\pm}_v, \mathbf{k^{\pm}}\right) \, ,
\end{equation}
with the condition that $k^{\pm}$ is null due to the GR equations of motion,
\begin{equation}
    g^{\alpha \beta}k^{\pm}_{\alpha}k^{\pm}_{\beta} = 0 \quad \implies \quad k^{\pm}_{u} = \frac{F (k^{\pm}_v)^2 - \mathbf{k^{\pm}}^2}{2k^{\pm}_v}.
\end{equation}
For a fixed ``mass" $k_v^+ = k_v^- = k_v$, the inner product reduces to
\begin{equation}
    k^+ \cdot k^- = -2 \Delta\mathbf{k}^2.
\end{equation}
With the Lorentz-invariant bound $k^+ \cdot k^- \ll \Lambda^2$, we get the desired result $\abs*{\Delta \mathbf{k}} \ll \Lambda$, or $\sigma \Lambda \gg 1$.

\section{Expansion of the Wigner-Smith time delay}\label{app: expressions for a's}
The expressions \eqref{eq: parameter dependence of first three orders in time delay} give the parameter dependence of the first three orders of the Wigner-Smith time delay expanded in 1) $S$-matrix perturbation theory, 2) small Gaussian width, and 3) early times.
Four dimensionless numbers
$\{a^{(2)},\tilde{a}^{(2)},a^{(3)},\tilde{a}^{(3)}\}$ feature in these equations and are given below in terms of integrals, where $\hat{\mathbf{b}} = \mathbf{b}/b$ is the unit vector in the direction of $\mathbf{b}$.
\begin{subequations}
\begin{align}
    &\begin{aligned}
        a^{(2)} = \,\,&-2\left(-\frac{2 \pi^{\frac{d}{2}}}{\Gamma\left(\frac{d-2}{2}\right)}\right)^2 \int_0^1 \text{d}w_1 \text{d}w_2 \Theta(w_1 - w_2) \int  \frac{\text{d}^d p_1}{(2 \pi)^d} \frac{\text{d}^d p_2}{(2 \pi)^d}\\
        &\frac{e^{i \mathbf{p}_1\cdot\hat{\mathbf{b}}}+\Theta_{\text{bal}}e^{-i\mathbf{p}_1\cdot \hat{\mathbf{b}}}}{\mathbf{p}_1^2} \frac{e^{i \mathbf{p}_2\cdot\hat{\mathbf{b}}}+\Theta_{\text{bal}}e^{-i\mathbf{p}_2\cdot \hat{\mathbf{b}}}}{\mathbf{p}_2^2}\left(\mathbf{p}_1+\mathbf{p}_2\right)^2 (w_1-w_2)\mathbf{p}_1 \cdot \mathbf{p}_2 \, ,
    \end{aligned}\\[\baselineskip]
    &\begin{aligned}
        \tilde{a}^{(2)} = &-2\left(-\frac{2 \pi^{\frac{d}{2}}}{\Gamma\left(\frac{d-2}{2}\right)}\right)^2 \int_0^1 \text{d}w_1 \text{d}w_2 \Theta(w_1 - w_2) \int  \frac{\text{d}^d p_1}{(2 \pi)^d} \frac{\text{d}^d p_2}{(2 \pi)^d}\\
        &\frac{e^{i \mathbf{p}_1\cdot\hat{\mathbf{b}}}+\Theta_{\text{bal}}e^{-i\mathbf{p}_1\cdot \hat{\mathbf{b}}}}{\mathbf{p}_1^2} \frac{e^{i \mathbf{p}_2\cdot\hat{\mathbf{b}}}+\Theta_{\text{bal}}e^{-i\mathbf{p}_2\cdot \hat{\mathbf{b}}}}{\mathbf{p}_2^2}\left(w_1\mathbf{p}_1 + w_2\mathbf{p}_2\right)^2(w_1-w_2)\mathbf{p}_1 \cdot \mathbf{p}_2 \, ,
    \end{aligned}\\[\baselineskip]
    &\begin{aligned}
        a^{(3)} = &2\left(-\frac{2 \pi^{\frac{d}{2}}}{\Gamma\left(\frac{d-2}{2}\right)}\right)^3 \int_0^1 \text{d}w_1 \text{d}w_2 \text{d}w_3 \int  \frac{\text{d}^d p_1}{(2 \pi)^d} \frac{\text{d}^d p_2}{(2 \pi)^d} \frac{\text{d}^d p_3}{(2 \pi)^d}\\
        &\frac{e^{i \mathbf{p}_1\cdot\hat{\mathbf{b}}}+\Theta_{\text{bal}}e^{-i\mathbf{p}_1\cdot \hat{\mathbf{b}}}}{\mathbf{p}_1^2} \frac{e^{i \mathbf{p}_2\cdot \hat{\mathbf{b}}}+\Theta_{\text{bal}}e^{-i\mathbf{p}_2\cdot \hat{\mathbf{b}}}}{\mathbf{p}_2^2} \frac{e^{i \mathbf{p}_3\cdot \hat{\mathbf{b}}} + \Theta_{\text{bal}}e^{-i\mathbf{p}_3\cdot \hat{\mathbf{b}}}}{\mathbf{p}_3^2}\left(\mathbf{p}_1+\mathbf{p}_2+\mathbf{p}_3\right)^2\\
        &\left(2\Theta(w_1 - w_2)\Theta(w_2-w_3)-\Theta(w_2 - w_1)\Theta(w_2-w_3)\right) \, ,
    \end{aligned}\\[\baselineskip]
    &\begin{aligned}
        \tilde{a}^{(3)} = &2\left(-\frac{2 \pi^{\frac{d}{2}}}{\Gamma\left(\frac{d-2}{2}\right)}\right)^3 \int_0^1 \text{d}w_1 \text{d}w_2 \text{d}w_3 \int  \frac{\text{d}^d p_1}{(2 \pi)^d} \frac{\text{d}^d p_2}{(2 \pi)^d} \frac{\text{d}^d p_3}{(2 \pi)^d}\\
        &\frac{e^{i \mathbf{p}_1\cdot \hat{\mathbf{b}}} + \Theta_{\text{bal}}e^{-i\mathbf{p}_1\cdot \hat{\mathbf{b}}}}{\mathbf{p}_1^2} \frac{e^{i \mathbf{p}_2\cdot \hat{\mathbf{b}}} + \Theta_{\text{bal}}e^{-i\mathbf{p}_2\cdot \hat{\mathbf{b}}}}{\mathbf{p}_2^2} \frac{e^{i \mathbf{p}_3\cdot\hat{\mathbf{b}}}+\Theta_{\text{bal}}e^{-i\mathbf{p}_3\cdot \hat{\mathbf{b}}}}{\mathbf{p}_3^2}\left(w_1\mathbf{p}_1+w_2\mathbf{p}_2+w_3\mathbf{p}_3\right)^2\\
        &\left(2\Theta(w_1 - w_2)\Theta(w_2-w_3)-\Theta(w_2 - w_1)\Theta(w_2-w_3)\right) \, .
    \end{aligned}
\end{align}
\end{subequations}


\bibliographystyle{JHEP}
\bibliography{references}

\end{document}